\documentclass{article}

\usepackage{arxiv}
\usepackage{amsmath}
\usepackage[utf8]{inputenc}
\usepackage[T1]{fontenc}  
\usepackage{hyperref}      
\usepackage{url}          
\usepackage{booktabs}     
\usepackage{amsfonts}      
\usepackage{nicefrac}      
\usepackage{microtype}     
\usepackage{cleveref}      
\usepackage{lipsum}
\usepackage{graphicx}
\usepackage{natbib}
\usepackage{doi}
\usepackage{siunitx}
\usepackage{float}
\usepackage{enumitem}
\usepackage[export]{adjustbox}
\usepackage{xcolor}
\usepackage{booktabs}
\usepackage{subcaption}
\usepackage[vlined,ruled,linesnumbered,noend]{algorithm2e}
\usepackage[normalem]{ulem}
\usepackage{listings}
\usepackage{graphicx}
\usepackage{caption}
\usepackage{adjustbox}
\usepackage{tikz}
\usepackage{xfakebold}
\usepackage{mathtools}
\usepackage{enumitem}
\usepackage{breakcites}

\SetKwProg{myproc}{Procedure}{}{}

\DontPrintSemicolon

\let\oldnl\nl
\newcommand{\nonl}{\renewcommand{\nl}{\let\nl\oldnl}}

\makeatletter
\newcommand{\removelatexerror}{\let\@latex@error\@gobble}
\makeatother

\makeatletter
\lst@Key{countblanklines}{true}[t]%
{\lstKV@SetIf{#1}\lst@ifcountblanklines}

\lst@AddToHook{OnEmptyLine}{%
	\lst@ifnumberblanklines\else%
	\lst@ifcountblanklines\else%
	\advance\c@lstnumber-\@ne\relax%
	\fi%
	\fi}
\makeatother

\newcommand{\ES}{\mbox{\sc equally-split}}
\newcommand{\Full}{\mbox{\sc full}}
\newcommand{\Partial}{\mbox{\sc partial}}

\newcommand{\DA}{\mbox{\sc density-aware}}
\newcommand{\St}{\mbox{\sc static}}
\newcommand{\Dn}{\mbox{\sc dynamic}}
\newcommand{\PSU}{\mbox{\sc predict-st-unsorted}}
\newcommand{\PS}{\mbox{\sc predict-st}}
\newcommand{\PD}{\mbox{\sc predict-dn}}
\newcommand{\WS}{\mbox{\sc work-steal}}
\newcommand{\WSP}{\mbox{\sc work-steal-predict}}
\newcommand{\DPISAX}{\mbox{\sc dpisax}}
\newcommand{\DMESSI}{\mbox{\sc dmessi}}
\newcommand{\DMESSIBSF}{\mbox{\sc dmessi-sw-bsf}}

\SetKw{Continue}{continue}
\SetKw{Break}{break}
\SetKw{WaitUntil}{wait until}
\SetKw{And}{and}
\SetKw{Or}{or}
\SetKw{Mod}{mod}
\SetKw{Allocate}{allocate}
\SetKw{Execute}{execute}
\SetKw{Using}{using}
\SetKw{Replacing}{replacing}
\SetKw{Adding}{adding}
\SetKw{Add}{add}
\SetKw{To}{to}
\SetKw{Each}{each}
\SetKw{Do}{do}
\SetKw{CASS}{CAS}

\newcommand{\comnospace}{\mbox{$\triangleright$}}
\newcommand{\com}{\mbox{\comnospace\ }}

\newcommand{\ignore}[1]{}
\newcommand{\remove}[1]{}

\definecolor{veryperi}{RGB}{104, 104, 167}
\definecolor{burnishedlilac}{RGB}{193, 171, 175}
\definecolor{driedmoss}{RGB}{201, 186, 132}
\definecolor{granitegree}{RGB}{138, 159, 150}
\definecolor{grey199}{RGB}{199, 199, 199}

\usetikzlibrary{tikzmark}

\newcommand{\fbseries}{\unskip\setBold\aftergroup\unsetBold\aftergroup\ignorespaces}
\makeatletter
\newcommand{\setBoldness}[1]{\def\fake@bold{#1}}
\makeatother

\title{Odyssey: A Journey in the Land of Distributed Data Series Similarity Search}

\date{}

\author{ 
	Manos Chatzakis\\
	EPFL \\
	\texttt{emmanouil.chatzakis@epfl.ch} \thanks{Work conducted %
		in collaboration with ICS-FORTH and Universit{\'e} Paris Cit{\'e}}\\
	\And
	Panagiota Fatourou \\
	FORTH, ICS \& University of Crete, CSD \\
	\texttt{faturu@ics.forth.gr}  \thanks{Part of work performed while P. Fatourou was working as a Marie Sklodowska-Curie Individual Fellow (project PLATON, GA No 101031688) at Universit{\'e} Paris Cit{\'e}, France.}\\
	\And
	Eleftherios Kosmas \\
	FORTH, ICS \& Hellenic Mediterranean University \\ \& University of Crete, CSD \\
	\texttt{ekosmas@csd.uoc.gr} \\
	\And
	Themis Palpanas \\
	Universit{\'e} Paris Cit{\'e} \& IU \\
	\texttt{themis@mi.parisdescartes.fr} \\
	\And
	Botao Peng \\
	Institute of Computing Technology \\
	Chinese Academy of Sciences \\
	\texttt{pengbotao@ict.ac.cn} \\
}

\renewcommand{\shorttitle}{}

\hypersetup{
pdftitle={A template for the arxiv style},
pdfsubject={q-bio.NC, q-bio.QM},
pdfauthor={David S.~Hippocampus, Elias D.~Striatum},
pdfkeywords={First keyword, Second keyword, More},
}

\begin{document}
\maketitle

\begin{abstract}
This paper presents Odyssey, a novel {\it distributed} data-series processing framework
that efficiently addresses 
the critical challenges of exhibiting good speedup and ensuring  high scalability 
in data series processing by  
taking advantage of the full computational capacity of modern distributed systems comprised of multi-core servers. 
%
Odyssey addresses a number of challenges in designing efficient and 
highly-scalable {\it distributed} data series index, including efficient scheduling, and load-balancing
without paying the prohibitive cost of moving data around. 
It also supports a flexible
partial replication scheme, which enables Odyssey to navigate through a fundamental trade-off between data scalability 
and good performance during query answering. 
Through a wide range of configurations and  
using several real and synthetic datasets, our experimental analysis demonstrates that 
Odyssey achieves its challenging goals.

This paper appeared in PVLDB 2023, Volume 16.
\end{abstract}

\section{Introduction}
\label{sec:intro}

\noindent\textbf{Motivation.}
Processing large collections of real-world data series is nowadays one of the most challenging 
and critical problems for a wide range of diverse application domains, including  finance, astrophysics, neuroscience, engineering, and 
others~\cite{DBLP:journals/sigmod/Palpanas15, fulfillingtheneed,Palpanas2019}. 
Such applications produce big collections of ordered sequences of data points, 
called {\em data series}. 
When data series collections are generated, they need to be analyzed in order to extract 
useful knowledge~\cite{kashino1999time,ye2009time,huijse2014computational,raza2015practical,norma,series2graph,valmodjournal,iedeal}.
This analysis usually encompasses answering {\em similarity search} queries~\cite{DBLP:journals/sigmod/Palpanas15,lernaeanhydra,lernaeanhydra2},
which are useful in a variety of downstream analysis tasks~\cite{DBLP:conf/wims/EchihabiZP20,DBLP:conf/edbt/EchihabiZP21,DBLP:journals/pvldb/EchihabiPZ21}.
Moreover, several applications across domains are very sensitive to the accuracy 
of the results~\cite{Palpanas2019,DBLP:journals/dagstuhl-reports/BagnallCPZ19}, and thus, require exact query answering~\cite{lernaeanhydra},
which is our focus.

As the size of the data series collections grows larger~\cite{DBLP:journals/sigmod/Palpanas15,DBLP:conf/ieeehpcs/Palpanas17,Palpanas2019}, 
recently proposed State-of-the-Art (SotA) data series indexes 
exploit parallelism 
through the use of multiple threads 
and the utilization of the SIMD capabilities of modern hardware~\cite{peng2018paris,parisplus,hercules}.  
However, the unprecedented growth in size that
data series collections experience nowadays, 
renders even SotA parallel data series indexes inadequate~\cite{DBLP:conf/ieeehpcs/Palpanas17,lernaeanhydra,
lernaeanhydra2,Palpanas2019,DBLP:journals/dagstuhl-reports/BagnallCPZ19,gogolou2019progressive,pros}, mainly due to the large number of random disk page reads required for exact query answering~\cite{lernaeanhydra}. 
To address these issues, fast in-memory solutions have been proposed~\cite{peng2020messi,messijournal,peng2021sing}. 
However, these solutions do not take advantage of distributed systems, and hence, are limited by the amount of memory of a single machine.
This is the limitation we address, thus allowing the above SotA solutions to handle datasets that far exceed the main memory capacity of any single node.

\noindent\textbf{Challenges.} 
In the context of data series similarity search, exact query answering is very demanding in terms of resources, even when using a data series index.
We need to either prune, or visit every 
leaf of the index. 
Previous works~\cite{lernaeanhydra,gogolou2019progressive} though, have shown that pruning is not very effective, 
especially for some hard datasets. 

The main goal we need to satisfy is (naturally) \emph{scalability}. 
That is, increasing the available hardware resources (e.g., the number of nodes) should decrease the time cost,
ideally by an equivalent amount, or should enable to process an equivalent amount of additional data (at about the same time cost).
In order to meet this goal, we need to ensure that 
all nodes of the distributed system equally contribute 
to completing the work, during the entire duration of the execution.
In turn, this translates to producing effective solutions to the following two problems: 
(i) {\em query scheduling}: given a query workload, decide which queries to assign to each system node; 
and (ii) {\em load-balancing}: devise mechanisms so that system nodes that have finished their work can help other system nodes finish theirs.

The challenges in this context are the following. 
First, to achieve effective query scheduling, we need to come up with mechanisms for estimating 
the execution cost of data series similarity search queries, which do not currently exist. 
Second, this observation renders a load balancing scheme necessary, 
yet, this also means that we need to replicate data in order to make such a mechanism viable,
as moving big volumes of data series around would be prohibitively expensive.
Data replication works against data scalability and is more costly in whatever regards
index creation time, but results in better query answering times, thus leading
in interesting trade-offs through which an effective solution should navigate. 
Third, along with all the above considerations, we also need to ensure that our solutions 
will still maintain their good parallelization properties for efficient execution 
in multi-core CPUs inside each system node, 
and also achieve high pruning power during query answering.

\noindent\textbf{Our Approach.} 
We propose a novel {\it distributed} data-series (DS) indexing and processing framework,
called {\em Odyssey}, that efficiently addresses 
the high scalability objective by  
taking advantage of the full computational capacity of the computing platform. 
%

To come up with an appropriate scheduling scheme for Odyssey, 
we performed a query analysis that shows correlation between the total execution 
time and a parameter
of the category of the single-node data series indexes we consider.
This analysis drove the design of efficient scheduling schemes, 
by generating an execution time prediction for each query of the input query batch. 

To achieve Load Balancing (LB) even in settings where predictions may not be accurate, 
Odyssey provides a LB mechanism, 
which ensures 
that nodes sitting idle can take away (or {\em steal}) work from other nodes which have still work 
to do (provided that these nodes store similar data). 
Combining Odyssey scheduler with this LB technique
results in very good performance and high scalability for all 
query batches we experiment with. 

Ensuring data scalability and, at the same time, good performance for query answering
are contradicting goals. A scheme where data are not replicated 
would result in the lowest space overhead, but experiments show that 
%
this technique does not ensure the best performance during query answering, because no data replication means that Odyssey's LB mechanism cannot be used. 
%

Odyssey manages to effectively unify these two contradicting goals by supporting a flexible
{\it partial replication scheme}. 
This way, it navigates through the fundamental trade-off between data scalability 
and good performance during query answering. 
The degree of replication is one of Odyssey's parameters. 
By specifying it appropriately,
users can choose the time-space trade-off that best suits their application and setting. 
%
Experiments show that Odyssey achieves good performance even for small replication degrees.


Supporting the components for efficient distributed computation 
that Odyssey provides, on top of an index that exploits the 
computation power of a single node as efficiently as SotA parallel 
indexes~\cite{peng2020messi,messijournal,peng2021sing}, was one more challenging task we undertook 
while designing Odyssey. 
A simple approach of using an instance of the SotA MESSI  index~\cite{messijournal} 
in each node did not result in good performance 
mainly due to two reasons.
First, different data series queries may exhibit variable degrees of locality (revealed only at runtime), resulting in low pruning in some of the nodes, and thus, in severe load balancing
problems and performance degradation. 
Second, supporting load-balancing on top of such a simple approach 
would require moving data around, which is often prohibitively expensive. 
Odyssey {\em single-node} indexing scheme borrows 
some techniques from SotA indexes~\cite{peng2018paris,parisplus,peng2020messi,messijournal,peng2021sing}, and couples those 
with  
new components and mechanisms, 
to achieve load balancing and come up with a scheme 
in which work from an overloaded node can be given away to idle nodes
without having to pay the prohibited cost of moving any data around.
%

Odyssey is innovative in different ways.
First, it employs a different pattern of parallelism from all existing approaches in traversing the index tree to produce the set of
data series that cannot be pruned. 
Second, it presents new implementations for populating and processing the
data structures needed for efficient query answering. 
To achieve load balancing among the threads, it is critical to choose
an appropriate {\em threshold} on the size of these data structures, and 
Odyssey proposes an effective mechanism for 
predicting a good threshold. 
Additionally, Odyssey
provides efficient communication and book-keeping mechanisms, 
to enable fast exchange of information among nodes to ensure good pruning degrees 
in all of them. 

Odyssey 
is up to $6.6$x faster than its competitors and more than
$3.5$x better than its best competitor. 
Additionally, Odyssey's index creation
perfectly scales with both the dataset size and the number of node. 
Moreover, Odyssey's best performing scheduling strategy is more than $2.5$x  faster than its 
initial one.

\noindent\textbf{Contributions.}
The main contributions of the paper are as follows: 

\noindent$\bullet$
We describe Odyssey, a scalable framework for distributed data series similarity search in clusters with multi-core servers. 
This makes our approach the first customized data series solution that exploits parallelization both inside and across system nodes.

\noindent$\bullet$
We develop a scheduling algorithm for assigning queries to the nodes of the cluster, which tries to balance the workload
across the nodes by computing a (good-enough) estimation of the execution time of each query.

\noindent$\bullet$
We present a novel exact search algorithm that supports {\em work-stealing} 
between nodes that share the same index (full replication).
Thus, our approach leads to high performance, even when the work is not (or cannot be) equally distributed over the nodes of the cluster.
We further extend our solution to work even when only a part of the index is shared among nodes (partial replication).

\noindent$\bullet$
Our approach supports different {\em replication} 
degrees among the nodes, allowing users to navigate the entire spectrum of solutions, trading space (replication degree) for speed (query answering time).

\noindent$\bullet$
We also present a density-aware data partitioning method that can efficiently partition data in a way that improves the work balancing capabilities of our approach. 

\noindent$\bullet$
Finally, we conduct an experimental evaluation (code and data available online~\cite{codedata}) with a wide range of configurations, 
using real and synthetic datasets. 
The evaluation demonstrates the efficiency of Odyssey, which exhibits an almost linear scale-up, and up to 6.6x 
times faster exact query answering times than the competitors.

\section{Preliminaries and Related Work}
\label{sec:prelim}

\noindent\textbf{Data Series.} 
A {\em data series}, denoted as $ S = \{ p_1, ... , p_n \}$,  
is a sequence of points, where each point $p_i$ is a pair $(u_i,t_i), 
1 \leq i \leq n $, of a real value $u_i$ and the position $t_i$ of $p_i$ in the sequence; $n$ is the {\em size} 
(or {\em dimensionality}) of the sequence.
When $t_i$ represents time, we talk about {\em time series}. 
In several cases, we omit the $t_i$, e.g., when they are equally spaced, or only play the role of an index for the values 
$u_i$~\cite{lernaeanhydra}; for simplicity, we omit them, as well.

\begin{figure}
	\centering
	\begin{subfigure} [b]{0.27\textwidth}
		\centering
		\includegraphics[width=\textwidth]{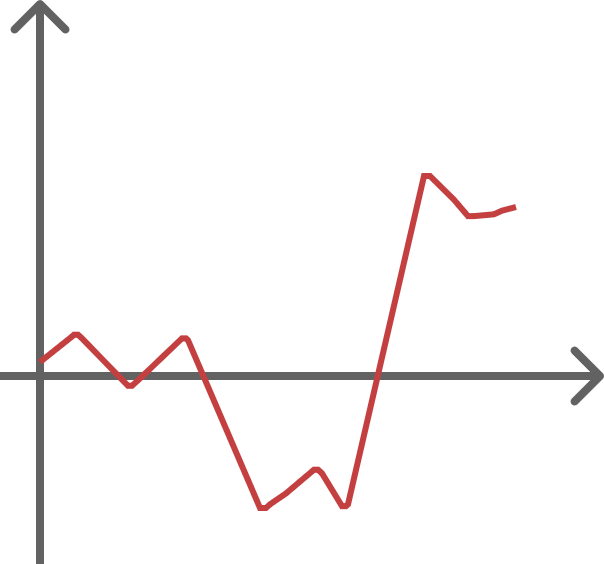}
		\caption{Data Series}
	\end{subfigure}
	\begin{subfigure} [b]{0.27\textwidth}
		\centering
		\includegraphics[width=\textwidth]{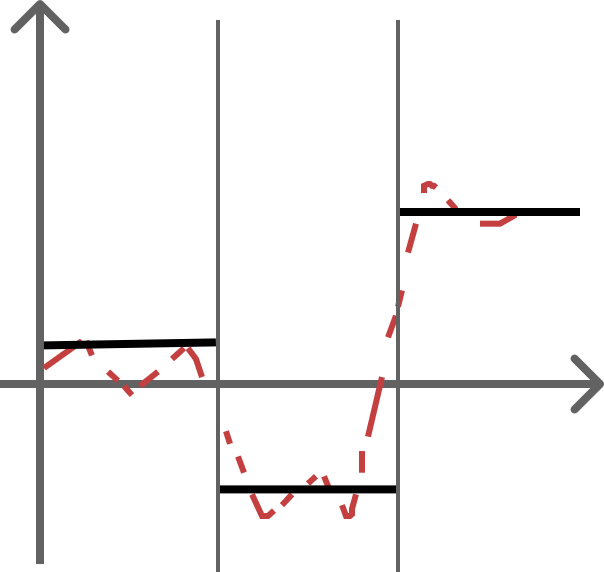}
		\caption{PAA Summary}
		\label{}
	\end{subfigure}
	\begin{subfigure} [b]{0.35\textwidth}
		\centering
		\includegraphics[width=\textwidth]{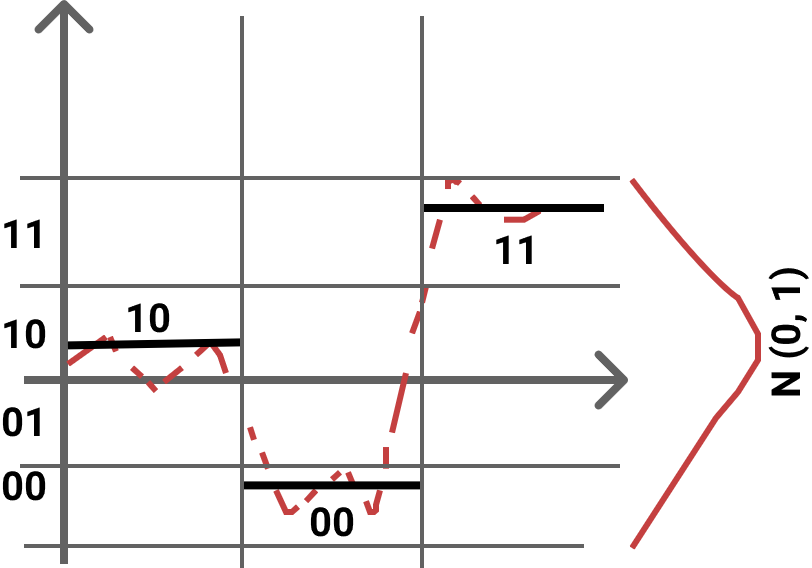}
		\caption{iSAX Summary}
		\label{}
	\end{subfigure}
	\begin{subfigure} [b]{0.6\textwidth}
		\centering
		\includegraphics[width=0.8\textwidth]{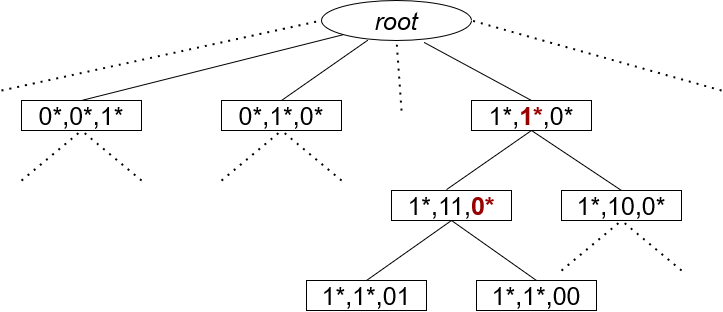}
		\caption{iSAX Tree}
		\label{}
	\end{subfigure}
	\caption{From data series to iSAX index}
	\label{fig:from_ds_to_iSAX}
\end{figure}

\noindent\textbf{iSAX Summary.} 
The {\em iSAX summary}~\cite{iSAX} of a data series splits the x-axis 
in equal segments and represents each segment with the mean value of the points of the data series that it contains
(see Figure~\ref{fig:from_ds_to_iSAX}). 
Then it partitions the y axis into regions of sizes determined by the normal distribution 
and represents each region using a number of bits ({\em cardinality}). 
The number of bits can be different for each region, and this enables 
the creation of a hierarchical index tree ({\em iSAX-based index tree}~\cite{isaxfamily}; 
see Figure~\ref{fig:from_ds_to_iSAX}). 

\noindent\textbf{Similarity Search.} Given a collection of data series $\mathcal{C}$ and an input data series $S$, called the {\em query},
{\em similarity search} is the task of finding the data series in $\mathcal{C}$ which are most similar to $S$. 
We focus on finding a single best answer, known as the {\em 1-NN} problem.
We also focus on Euclidean Distance (ED). 
The {\em euclidean distance} (or {\em real distance}) 
between two time series $T = \{t_1, ... , t_n\}$ and $S = \{s_1, ... , s_n\}$ 
is defined as $ ED(T,S) = \sqrt{\sum_{i=1}^{n} (t_i - s_i)^2}$. 
We call the distance between the {\em iSAX summaries} of $T$  and $S$, {\em lower-bound distance}.
The lower-bound distance between any two data series is always smaller than or equal to the real distance between them. 

\noindent\textbf{Single-Node Parallel Summary-Based DS Indexing.} 
Such indexes~\cite{peng2018paris,parisplus,peng2020messi,messijournal,peng2021sing,hercules,elpis} exploit multiple threads (and SIMD) 
to create an index tree and answer queries on top of this tree. 
They are usually comprised of two main phases, the {\em index tree construction} and the {\em query answering} phases. 
In the index tree construction phase, they first calculate, in parallel, summarizations of 
all data series in the collection. If the summarizations are iSAX summaries, we talk about
{\em iSAX-based DS indexing}.
To achieve a good degree of locality and low synchronization overheads, 
they store these summaries into a set of {\em summarization buffers}. 
Data series that have similar summarizations are placed into the same buffer. 
Subsequently, the data series of each of these buffers are
stored into each of the subtrees of the index tree that they construct. 
These design decisions allow them to build the index tree in an almost embarrassingly parallel 
way (thus, without incurring synchronization overheads), and achieve locality in accessing 
the data during tree construction. 
They thus respect crucial principles for achieving
good performance that should be respected when designing a parallel index. 

\begin{figure}[tb]
	\centering
	\begin{subfigure} [b]{0.5\textwidth}
		\centering
		\includegraphics[width=\textwidth]{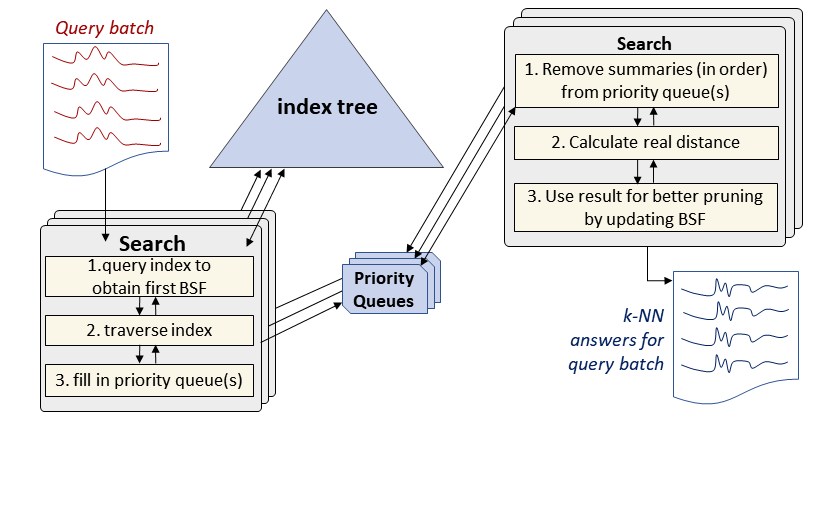}
	\end{subfigure}
	\caption{Algorithm Outline of Parallel DS Indexes.}
	\label{fig:messi_outline}
\end{figure}

To answer a query, these indexes first calculate the summarization of the query. Subsequently, 
they traverse the index tree to find the most appropriate data series based on the iSAX summary lower bound distances. 
The distance of these data series from the query series is stored in a variable called best-so-far (BSF),
and serves as an initial approximate answer to the active query. 
Then, BSF is used to prune data series from the initial collection. 
A data series $S$ is pruned when the lower bound distance between $S$ and the query 
is higher than the current value of the BSF. This process outputs a hopefully small
subset of the initial DS collection, containing series that need to be further examined. 
These series are often stored in (one or more) priority queues~\cite{peng2020messi,messijournal,peng2021sing}. 
Multiple threads process, concurrently, the elements of the priority queues,
calculating real distances (if needed), and updating the BSF each time 
a new minimum is met (see Figure~\ref{fig:messi_outline}).
Once this process completes, the distance to the answer is contained in BSF. 

\noindent\textbf{Multi-node Systems and Query Processing.} 
The system consists of a number of asynchronous nodes 
which communicate by exchanging messages. Each node is a multi-core machine, 
capable to support multiple threads (and possibly SIMD computation). 
Threads communicate by accessing shared variables. 
A shared variable can be atomically read and written. 
Stronger primitives, such as Fetch\&Add may also be provided. 
Fetch\&Add($V,val$) atomically adds the value $val$ to the current value of variable $V$
and returns the value that $V$ had before this update.    

An arbitrarily large batch of queries is provided in the system as input. 
The goal is to utilize the system's computational power 
to execute these queries in a way that minimizes the {\em makespan},
i.e., the length of time that elapses from the time that any node starts processing
a query of the batch to the first point that all nodes have completed their computation.
Our techniques can easily be adjusted to work with 
queries that arrive in the system dynamically. 

The data series in the initial collection can be stored in all nodes ({\em full replication}),
or may be scattered to the different nodes so that nodes store disjoint subsets of the data
({\em no replication}). A {\em partial replication} scheme is also possible, where
nodes store subsets of the data which are not necessarily pairwise disjoint
(e.g., more than one node may store the same subset of data series). 
A {\em data partitioning} mechanism determines how to split and distribute the data 
of the initial data-series collection to nodes.


%

Query scheduling algorithms aim to schedule the input queries 
to nodes in a way that each node has approximately 
the same amount of work to do. 
Considering full replication, a {\em Static Query Scheduler (SQS)} partitions 
the sequence of queries into $N$ subsequences and each node gets one of these subsequences to answer. 
A {\em Dynamic Query Scheduler (DQS)} employs a coordinator node, and has other nodes requesting
queries to execute from the coordinator. 
The coordinator may serve requests by assigning the next unprocessed query to a worker when it receives its request, 
or it may preprocess the sequence of queries (e.g., by re-arranging the queries based on some property) 
before it starts assigning queries to nodes. 
To avoid loosing computational power, the coordinator can answer queries itself 
between serving requests from other nodes. 

\subsection{Related Work}
\label{section:related}

Data series similarity search queries require the use of specialized index structures in order to be executed fast on very large collections of data sequences. 
In general, data series indexes operate by pruning the search space based on the summarizations of the series and corresponding lower bounds, 
and only use the raw data of the series in order to filter out the false positives. 

\noindent\textbf{Data Series Indexes.}
Agrawal et al.~\cite{DBLP:conf/fodo/AgrawalFS93} presented the first work that argued for the use of a spatial indexing structure for indexing data sequences, 
based on the R-Tree~\cite{DBLP:conf/vldb/SellisRF87}, 
and was later optimized~\cite{DBLP:conf/fodo/RafieiM98}. 
Various indices, specific to data sequences, have been proposed in the literature~\cite{DBLP:conf/edbt/EchihabiZP21}. 
DSTree~\cite{DBLP:journals/pvldb/WangWPWH13} is an index based on the APCA summarization~\cite{DBLP:conf/sigmod/KeoghCMP01}. 
The DSTree can adaptively perform split operations by increasing the detail of APCA as needed. 
The iSAX index is based on the SAX summarization, and its extension, iSAX~\cite{iSAX}. 
In this case, the data series summarization is bitwise, leading to a concise representation and overall index. 
Several other iSAX-based indices have been proposed in the literature~\cite{journal/kais/Camerra2014,DBLP:conf/sigmod/ZoumpatianosIP14,zoumpatianos2016ads,ulissevldb,ulissejournal,isaxfamily,DBLP:conf/kdd/WangP21,c23-sigmod-wang-dumpy}. 
These indexes are among the SotA solutions in this area~\cite{lernaeanhydra}, 
including MESSI~\cite{peng2020messi,messijournal}, an in-memory, multi-core and SIMD-enabled version of the iSAX index. 

\noindent\textbf{Data Series Management Systems.}
Several data series management systems have been developed in the last few 
years~\cite{DBLP:journals/tkde/JensenPT17,DBLP:conf/edbt/EchihabiZP21}.
Beringei~\cite{DBLP:journals/pvldb/PelkonenFCHMTV15} has a custom in-memory storage engine. It compresses and organizes data in a series per series scheme. 
CrateDB~\cite{Crate} partitions data in chunks, stores them in a distributed file system, and indexes them using Apache Lucene.
InfluxDB~\cite{InfluxDB} uses Time-Structured Merge Trees (LSM tree variant). 
Prometheus~\cite{Prometheus} is based on the Beringei ideas.
QuasarDB~\cite{Quasar} utilizes either RocksDB or Hellium~\cite{Hellium}.  
Riak TS~\cite{RiakTS} supports both LevelDB or Bitcask, which is a custom log structured hash table.
Timescale~\cite{timescale} is a Postgres extension. 
IoTDB~\cite{DBLP:journals/pvldb/0018HQ00MFZ0ZKJ20} is geared towards streaming data series.
Finally, various systems such as OpenTSDB~\cite{OpenTSDB}, Timely~\cite{Timely} (concentrated on security) and Warp10~\cite{Warp10} are developed on top of HBase.
All the aforementioned systems support range scans in the positions, aggregation functions and filtering.
InfluxDB supports queries like moving averages, prediction, transformations, etc, and
Timescale supports gap filling.
Nevertheless, none of the above systems supports exact whole-matching similarity search queries.

\noindent\textbf{Distributed Data Series Indexes.}
KV-Match~\cite{DBLP:conf/icde/WuWPWWW19} and its improvement, L-Match~\cite{DBLP:journals/access/FengWWW20}, are index structures that can support similarity search. 
These indices can be implemented on top of Apache HBase, and operate in a distributed fashion within Apache Spark.
We note that these solutions only support subsequence similarity search, and not whole-matching~\cite{lernaeanhydra}, which is the focus of our paper.
TARDIS~\cite{DBLP:conf/icde/ZhangAER19} is an Apache Spark system for similarity search. 
It supports approximate queries, as well as \emph{exact match} queries, where we want to know if the query appears \emph{exactly the same} within the dataset, or not. 
This query type is much easier than the exact queries we consider in our work, and cannot be efficiently transformed to exact querying. 
Finally, DPiSAX~\cite{DPiSAX,dpisaxjournal} is a distributed solution for data series similarity search, developed for Apache Spark using Scala. 
It was designed for answering batches of approximate search queries, but also supports exact search. 
DPiSAX exploits the iSAX summaries of a small sample of the dataset, in order to distribute the data to the nodes equally. 
Then, an iSAX index is built in each node on the local data, and is used to perform query answering.
In order to produce the exact search results, all nodes need to send their partial results to the coordinator, which merges them and produces the final, exact answer.
Note that DPiSAX was not explicitly designed for intra-node parallelization, but is 
the only distributed data series index in the literature that supports exact search.

Work-stealing was employed in the Cilk framework~\cite{Cilk}. The work-stealing approach 
was formally studied and analyzed in~\cite{BL99,FS99,FS00}. 
Lots of work has been done on this topic (e.g.,~\cite{BG99,10.1145/258492.258494,10.1145/378580.378639}). 

\section{The Odyssey Framework}

\begin{figure*}[tb]
	\centering
		\centering
		\includegraphics[width=0.9\textwidth]{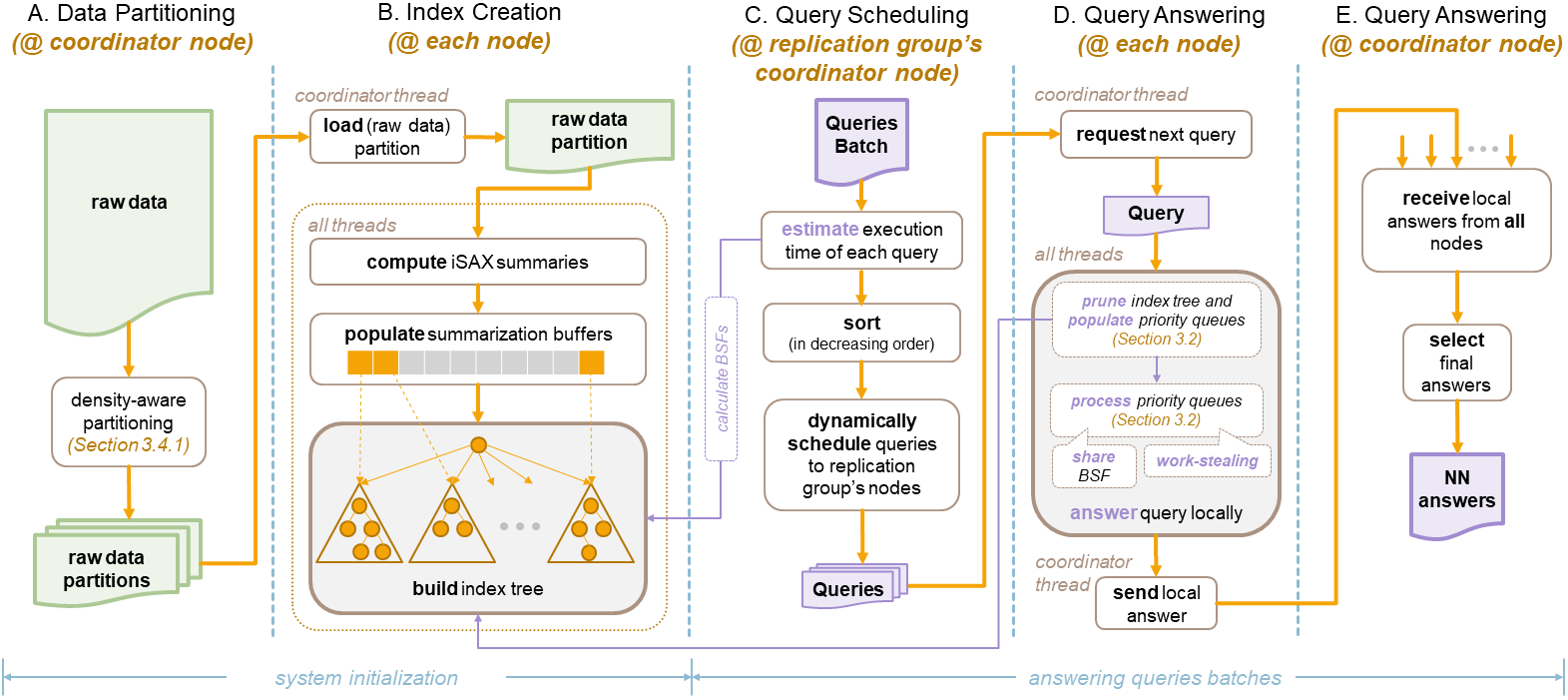}
	\caption{Odyssey flowchart.}
	\label{fig:odyssey-flow}
\end{figure*}

We start with a high level overview of the Odyssey flowchart, 
which comprises of five stages (see Figure~\ref{fig:odyssey-flow}).

In the first stage, 
a {\em coordinator node} partitions the raw data-series collection
to as many chunks as the number of system nodes, and assigns 
a chunk to each node (including itself). 
(Section~\ref{sec:data-part} details
Odyssey's partitioning schemes.)
In the second stage,  
each node (i) loads its chunk of data in memory, (ii) computes their iSAX 
summaries and stores them into a number of {\em summarization buffers}, for achieving locality,
and (iii) builds its {\em index tree}. 
To enhance performance at query answering, Odyssey employs data replication. 
It 
forms groups of nodes ({\em replication groups}, described in Section~\ref{sec:data-repl}), where all nodes of each group store the same chunk of data. 
Each replication group has a coordinator node, called {\em group coordinator}, 
which schedules queries to the group's nodes. 
A batch of queries (e.g., originating from a \emph{k-NN} classification task) to execute is submitted to all group coordinators
(as different groups store different data chunks). 
%
%
%
%
In the third stage, the group coordinators start by estimating the
execution time of each query, 
then sort queries in descending order of estimated execution times, 
and {\em dynamically} schedule them to the group's nodes (Section~\ref{sec:sched} describes query scheduling).  
In the fourth stage, each 
node processes the queries assigned to it. 
It first calculates an
initial BSF, and then prunes the index tree using this BSF, populating the priority queues with leaves that cannot be
pruned. 
Finally, it processes the elements of the priority queues to find the best local answer (corresponding to its data chunk).
In this stage, Odyssey supports {\em BSF-sharing} and {\em work-stealing} (detailed in Section~\ref{sec:load-balancing}). 
In the last stage, the coordinator node collects the local answers from the group coordinators, 
and produces the final answers.

\subsection{Query Scheduling}
\label{sched}
\label{sec:sched}

To correctly answer a query, it should be forwarded to at least one set of system nodes 
that collectively store all the data. We call such sets node \emph{clusters} in Section~\ref{sec:data-repl}. 
Thus, in the no-replication case
this set contains all system nodes, 
so 
a scheduling algorithm should forward all queries 
to all nodes. Other replication settings (and especially full replication) 
are more interesting,
as they enable the utilization of different scheduling techniques. 

To come up with Odyssey scheduler, 
we experimented with a collection of
scheduling techniques, including the simple static and dynamic schemes (SQS and DQS)
for full replication settings, 
discussed in Section~\ref{sec:prelim}. 
Unfortunately, these schemes suffer from severe load 
imbalance problems for many 
categories of query batches. 
For the static case, consider for example, a query sequence 
which consists of progressively more {\em difficult} queries (i.e., 
of queries that each requires less time to run 
than the next one). 
SQS will assign to the first system nodes easy queries, 
while the last nodes will get more work to do. 
The dynamic method (DQS) 
may also result in
load imbalances: even in simple cases where e.g., a query batch includes a single difficult
query at the end, most nodes may be sitting idle, while a single node is running the difficult query.
This may significantly degrade performance. 

\begin{figure}[tb]
	\centering
	\begin{subfigure} [b]{0.5\textwidth}
		\centering
		\includegraphics[width=0.9\textwidth]{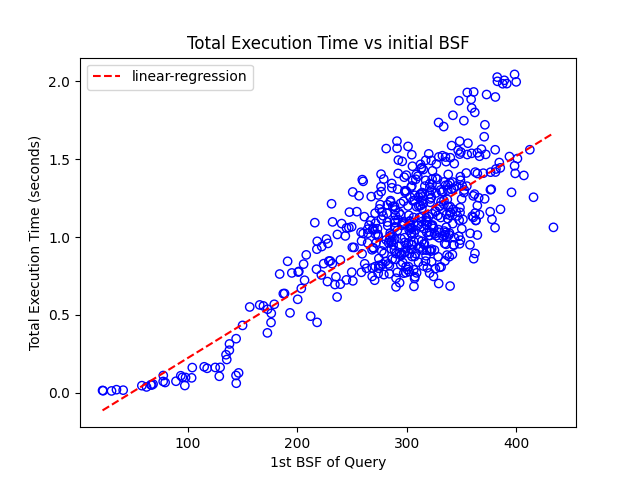}
	\end{subfigure}
	\caption{Linear regression for Seismic queries prediction.}
	\label{fig:linear_regression}
\end{figure}

Some of these load imbalances could be avoided, if we knew the
execution time of each query. Recent work~\cite{progressiveISAX,pros}
illustrated that there exists a correlation
between the initial 
BSF and the number of vertices
visited in a single-node index tree. 
We performed a corresponding query analysis which showed 
that similarity search queries, for which the {\em initial BSF} is high, 
tend to also have high execution times. 
In this work, we use a linear regression model (other pediction schemes can be used, as well) to produce estimates for each query.
An example of this outcome is shown in Figure~\ref{fig:linear_regression} (for Seismic; we follow the same process for the other datasets).
%

These observations led us to design two scheduling algorithms. 
The first, {\em static prediction-based scheduling}, 
statically allocates the queries to nodes based on their estimations. 
Each node maintains a $\mathit{load}$ variable, 
which stores the sum of the estimations of the queries that are assigned 
to it. The algorithm uses a greedy approach to assign queries to nodes
so that load balancing is achieved. 
There are two variations of the algorithm: 
the first ({\em unsorted}) schedules the queries using their order in the sequence, and the second ({\em sorted}) sorts 
the sequence 
based on decreasing execution time estimations. 
.
The second scheduling algorithm, called {\em dynamic prediction-based scheduling}, is an enhanced version of
DQS, where queries are assigned to nodes after sorting the entire query batch, 
based on estimations (in decreasing order). 

Consider a system of two nodes, $\mathit{sn_1}$ and $\mathit{sn_2}$,
and let 
$\mathcal{Q} = \{q_1, q_2, q_3, q_4, q_5\}$ be a query batch to execute. 
Assume that $\mathcal{ES} = \{100, 50, 200, 250, 80\}$ is the set of
the estimated execution times,
where the $i$-th element of $\mathcal{ES}$ is the estimated execution time for $q_i$, $1 \leq i \leq 5$.
Unsorted static prediction-based scheduling,
with load variables $l_1$ and $l_2$ (for $\mathit{sn_1}$ and $\mathit{sn_2}$, respectively), proceeds as follows: 
$q_1$ is assigned to $\mathit{sn_1}$ (so, $l_1 = 100$), 
and $q_2$ is assigned to $\mathit{sn_2}$ (so, $l_2 = 50$). 
Since $l_2 < l_1$, $q_3$ is assigned to $sn_2$ (thus, $l_2 = 250$).
Following a similar strategy, $q_4$ is assigned to $\mathit{sn_1}$,
and $q_5$ is assigned to $\mathit{sn_2}$.
So, $sn_1$ receives $\{q_1, q_4\}$ and $sn_2$ receives $\{q_2, q_3, q_5\}$. 
In sorted static prediction-based scheduling, the queries of $\mathcal{Q}$ are first sorted 
in decreasing order of their estimated times, resulting in  
$Q' = \{q_4, q_3, q_1, q_5, q_2\}$ (which corresponds to $\mathcal{ES}' = \{250, 200, 100, 80, 50\}$). 
After applying the 
static prediction-based scheduling algorithm (as above) on these sets,
$\{q_4, q_5\}$ is assigned to $sn_1$ and $\{q_3, q_1, q_2\}$ is assigned to $sn_2$. 
Finally, dynamic prediction-based scheduling also sorts 
the queries of $\mathcal{Q}$. In this case, $q_4$ is assigned 
to $\mathit{sn_1}$, $q_3$ to $\mathit{sn_2}$, while the rest of the queries are dynamically
assigned to nodes (in order) upon request (thus, based on actual execution times). 

The Odyssey framework supports all of the above scheduling algorithms. 
The Odyssey index utilizes dynamic prediction-based scheduling, which turned out to be the best
approach in most cases. 

\remove{
\begin{figure}
	\centering
	\begin{subfigure} [b]{0.1\textwidth}
		\centering
		\includegraphics[width=\textwidth]{../resources/plots_chatzakis/theory/isax-static-qa-drawio.png}
		\caption{Static Scheduling}
	\end{subfigure}
	\begin{subfigure} [b]{0.15\textwidth}
		\centering
		\includegraphics[width=\textwidth]{../resources/plots_chatzakis/theory/isax-greedy-qa-drawio.png}
		\caption{Static Predictions Scheduling}
	\end{subfigure}
	\begin{subfigure} [b]{0.15\textwidth}
		\centering
		\includegraphics[width=\textwidth]{../resources/plots_chatzakis/theory/isax-dynamic-qa-drawio.png}
		\caption{Dynamic Scheduling}
	\end{subfigure}
	\caption{Different Distributed Query Scheduling Methods}
	\label{fig:scheduling_methods}
\end{figure}
}


\subsection{Load Balancing}
\label{load-balancing}
\label{sec:load-balancing}

Odyssey provides a load balancing (LB) mechanism, which can be applied on top
of any of the scheduling schemes described in Section~\ref{sched}. Specifically,
idle nodes 
can {\em steal} work  
from other nodes which still have work to do (provided that they
store similar data). 

This is necessary as predictions may not always be accurate, or
the query batch may be produced dynamically at run time, 
in which case sorting of the entire query batch is not possible.
It is also necessary
for achieving high scalability. 
As the number of utilized nodes increases, the number of batch queries that each node
has to process becomes smaller and smaller. Thus, problematic scenarios as those
described in Section~\ref{sched}, may appear, where just one or a few nodes work on difficult 
queries, while others 
are sitting idle. 

\noindent
{\bf Overview of our approach.}
We performed a number of experiments to get a break-down of the query answering time.
This break-down illustrated that the biggest part of the time for query answering
goes to priority queues' processing. We thus focus on designing a method that allows nodes to steal work during the execution of that phase. 
For simplicity, we first focus on the full-replication case,
where the initial collection of data is available in every node; partial replication is then discussed in Section~\ref{pr}. 

A simple work-stealing scheme~\cite{BL99,Cilk} would not work, mainly because
moving data (stored in priority queues) 
around from one node to another is expensive and should be avoided.
Thus, the main challenge in our setting is to 
take work away from one node and assign it to another
without ever moving any data around. 

Odyssey's load-balancing mechanism works as follows. 
An idle system node $sn$ randomly chooses another node $sn'$
and sends it a steal request. 
If $sn'$ has still work to do, 
it chooses a number of priority queues to give away to $sn$. 
To avoid paying the cost of transferring data around, Odyssey
employs a technique that informs $sn$ on how to locally build the priority queues
to work on, based on its own index. 
%
Node $sn$ traverses the identified part 
of its index tree and re-constructs these priority queues.
As the time to create the priority queues is relatively small
in comparison to that for processing them, this scheme
works quite well. 

Note that the approaches followed by existing SotA indexes~\cite{peng2020messi,messijournal,peng2021sing} 
for creating and processing the priority queues
are too naive to support work-stealing without moving 
any data around.
In Odyssey, we propose (in Section~\ref{sni}) a new implementation of a single-node, multi-threaded index,
which respects the good design principles described for parallel indexes in Section~\ref{sec:prelim},
while it simultaneously copes with the problem mentioned above. 


\subsubsection{Single-Node Query Answering}
\label{sni}

\begin{figure}[tb]
	\centering
	\hspace*{4cm}
	\includegraphics[scale=0.4]{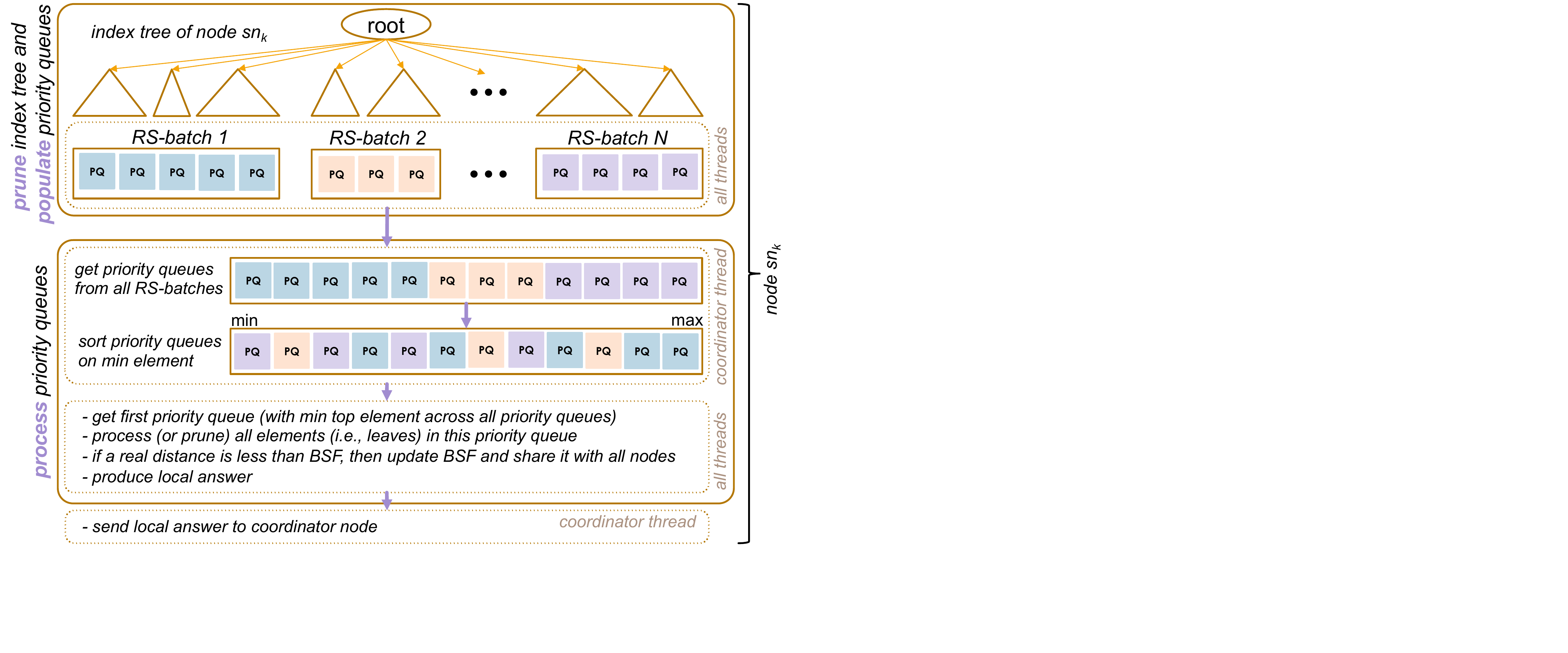}
		\caption{Outline of the Odyssey single-node query-answering process.}
	\label{fig:ws-messi}
\end{figure}

Consider any system node $\mathit{sn}$ and
assume that an iSAX-based index tree has been created and an initial value for the BSF has been computed in $sn$. 
An outline of the single-node query answering algorithm of Odyssey is depicted in Figure \ref{fig:ws-messi}.
The pseudocode is provided in Algorithms~\ref{eshq} and~\ref{workstealing_supporting_algorithm}. 

\noindent 
{\bf Description.}
Node $sn$ executes each of the queries in the query batch assigned to it 
one by one 
(Algorithm~\ref{eshq}). For each such query $Q$,
it creates a number of search workers to execute it (line~\ref{eshq:cw}). 
As soon as, all queries in $sn$'s query batch have been processed,
$sn$ informs other nodes that it has completed 
(line~\ref{eshq:done}). 
Then, it tries to help other active nodes by executing 
PerformWorkStealing (line~\ref{eshq:workstealing}). 
Each node allocates a thread to play the role of the work-stealing manager (line~\ref{eshq:ws}).
This thread simply processes all work-stealing requests that the node
will receive (Algorithm~\ref{WorkStealingManager}). 
(Work-stealing is discussed in Section~\ref{sec:wsalg}.) 

The query answering algorithm in $\mathit{sn}$ splits the tree 
into {\em root subtree (RS) batches}, i.e., sets of consecutive root subtrees (see Figure \ref{fig:ws-messi}), 
and allocates a number of threads to work on them. 
Each thread begins by getting an RS-batch to work on using Fetch\&Add (Algorithm~\ref{workstealing_supporting_algorithm}). 
Then, the thread executes the $\mathit{ProcessBatch}$ routine, 
which traverses the tree recursively and inserts the leaves 
that cannot be pruned into one of a set of priority queues that belong to the RS-batch.
%

%
%
For every RS-batch, there exists one active priority queue at each point in time. 
When the size of this priority queue surpasses a threshold, 
this queue is abandoned and another one is initialized for the RS-batch.

As soon as an idle thread $\mathit{th}$ discovers that all RS-batches have been assigned for processing, 
it tries to help some other still active thread, $\mathit{th'}$, to complete 
processing its assigned RS-batch (lines~\ref{workstealing_supporting_algorithm:help_start}-\ref{workstealing_supporting_algorithm:help_end}, 
Algorithm \ref{workstealing_supporting_algorithm}). 
\remove{
Coordination between $\mathit{th'}$ and its helpers is achieved using 
a Fetch\&Add object. This way,
each root subtree of an RS-batch is assigned to just one thread. Synchronization in accessing a priority queue
is achieved using locks. }
To reduce the synchronization cost, there is a threshold, $\mathit{HelpTH}$, on the
number of threads that help on each RS-batch (line~\ref{workstealing_supporting_algorithm:helpTH}). 
This phase ends when the subtrees of all RS-batches have been traversed
and all priority queues have been populated. 
Experiments showed that we get the best performance when 
the number of RS-batches, 
$N_{sb}$, equals the number of worker threads.

As soon as this {\em tree traversal phase} is over, we have a set of priority queues for each RS-batch, 
stored in an array. For performance reasons, this array is sorted by the priority of the top element of each priority queue.
This comprises the {\em priority queue preprocessing} phase (lines~\ref{workstealing_supporting_algorithm:prepro_start}-\ref{workstealing_supporting_algorithm:prepro_end}). 
This way, the algorithm processes the priority queues with the smallest lower bound distances to the query first.
These queues contain data series that  are more probable to be in closer real distance to the query, thus enabling further pruning. 

Then, the {\em priority queue processing} phase starts 
(lines~\ref{workstealing_supporting_algorithm:pq_pro_phase_start}-\ref{workstealing_supporting_algorithm:pq_pro_phase_end}). 
Every thread gets a priority queue from the PQueues array to process (using Fetch\&Add). 
Routine $\mathit{ProcessPriorityQueue}$ processes those data series 
stored in the priority queue, 
which cannot be pruned. Whenever a lower real time distance between any of these series and the query series
is calculated, the BSF is updated to contain this distance. 
This improved BSF is submitted to all nodes of the system. 
Finally, all answers are transmitted to the coordinator node, and the globally smallest value of the BSF is the response to the query.


\begin{algorithm}[tb]
	{	
		{\footnotesize
		\SetAlgoLined
		\textbf{Shared Variables:}
		\textbf{Shared PointerToArray} $PQueues = NULL$\; 

		\KwIn{\textbf{QuerySeriesBatch} $QBatch$, \textbf{Index} $Index$, \textbf{Integer} $NThreads$ }

		\textbf{Array} $BSFArray[]$ \com{with size $|QBatch|$}\; 

		\For{every query series id $Q$ in $\mathit{QBatch}$ \label{eshq:scq}} {

			$iSAX_{Q}$ = calculate iSAX summary for $\mathit{Q}$\;\label{eshq:calcuqisax}
			$BSF$ = approxSearch($iSAX_{Q}$, $Index$)\;\label{eshq:appro}
			create a thread to execute an instance of WorkStealingManager(Q)\;\label{eshq:ws}
			\For{$i$ $\leftarrow$ $0$ \emph{\KwTo} $NThreads-1$\label{eshq:scw}} {
				create a thread to execute an instance of  		
				SearchWorker($Q$, $Index$, $N_{sb}$, $i$, $PQueues$ )\;\label{eshq:cw}
			}\label{eshq:ecw}
			Wait for all threads to finish\;\label{eshq:finish}
		        $\mathit{FinishFlag[Q]}$ := TRUE\;
			BSFArray[Q] := $BSF$;
		}
		send(DONE, $sn$) to all nodes\; \label{eshq:done}
		PerformWorkStealing()\; \label{eshq:workstealing}
		\Return ($BSFArray$)\;\label{eshq:return}
	}
} 
\caption{Odyssey Single-Node Query Answering - Code for node $sn$}
\label{eshq}
\end{algorithm}

\begin{algorithm}[t]
{
	{\footnotesize
\SetAlgoLined
		
\com{Shared Variables} \\
\textbf{Shared Integers} $BCnt$ = 0, $PQCnt$ = 0,  $TotPQ$ = 0; \\

\vspace*{.3cm}
\KwIn{\textbf{QuerySeries} $Q$, \textbf{Index} $Index$, \textbf{Integer} $N_{sb}$, \textbf{Integer} $tid$, \textbf{Queue} $PQueues[]$}
\vspace*{.1cm}
\vspace*{.1cm}

	\textbf{Integer} $bindex$, $pqindex$; \\

	\com{Tree Traversal Phase}	
		
	\While{$(TRUE)$}{
		$bindex$ $\gets$ Fetch\&Add($BCnt$,1);\\
		\If{$bindex \geq N_{sb}$}{break;}
		$ProcessRSBatch(Q, bindex, Index.RSBatches)$;\\
		$Index.RSBatches[bindex].complete \gets TRUE$; \label{workstealing_supporting_algorithm:bcom1}
	}
		
	\For{$bindex \gets 0$ \textbf{to} $N_{sb}$}{ \label{workstealing_supporting_algorithm:help_start}
		\If{$!Index.RSBatches[bindex].complete$ \textbf{AND} Fetch\&Add($Index.RSBatches[bindex].helped, 1) < HelpTH$}{ \label{workstealing_supporting_algorithm:helpTH}
			$ProcessBatch(Q, bindex, Index.RSBatches)$; \\
			$Index.RSBatches[bindex].complete \gets TRUE$;\label{workstealing_supporting_algorithm:bcom2}
		} 
	} \label{workstealing_supporting_algorithm:help_end}
		
	Barrier for all threads;  \label{workstealing_supporting_algorithm:prepro_start}

	\com{Priority Queue Preprocessing Phase}	

	\If{$tid == 0$}{
			Traverse all RS-batches and put their priority queues into $PQueues[]$; \\  \label{workstealing_supporting_algorithm:sharray}
			$SortByRootPriority(PQueues)$;  \\ \label{workstealing_supporting_algorithm:sort}
			$TotPQ \gets$ number of valid elements of $PQueues$; \label{workstealing_supporting_algorithm:pqnum}
	}
		
	Barrier for all threads; \label{workstealing_supporting_algorithm:prepro_end}
		
	\com{Priority Queue Processing Phase}	

	\While{$(TRUE)$}{ \label{workstealing_supporting_algorithm:pq_pro_phase_start}
		$pqindex$ $\gets$ Fetch\&Add($PQCnt, 1)$;  \\ \label{workstealing_supporting_algorithm:pqindex}
		\If{$pqindex \geq TotPQ$}{break;}
		\If{$PQueues[pqindex].stolen$}{continue;} \label{workstealing_supporting_algorithm:continuestolen}
		$ProcessPriorityQueue(PQueues[pqindex])$; \\ \label{workstealing_supporting_algorithm:process_pq}
	}
		 \label{workstealing_supporting_algorithm:pq_pro_phase_end}
}
} 
\caption{SearchWorker - Code for thread $tid$ }
\label{workstealing_supporting_algorithm}
\end{algorithm}

\begin{algorithm}[tb]
{
	{\footnotesize
\SetAlgoLined
\KwIn{\textbf{Integer} $N_B$}
\vspace*{.1cm}
\vspace*{.1cm}
{\bf Upon Receiving a message of type StealingRequest from node $sn'$:} \\
		S := Set of at most $N_{send}$ ids of RS-batches that satisfy the Take-Away Property\; \label{WorkStealingManager:start}
		send($S$, $Q$ of $sn$, $Q$'s current BSF) to $sn'$\;
		Mark the priority queues of the RS-batches with ids in $S$ as stolen\; \label{WorkStealingManager:end}

\vspace*{.3cm}
\com{Always-enabled event: it is executed repeatedly}\\
{\bf Upon receiving no message:} \\
	\If {$\mathit{FinishFlag[Q]}$ in $sn$ is set}{ 
		Terminate\;	
	}
}
} 
\caption{WorkstealingManager - Code for node $sn$}
\label{WorkStealingManager}
\end{algorithm}

\begin{algorithm}[tb]
{
	{\footnotesize
\SetAlgoLined
\KwIn{\textbf{Index} $index$, \textbf{Function} $exact\_search\_workstealing\_func$, \textbf{QuerySeries} $queries[]$, \textbf{Integer} $total\_nodes\_per\_nodegroup$}
\vspace*{.1cm}
\vspace*{.1cm}
{\bf Upon Receiving a DONE message from node $sn'$:} \\
		add $sn'$ in set $DoneNds$\; 
		\If {$DoneNds$ contains all system's nodes} {
			Terminate\;
		}

\vspace*{.3cm}
{\bf Upon Receiving a $msg = \langle S, Q_s, BSF_s\rangle$ from node $sn'$:} \\
	\If {|S| > 0} {
		Create threads to traverse the RS-batches with ids in $S$\; 
		Populate and process the corresponding priorities queues\;  
		BSFArray[$Q_s$] := $BSF_s$; \com{computed by threads above}\; 
		Wait all threads to complete\; 
	}
	ResponseFlag := 0\;  
\vspace*{.3cm}
\com{Always-enabled event: it is executed repeatedly} \\
{\bf Upon receiving no message:} \\
		\If {!(ResponseFlag)} {
		$sn'$ := choose randomly a node not in $DoneNds$\; \label{PerformWorkStealing:prot_start} \label{PerformWorkStealing:choose_node}
			send(StealingRequest, $sn$) to $sn'$\; \label{PerformWorkStealing:send_stealing_req}
			ResponseFlag := 1\;   \label{PerformWorkStealing:prot_end}
		}
}
} 
\caption{PerformWorkStealing - Code for node $sn$}
\label{PerformWorkStealing}
\end{algorithm}

\noindent 
{\bf Size of Priority Queues.}
The size of each priority queue cannot be larger than a specific threshold, $\mathit{TH}$. 
If adding an element in a priority queue results the size of the queue to reach $\mathit{TH}$,
then the thread gives up this priority queue and initiates a new one for the RS-batch. 
This way, each priority queue does not contain leaves from more than one RS-batch, 
and contains at most $\mathit{TH}$ leaves from the tree part that corresponds to the RS-batch. 

\begin{figure}
	\centering
	\begin{subfigure} [b]{0.45\textwidth}
		\centering
		\includegraphics[width=\textwidth]{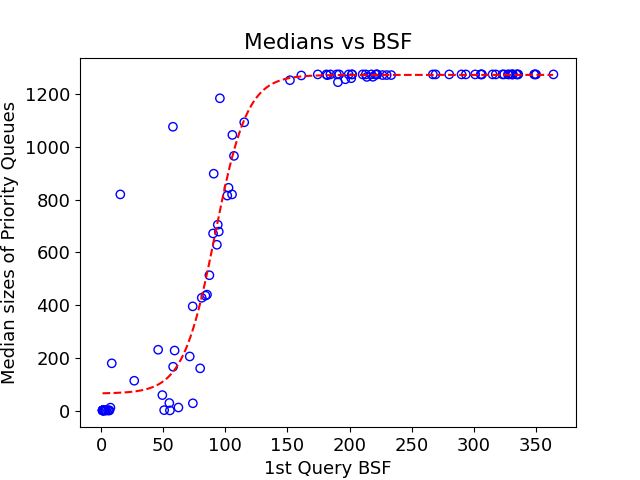}
		\caption{Sigmoid function fitting for determining $\mathit{TH}$.}
		\label{fig:ws-messi-fitting}
	\end{subfigure}
	\begin{subfigure} [b]{0.45\textwidth}
		\centering
		\includegraphics[width=\textwidth]{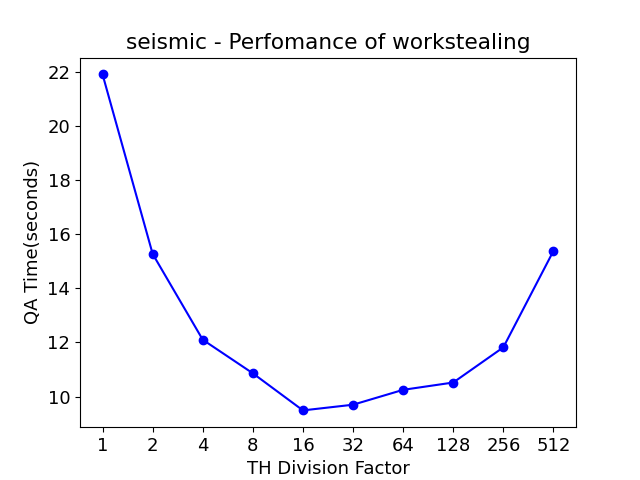}
		\caption{Performance for different Threshold division factors.}
		\label{fig:ws-messi-th}
	\end{subfigure}
	\caption{Odyssey Single-Node Query-Answering Algorithm Configuration.}
	\label{fig:ws-algorithm-figs}
\end{figure}

Choosing the appropriate value for the threshold, $\mathit{TH}$, is important for achieving load balancing
among the different threads. Our goal is to develop a method for determining a threshold value 
which will result with a set of priority queues that have about the same size. 
\remove{
\begin{center}
	\begin{adjustbox}{center}
		\begin{tabular}{||c | c||} 
			\hline
			Parameter & Notation\\ [0.5ex] 
			\hline\hline
			Number of {\em root subtree batches} & $N_{sb}$  \\ 
			\hline
			Threshold & $th$\\ 
			\hline
			{\em Root subtree batches to send} &  $N_{send}$ \\ 
			\hline
		\end{tabular}
	\end{adjustbox}
\end{center}
}
The threshold is determined and configured for every dataset we use, 
based on the queries we run. We explain the process of determining $\mathit{TH}$ 
for the Seismic real dataset~\cite{iris}, 
but the process is similar for all other datasets (real or synthetic) we experimented with. 
After running multiple queries of varying difficulty, we figured out that 
there exists again a correlation between the initial BSF that is computed for the query 
and the median size of the priority queues produced for answering it.
Then we performed a sigmoid function fitting using the following parameterized formula:
{\small
\[ f(Z) = m + (M-m) \frac{1}{1+b \cdot exp(-c(Z-d))} \]
where $M\in[0,1], m \leq M, b,c\in \mathbf{R}^{*},$ and $d\in \mathbf{R} $ 
} 
are the parameters of the sigmoid function (Figure~\ref{fig:ws-messi-fitting}). 
The final threshold value for each query is the median value estimation 
as it comes from the sigmoid function, 
divided by a factor (e.g. for seismic this factor has to be 16, based on the diagram shown in Figure \ref{fig:ws-messi-th}).

Experiments show that after the tree traversal phase is completed, 
we end up with a set of RS-batches that have a number of priority queues with most of them being the same size. 
This results in load balancing among the threads when processing priority queues.

\subsubsection{Work-Stealing Algorithm}
\label{sec:wsalg}

If a system node $\mathit{sn}$ becomes idle, 
$\mathit{sn}$ initiates the work-stealing protocol (Algorithm~\ref{PerformWorkStealing}, lines~\ref{PerformWorkStealing:prot_start}-\ref{PerformWorkStealing:prot_end}). 
It randomly chooses a system node $\mathit{sn'}$ from the set of those nodes that $sn$ knows to be still active 
and sends a steal request to it\footnote{The codes for Algorithms~\ref{WorkStealingManager} and~\ref{PerformWorkStealing} 
are written in an event-driven style~\cite{HagitAttiyaBook,Lynchbook}}. 
\remove{The use of the ResponseFlag is necessary
for avoiding sending more than one steal requests
\footnote{The codes for Algorithms~\ref{WorkStealingManager} and~\ref{PerformWorkStealing} 
are written in an event-driven style~\cite{HagitAttiyaBook,Lynchbook}. 
The code under each {\it upon event E takes place} clause is executed atomically each time event E is enabled.
The {\it upon receiving no message} event is always enabled, meaning 
that the code under it may be executed an arbitrarily large (or even infinite) number of times. 
The use of ResponseFlag ensures that this code will run only once for each target node.}. }
A thread in each node acts as the work-stealing manager (Algorithm~\ref{WorkStealingManager}). 
As soon as the work-stealing manager of $\mathit{sn'}$ receives the request, it tries to give away work to $\mathit{sn}$
(lines~\ref{WorkStealingManager:start}-\ref{WorkStealingManager:end} of Algorithm~\ref{WorkStealingManager}).

Earlier work 
has demonstrated that
a large amount of the query answering execution time is devoted to verifying that there is no better answer after the correct answer has been processed~\cite{gogolou2019progressive,progressiveISAX,pros}.
Based on these findings, Odyssey's work-stealing mechanism 
chooses to give away an RS-batch $B$ which satisfies the {\em Take-Away Property},
namely that $B$ is not yet stolen 
and its first priority queue is located in the rightmost 
possible index of the $\mathit{PQueue}$ array. 
This priority queue is then marked as stolen. 
If more than one batches are to be given away, this process is applied repeatedly
to choose additional RS-batches. 
Recall that the $PQueue$ array is sorted by the priority of the top element of each priority queue.
Thus, by giving away batches in this way, $sn'$ assigns to helpers priority queues that may still contain work. 
Additionally, it gives away RS-batches that have the highest probability to be unprocessed. 
Throughout the process, the current $BSF$ is shared among the nodes, 
every time it is updated, as a helper may steal a priority queue that contains a better answer
(or the owner may compute a better BSF later).

The number, $N_{send}$, of RS-bathes that a node gives-away during stealing affects performance. 
Theoretically, we would like to give away a number of RS-batches 
which on the one hand, it will enable the stealing node to do a noticeable amount of work, 
but on the other, the work to be given away should not result in higher query answering times. 
Experiments show that fixing $N_{send}$ to 4 was the best choice
(so $N_{send} = 4$ in Odyssey).

\remove{
\begin{figure}
	\centering
	\begin{subfigure} [b]{0.5\textwidth}
		\centering
		\includegraphics[width=\textwidth]{../resources/plots_chatzakis/plots/ws_comparison_bsf_manybatches.png}
	\end{subfigure}
	\caption{Single-Node Query-Answering Algorithm Configuration for different number of RS-batches}
	\label{fig:ws-worktealing_batches}
\end{figure}
}



\subsection{Data Replication}
\label{pr}
\label{sec:data-repl}

Odyssey aims at ensuring 
data scalability and, at the same time, good performance for query answering. 
Optimal data scalability requires to follow a no replication approach,
but experiments show that  the best query answering performance is noticed
for fully replicated settings. 
Odyssey manages to effectively navigate through
this trade-off between data scalability 
and good performance during query answering,
by providing a flexible {\it partial replication scheme}.

\begin{figure}[tb]
	\centering
	\includegraphics[width=0.7\textwidth]{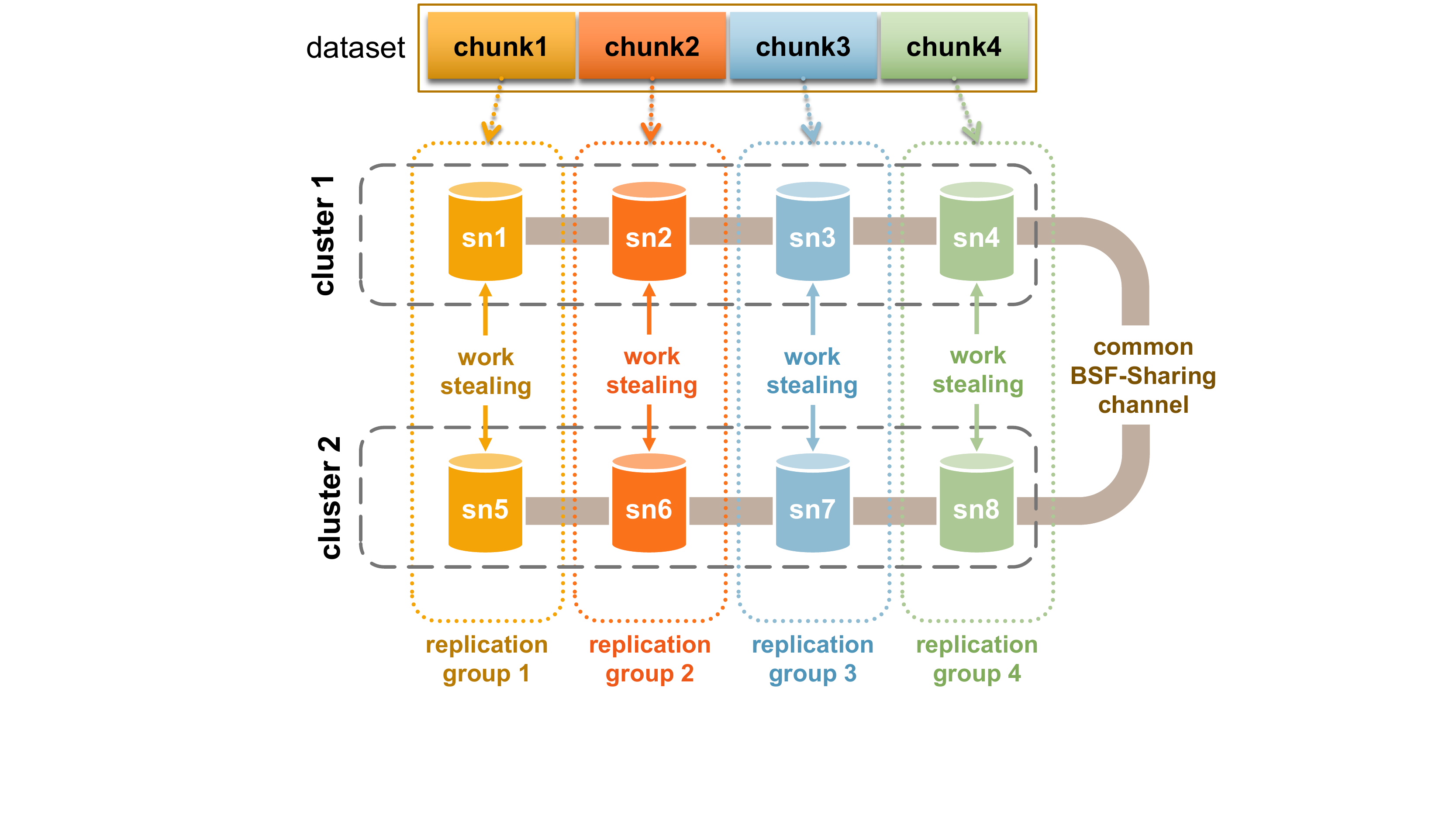}
	\caption{Data replication \Partial-$4$ ($N_{sn}=8$, data size $80$GB).}
	\label{fig:pdr_4_groups}
\end{figure}

The idea is to split the set of system nodes 
into {\em clusters}, where each cluster collectively stores the entire dataset (see Figure~\ref{fig:pdr_4_groups}).
Each cluster node stores (and indexes) a chunk of the dataset. The chunks stored in each node of a cluster 
are mutually disjoint. 
A {\em replication group} is a group of nodes such that each node stores the same dataset as every other node in the group. 
(We experimented with replication groups of the same size, but 
Odyssey can operate with replication groups of different sizes, as well.) 
The nodes of a replication group build their iSAX indices from the same data chunk. 
Thus, inside every replication group, we can apply the scheduling and load-balancing schemes 
described in Sections~\ref{sched} and~\ref{load-balancing}, respectively. 
We call the number of clusters the {\em replication degree} of the system.  

Consider a system with $N_{sn}$ system nodes. 
We call \Partial-$k$, $k \in \{1, 2, 4, \ldots, N_{sn}\}$, a replication setting with $k$ replication groups and $N_{sn}/k$ clusters. 
Observe that \Partial-$N_{sn}$, or \ES\, corresponds to no replication (each node stores a disjoint chunk of the dataset), and \Partial-$1$, or \Full\, corresponds to full replication (each node stores the full dataset).
Note that Odyssey's data replication scheme supports $1+\log N_{sn}$ different {\em replication degrees}.
Smaller replication degrees lead to smaller space overheads (and thus better data scalability). 
Thus, Odyssey's data replication scheme allows us to tackle memory limitation problems. 
Moreover, more replication groups lead to scalability in index creation.

A system with $8$ nodes supports $1+\log 8 = 4$ different replication degrees: \Full\ (\Partial-$1$), \Partial-$2$, \Partial-$4$, and \ES\ (\Partial-$8$). 
		Figure~\ref{fig:pdr_4_groups} illustrates the case of 
\Partial-$4$: we have 4 replication groups, organized in 2 clusters; replication degree is 2. 


%
%
%
%

\subsection{Data Partitioning}
\label{sec:data-part}

Odyssey framework supports more than one partitioning schemes. 
Under \ES, each system node is assigned a discrete chunk and builds the corresponding index, resulting in a scheme where each node 
keeps a local index on its own part of the data. 
Queries are forwarded to all nodes. Each node produces an answer based on its local index and data. 
The minimum among them is the final answer. 
Before distributing the data, {\em random shuffling (RS)} 
can be applied to randomly rearrange the series of the initial collection.

\remove{
This method may lead to load imbalances during query answering. 
For example, experiments on real data-sets showed that a query may have close answers in one of the local indexes, 
leading this node to have a good initial BSF and, therefore, prune more, 
while the other nodes get bad initial estimations, thus pruning less. 

Odyssey copes with this problem by employing a System-Wide Best-So-Far approach. 
Specifically, the nodes share the current best-so-far answer, each time it is locally updated, 
using message passing, so that every node is capable to prune based on the globally-smallest
BSF value at each point in time. 

However, this implementation supports sharing for the current query, 
meaning that it needs every node to answer the same query at each point of time.

Odyssey supports also appropriate scheduling and load balancing techniques
to further reduce load imbalances.

RF results in a better distribution of the data over the nodes of the system, 
but it has overhead when it comes to big collections.
}

To answer a query batch using partial data replication (or no replication), 
each query is sent to every replication group. 
Each node answers queries using its local data, 
and the partial answers for each query are gathered in the end to find the smallest answer. 
Very often for real data, the close answers to a query could be located into a small part of the dataset. 
The group that has these data will get a good initial answer, 
it will prune more and it will answer each query really fast, 
while other groups, will not necessarily compute good initial BSF values.
Thus, they will have more work to do leading to imbalances. 
For this reason, we enhance our distributed index with a 
book-keeping method that supports BSF sharing. 
When a node is processing a query and finds an improved value for BSF, 
it shares this value through a common BSF-Sharing channel 
(as illustrated in Figure \ref{fig:pdr_4_groups}). 
Every node periodically checks this channel to see if an answer for a query has arrived. 
Because this process runs in parallel, a node may receive a better answer for a 
query that will be encountered later on. 
Odyssey's book-keeping method solves such
synchronization problems. 
Each node holds an array that stores the improvements received from the channel for the BSF of each query, 
and before answering 
a query it checks the data held in this array. 
Thus, each node has the best answers 
extracted from all 
nodes, and our experimental evaluation shows that the use of this 
method is critical for performance.

In addition to these simple techniques, 
Odyssey also provides a sophisticated data partitioning scheme,
based on preprocessing of the initial data series collection, 
which provides a density-aware distribution of the data among the available nodes.
The required preprocessing incurs some time overhead.
However, it occurs only once for answering as many queries as needed, and
thus, as the number of queries to process increases, 
this overhead is amortized.
We describe this scheme in Section~\ref{sec:botao}.

\subsubsection{\DA\ Data Partitioning}
\label{sec:botao}

We observe that a good partitioning strategy should not assign all similar series to the same system node. 
In such a case, we risk to create work imbalance for the following reason. 
Assume that we need to answer a similarity search query, for which all candidate series from the dataset 
that are similar to the query are stored in one of the system nodes, while all other nodes are storing 
series that are not similar to the query. 
Then, during query answering, the node with the similar series will need to perform many (lower bound 
and real distance) computations in order to determine which of the candidate series is the nearest 
neighbor to the query, with essentially little pruning (if at all). 
On the other hand, all the other nodes that store dissimilar series will be able to prune aggressively, 
and therefore, finish their part of the computations much faster.

The above observations led us to the design of the \DA\ partitioning strategy, whose goal is to partition 
similar series across all system nodes, without incurring a high computational cost.
This is achieved by exploiting Gray Code~\cite{graycode} ordering for effectiveness (since it helps us 
split the similar series), and the summarization buffers of our index for efficiency (since we have to 
operate at the level of buffers, rather than individual series).

\begin{figure}[tb]
	\centering
	\begin{subfigure} [b]{0.6\textwidth}
		\centering
		\includegraphics[width=\textwidth]{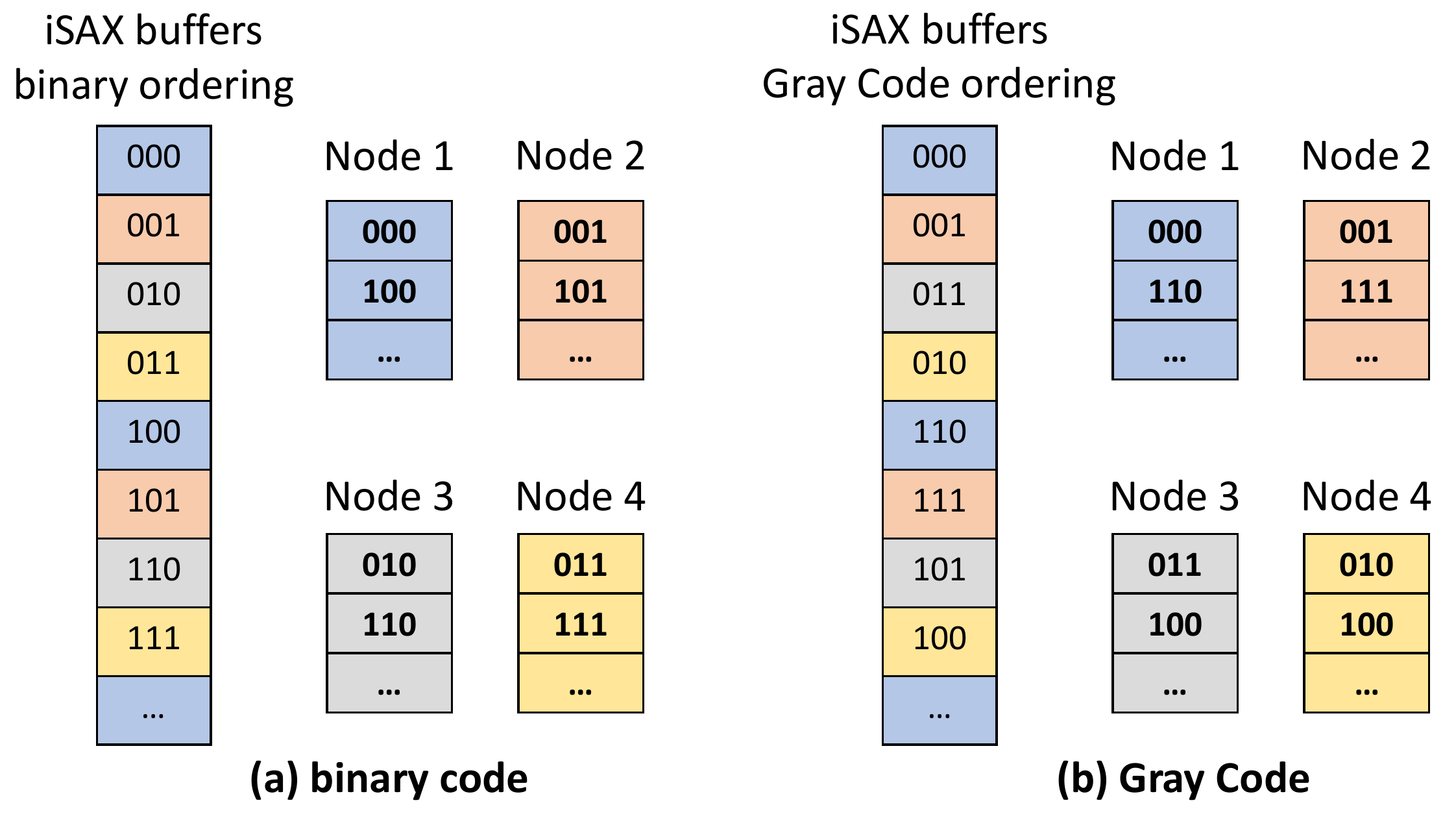}
	\end{subfigure}
	\caption{Examples of partitioning the iSAX buffers' data to 4 system nodes, based on (a) simple iSAX and (b) Gray Code.}
\label{fig:grayreplication}
\end{figure}

Figure~\ref{fig:grayreplication} shows an example of partitioning the data series in the summarization 
buffers according to a simple strategy using binary code, and to a strategy based on Gray Code.
In the former case, the buffers that end up in the same node contain similar series: their 
iSAX representations (the iSAX word of the buffer) are very close to one another, e.g., node 1 stores 
buffers "000" and 
"100", so series whose iSAX summaries only differ in one bit. 
In the latter case, this problem is addressed. 
The Gray Code ordering places similar buffers close to one another (by definition, two neighboring 
buffers in this order differ in only one bit), so it is then easy to assign them to different system 
nodes in a round-robin fashion.

\begin{figure}[tb]
	\centering
	\begin{subfigure} [b]{0.6\textwidth}
		\centering
		\includegraphics[width=\textwidth]{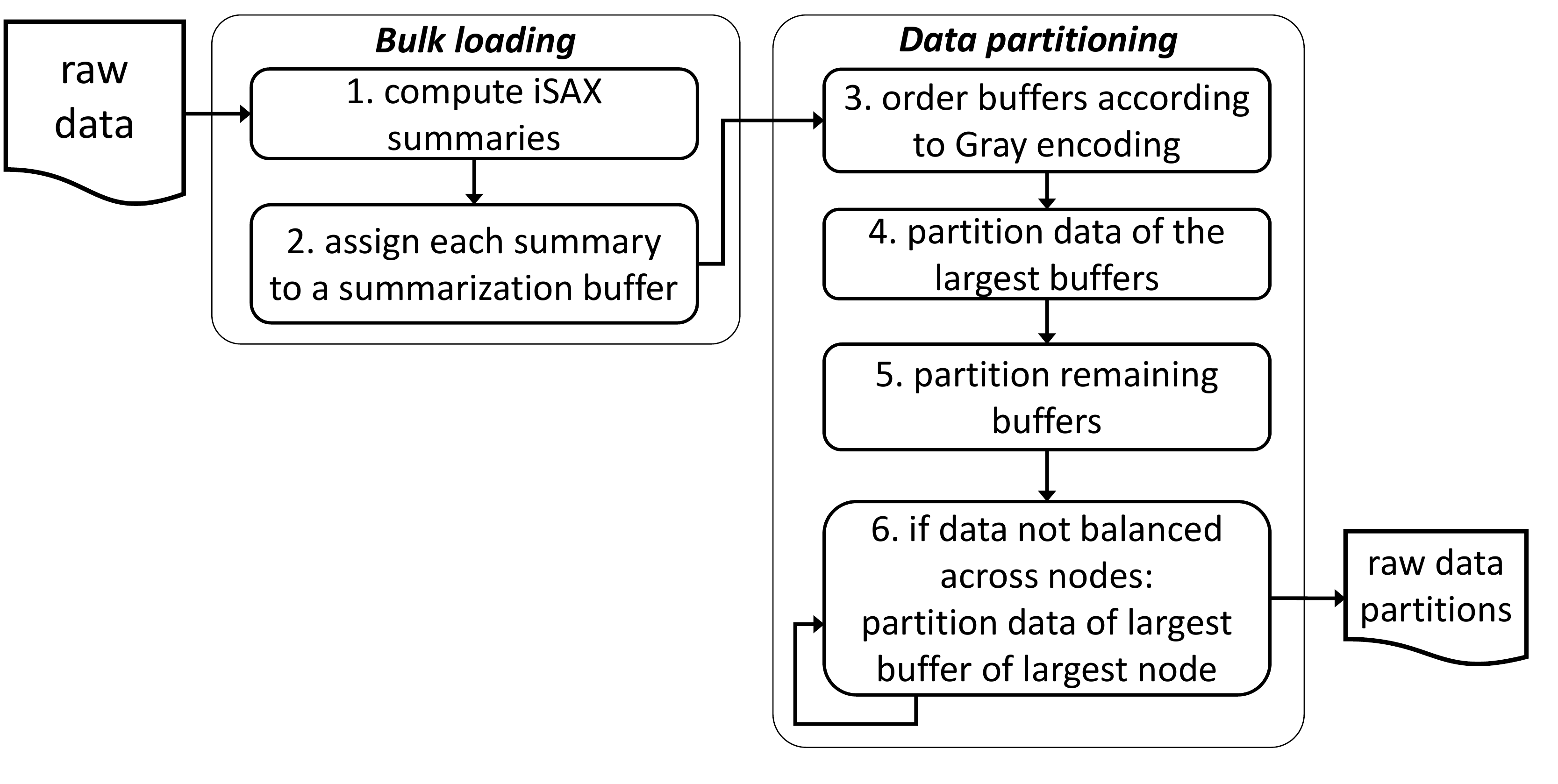}
	\end{subfigure}
	\caption{Flowchart of the \DA\ data partitioning.}
	\label{fig:grayreplicationflowchart}
\end{figure}

We depict the flowchart of the \DA\ partitioning strategy in Figure~\ref{fig:grayreplicationflowchart}.
We start by computing the iSAX summaries of the data series collection, and assigning each summary to the corresponding summarization buffer. 
These buffers are ordered according to Gray Code, and then the actual data partitioning starts (using round-robin scheduling). 
We first partition the series inside the $\lambda$ largest buffers; this is necessary, since often times a small 
number of buffers will contain an unusually large number of series (that we do not want to assign them all to 
the same system node).
Then, we partition the remaining buffers, and we check if the partitioning is balanced. 
If it is not, then we select the largest buffer of the largest node, and we partition the series inside this buffer.
Our experiments with several real datasets (omitted for brevity) showed that \DA\ exhibits a very 
stable behavior as we vary $\lambda$ from a few hundred to several thousands. 
In this study, we use $\lambda =400$.

\section{Extensions}
\label{sec:ext}

We now discuss two extensions of Odyssey, in order to support \emph{k-NN} search and the Dynamic Time Warping (DTW) distance.

\noindent{\bf \emph{k-NN} Search.}
Extending Odyssey to support \emph{k-NN} similarity search is straight-forward. 
Instead of computing a single BSF value, we simply need to keep track of the $k$ smallest BSF values.\\
\noindent{\bf DTW Distance.}
We also extend Odyssey to perform similarity search using Dynamic Time Warping (DTW), which is an elastic distance measure~\cite{keogh2005exact}.
Note that no changes are required in the index structure for this: the index we build can answer both Euclidean and DTW similarity search queries. 
Supporting DTW queries requires modifying the query answering algorithm only, and using LB\_Keogh~\cite{keogh2005exact}, which is a tight lower bound of the DTW distance.
We note that a lower bound for the DTW distance between the query and a candidate series can be computed by considering the distances between the corresponding points of the candidate series and the points of the LB\_Keogh envelope of the query. 

\remove{
	
Botao's stuff goes here. 

\textcolor{blue}{The benefit of the distributed system is using a distributed environment to expand the system's scalability. 
	So we should consider the data stored separately and independently on different machines. 
	We can store almost unlimited data in the data management system by using more servers. 
	However, the different data on a different machine will result in imbalance and inefficiency 
	during the query processing between the nodes in the cluster. 
	Therefore, we should find an efficient way to manage data and make the distributed environment 
	not only store data distributed but also answer the query efficiently.}

\textcolor{blue}{Algorithm~\ref{createdindex} presents the pseudocode for the distribute index creation. 
	First, the nodes of the cluster get the $node\_ID$ of the MPI processing(line~\ref{ci:getnodeid}). 
	Then, the DataPartition function will be executed for all nodes and return the data belong this node(line~\ref{ci:DataPartition}). 
	In the end, all nodes will create the index based on its data (line~\ref{ci:CreateIndex}).}

\begin{algorithm}[tb]
	{
		\SetAlgoLined
		\KwIn{\textbf{DataSeries} $RawData$, \textbf{Integer} $N_n$,\textbf{Integer} $N_r$,\textbf{float} $P_b$	}	\vspace*{.1cm}
		\vspace*{.1cm}
		\vspace*{.1cm}
			$node\_id \gets get\_node\_id()$;\label{ci:getnodeid}\;
			$Data$=$Data Partition$($RawData$,$node\_id$,$N_n$,$N_r$,$P_b$)\label{ci:DataPartition}\; 		
			$index$=$CreateIndex$($Data$)\label{ci:CreateIndex}\;
	}
	\caption{Create Distribute Index}
	\label{createdindex}
\end{algorithm}

\textcolor{blue}{The data partitioning function is shown in Algorithm~\ref{DataPartition}. 
	Firstly, we  execute $DataDistribute$ function to decide the initial $RawData$ attribution(line~\ref{dp:DataDistribute}).  
	$DataDistribute$ can roughly distribute the data in $N_n$ part. 
	However, the distribution of the data in the different buffers is imbalanced. 
	We should also balance the data attribution by the function $BufferBalance$ at line~\ref{dp:BufferBalance}.
	$p_b$ is the balance rate to control the size difference between different partitions. 
	Finally, the data belonging to the this nodes will be collected(line~\ref{dp:keep}) and returned back(line~\ref{dp:re}).
}

\begin{algorithm}[tb]
	{
		\SetAlgoLined
		\KwIn{\textbf{DataSeries} $RawData$, \textbf{Integer} $node\_id$, \textbf{Integer} $N_n$,\textbf{Integer} $N_r$,\textbf{float} $P_b$}	
		\KwOut{\textbf{DataSeries} $Data\_partitioned$}
		\vspace*{.1cm}
		\textbf{Distribute Buffer} $Buf_d$\;
		\tcp*[f]{{\footnotesize Buffer initial distribute}}\\
		$Data Distribute$($RawData$,$Buf_d$,$N_n$)\label{dp:DataDistribute}\;
		$Buffer Balance$($RawData$,$Buf_d$,$p_b$,$N_n$)\label{dp:BufferBalance}\;

		$Data\_S\leftarrow$ collect data belong node $node\_id$\label{dp:keep}\;

		return $Data\_S$\label{dp:re}\;

	}
	\caption{Data Partition}
	\label{Data Partition}
\end{algorithm}

\textcolor{blue}{The critical problem of our index creation algorithm is how to distribute data based on its locality balance. 
	This implementation is shown in Algorithm~\ref{Datadistribute}. 
	We pass all data series in the dataset(line~\ref{Dd:for}). 
	Each of them computes its iSAX summary by calling the ConvertToiSAX function (line~\ref{Dd:cover}), 
	and attributes the result to the appropriate iSAX buffer of the index (line~\ref{Dd:mask}). 
	Next, call GreyEncode function to decide which node the data belong to(line~\ref{Dd:grey}). 
	Finally, store the data in the appropriate layer(line~\ref{Dd:insert}) and update the data counter(line~\ref{Dd:plus1}).
}

\textcolor{blue}{We distribute the entire iSAX buffer instead of allocating data randomly. 
	The reason is that we should keep the load balance during query answering. 
	When we distribute data, we should keep the data balanced and the query answer load balance and efficiently. 
	Be benefited by the lower bound distance between the query and the node. 
	We only need to do the lower bound distance calculation, 
	and the real distance calculation between the Data series belongs to the close leaf node to query. 
	Distributing similar data in the same node will result in a significant workload imbalance during query answering. 
	We need to distribute the unfamiliar data(iSAX buffer) in the same node to resolve this problem. 
	Grey code is a reflected binary code\cite{gray1953pulse}. 
	It is an ordering of the binary numeral system such that two successive values differ in only one bit (binary digit). 
	The similar iSAX buffer position will be close to the Gray code. 
	Table~\ref{tablegray} show the average unfamiliar segment between different buffers in a single distributed set with Gray and Binary codes.
	The average distance with Gray code in 1 distributed set is smaller than the distance with Binary code. 
	The result indicates that the buffers in a single distributed set are neighbours when using Gray code to pass all buffers.
	Moreover, if we distribute the close number of the buffer in Grey code in different nodes, 
	we can achieve that attribute the unfamiliar data in the same node.}

\begin{table}[tb]
	\centering
	\caption{Average distance between different buffers in the distributed set.}\label{tablegray}
	\hspace*{-0.3cm}	
	\begin{tabular}{|c|c|c|c|c|}
		\hline
		set size&2&4&8&16\\
		\hline
		Binary&1.999969&	2.166616& 2.464194&2.841494\\
		\hline
		Gray&1&	1.5& 1.892857&2.308333\\
		\hline
	\end{tabular}
\end{table}

\begin{algorithm}[tb]
	{
		\SetAlgoLined
		\KwIn{\textbf{DataSeries} $RawData[]$, \textbf{iSAXBuffer} $buf_d$,\textbf{Integer} $N_n$}
		
		\textbf{Integer} $b \leftarrow$ number of data in $RawData[]$\;  
		\vspace*{.1cm}

		\For{$i$ $\leftarrow$ $0$ \emph{\KwTo} $b$\label{Dd:for}}
		{
			$isax$ = ConvertToiSAX($RawData[i]$)\;\label{Dd:cover}
			$\ell$ = find appropriate buffer where $isax$ belone\;\label{Dd:mask}
			$C_n$=GreyEncode[$\ell$] mod $N_n$\label{Dd:grey}\;
			$Buf_d[\ell].node[C_n].data[k] = \langle isax, j \rangle$\;\label{Dd:insert}
			$Buf_d[\ell].node[C_n].datacounter+=1$\;\label{Dd:plus1}
		}
		
	}
	\caption{Data Distribute}
	\label{Datadistribute}
\end{algorithm}

\textcolor{blue}{
	Usually, the data distribution is not balanced. 
	Some big iSAX buffers will be attributed in the same node even if they are not similar. 
	Therefore, after the initial data distribution by Algorithm~\ref{Datadistribute}, 
	we need to execute Algorithm~\ref{bufferbalance} to keep the data balance in nodes. 
	Firstly, we select the largest $N_r$ buffers(line~\ref{bb:for2}-\ref{bb:big}). 
	Next, we distribute the data belonging to this buffer from the single node to each node (line~\ref{bb:distribute}). 
	After updating the node size(line~\ref{bb:nodesize}), 
	We distribute the largest buffer from the node with more data 
	until the balance rate satisfies our threshold(line~\ref{bb:rate}-\ref{bb:nodesize2}).}

\begin{algorithm}[tb]
	{
		\SetAlgoLined
		\KwIn{\textbf{DataSeries} $RawData$, \textbf{iSAXBuffer} $Buf_d$, \textbf{float} $P_b$,\textbf{Integer} $N_r$}	\vspace*{.1cm}
		\vspace*{.1cm}
		
		\For{$j$ $\leftarrow$ $0$ \emph{\KwTo} $N_r$\label{bb:for2}}
		{
			$Buf_m$$\leftarrow$ largest buffer in $Buf_d$[] \& $Buffer_d$.$distributed$==$false$ \label{bb:big}\;
			$Buffer Distribute$($Buf_m$)\label{bb:distribute}\;
		}
		
		update $Buf_d$[j].$nodesize$\label{bb:nodesize}\;
		\While{$MAX$($Buf_d$[].$nodesize$)-$MIN$($Buf_d$[].$nodesize$)$>$$N\_S$*$P_b$\label{bb:rate}}
		{	
			$Buffer_c \leftarrow$ largest buffer of lagest node $ \& Buf_m$.$distributed$==$false$\label{bb:big2}
			$Buffer Distribute$($Buf_m$)\;\label{bb:distribute2}
			<update $Buf_d$[].$nodesize$>\;\label{bb:nodesize2}
		}

	}
	\caption{BufferBanlance}
	\label{bufferbalance}
\end{algorithm}

%
%

\textcolor{blue}{Algorithm~\ref{bufferdistribute} illustrate how we distribute data from a single node to others. 
	We extract the data buffer belong this node(line~\ref{bd:ex}) at first and reset the counter(line~\ref{bd:re}). 
	Then, we distribute the data from the extracted buffer to the buffers from each node(line~\ref{bd:dis} and~\ref{bd:mv}). 
	In the meantime, we update the corresponding counter(line~\ref{bd:re2}). }

\begin{algorithm}[tb]
	{
		\SetAlgoLined
		\KwIn{\textbf{iSAXBuffer} $Buf_m$}	\vspace*{.1cm}
		\vspace*{.1cm}
		
		$Buf_t$=$Buffer_c$.$node$[$i$].$B_L$\;\label{bd:ex}
		$buffersize_t$=$Buf_m$.$node$[$i$].$B_{size}$\;
		$Buf_m$.$node$[$i$].$B_{size}$=0\;\label{bd:re}
		$tmp\_counter$=0;	
		\For{$i$ $\leftarrow$ $0$ \emph{\KwTo} $tmpbuffersize-1$}
		{
			$C_{n}$=$i$ mod $N_n$\;\label{bd:dis}
			$Buf_m$.$node$[$C_{n}$].$B_L$[$Buf_m$.$node$[$C_{n}$].$B_{size}$]=	$Buf_t$[$i$]\;\label{bd:mv}
			$Buf_m$.$node$[$C_{n}$].$B_{size}$++\;\label{bd:re2}
		}
	}
	\caption{Buffer Distribute}
	\label{bufferdistribute}
\end{algorithm}

Botao's experiments goes here.

\begin{figure}[tb]
	\centering
	\includegraphics[page=1,width=0.9\columnwidth]{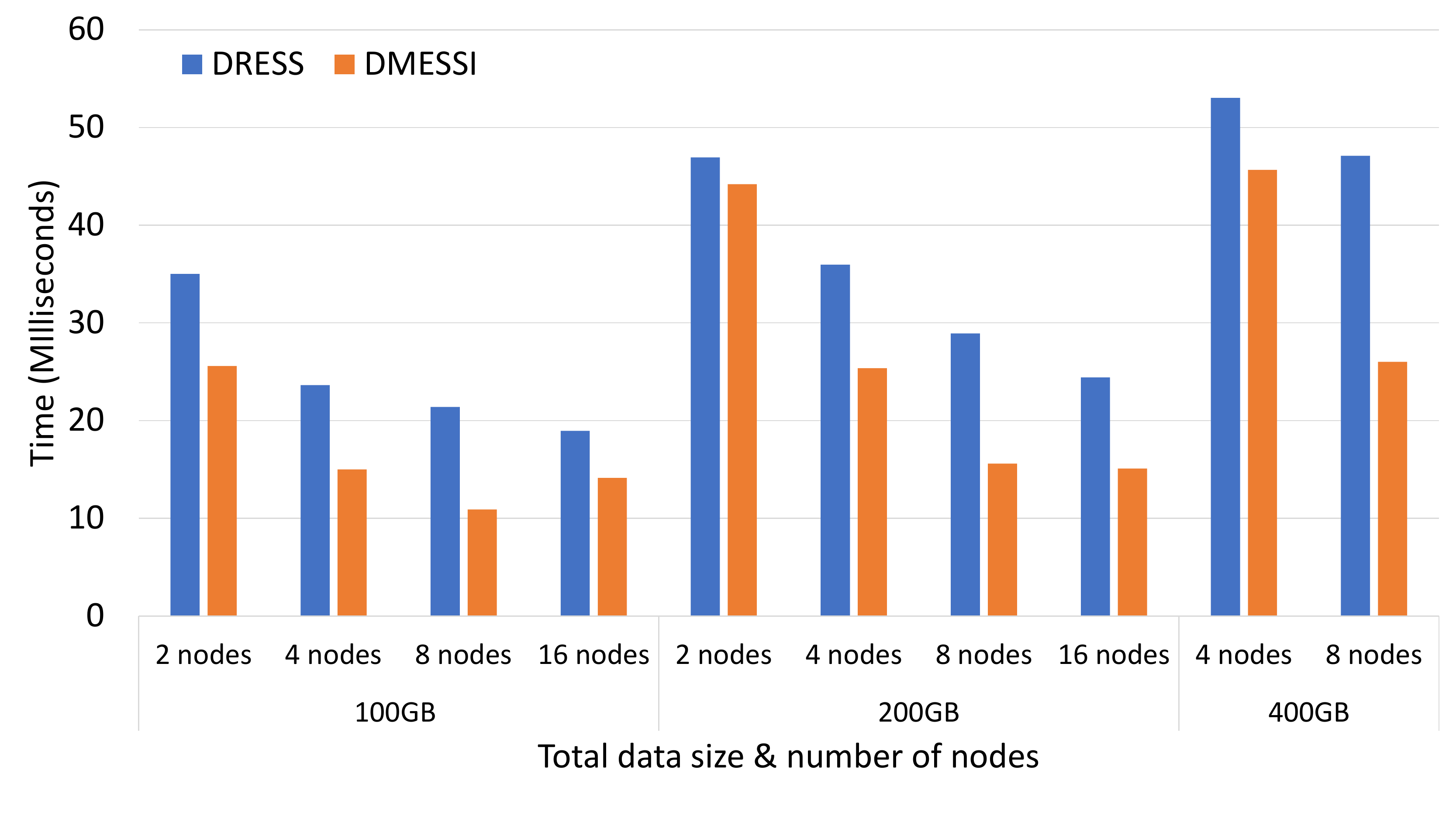}
	\caption{Query answering, vs. number of node and varying data size}
	\label{fig:scal}
\end{figure}
\textcolor{blue}{
Figure~\ref{fig:scal} illustrate the time consumption for query answering on synthetic dataset varying number of node and data size. DMESSI is up to 1.97x faster than DRESS. 
The experimental results show that as we increase the number of the nodes used, the query answering time are decreasing for both DRESS and DMESSI.
However, the Comparative advantage of DMESSI are larger along with the increasing of the node number we use. 
The query answering time of DRESS becuse stable when we use more nodes. 
Under the same situation, The query answering time of DRESS still keep the trend of decreasing.
The result indicate that DMESSI is better than DRESS with many calculation nodes in the distributed environment.
DMESSI also work well with 16 nodes. 
However, the data sizes for single nodes were too small when we distributed 100GB of data on 16 nodes. 
Therefore, the query answering time goes up because the index processing overhead is noticeable in this situation.}

\begin{figure}[tb]
	\centering
	\includegraphics[page=1,width=0.9\columnwidth]{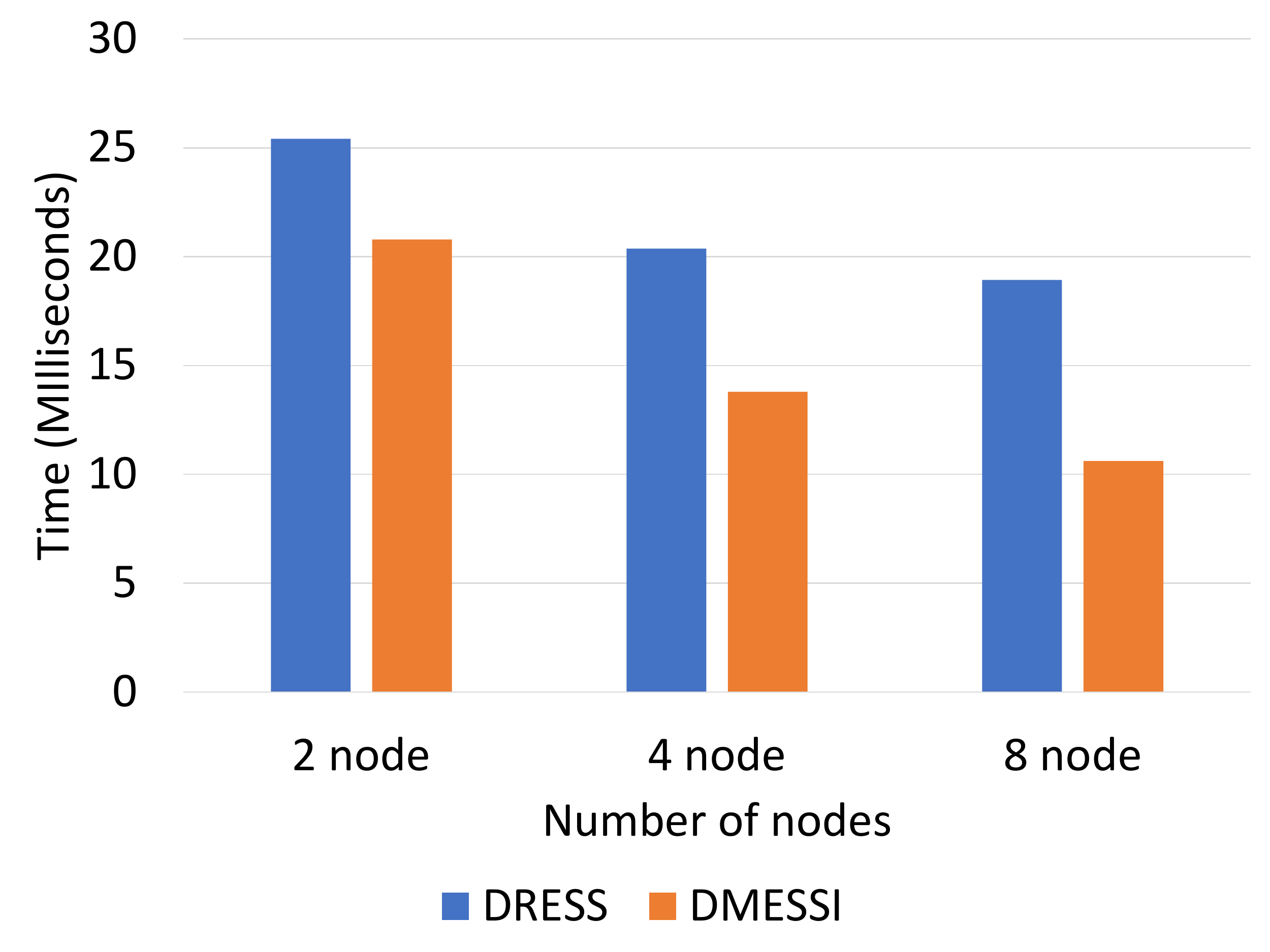}
	\caption{Query answering of Seismic data, vs. number of node }
	\label{fig:seismic}
\end{figure}

\textcolor{blue}{Figure~\ref{fig:seismic} and Figure~\ref{fig:sald} show the same trend. DMESSI is up to XXXx faster than DRESS, especially with many nodes.}

\begin{figure}[tb]
	\centering
	\includegraphics[page=2,width=0.9\columnwidth]{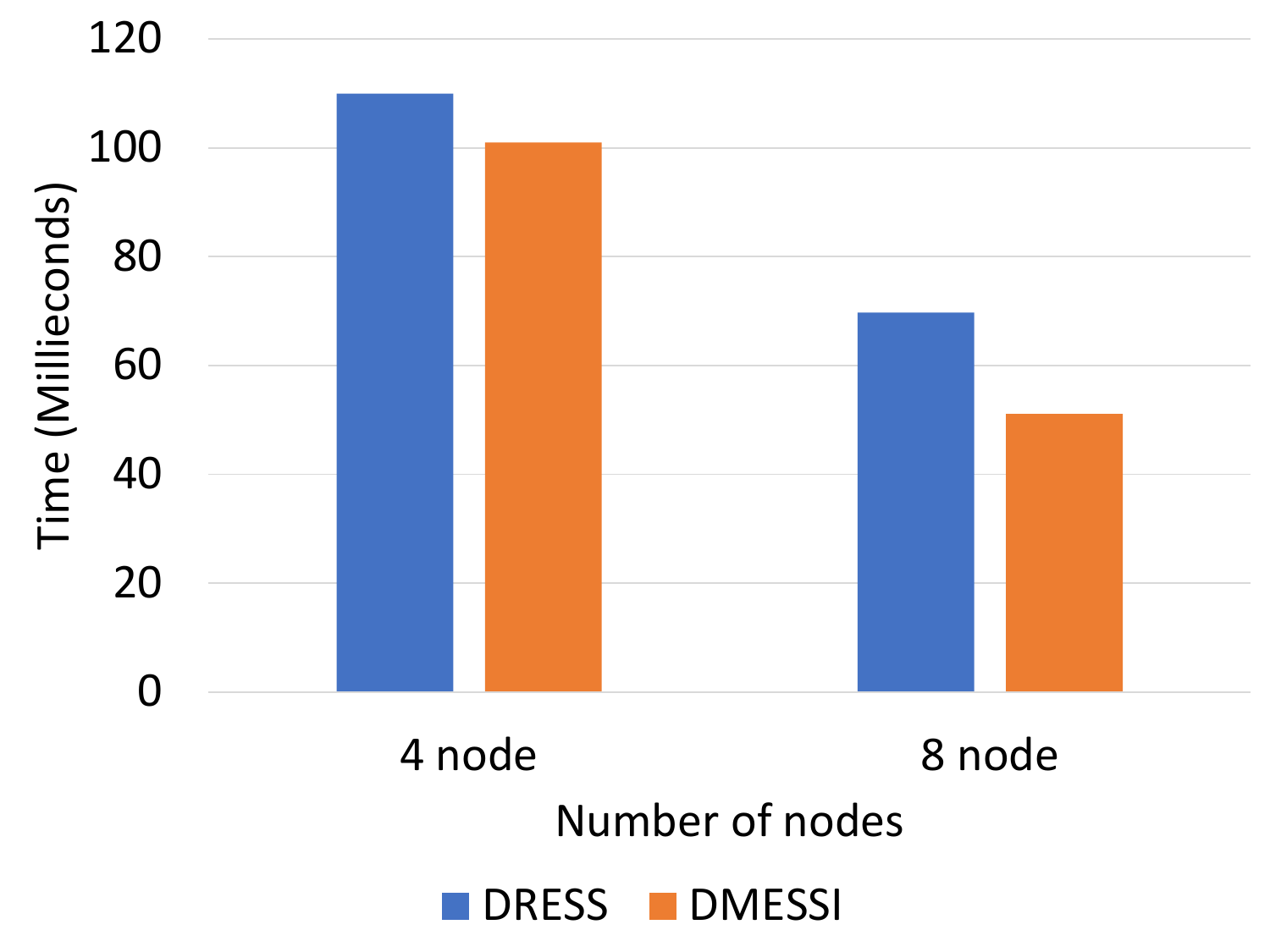}
	\caption{Query answering of SALD data, vs. number of node}
	\label{fig:sald}
\end{figure}

\textcolor{blue}{Figure~\ref{fig:disnode100} and~\ref{fig:disnode200} illustrate the impact of the number of the largest buffers distributed. 
	The query answering time goes down along with the increased distributed buffer number when the number is small. 
	However, when we distributed lots of buffers, the benefit from the data distribution can not pay the cost of the more buffer pass overhead. 
	The query answering goes up when the number of the buffer distributed is large. 
	The results illustrate that the best setting is around 8000 buffers. }

\begin{figure}[tb]
	\centering
	\includegraphics[page=1,width=0.9\columnwidth]{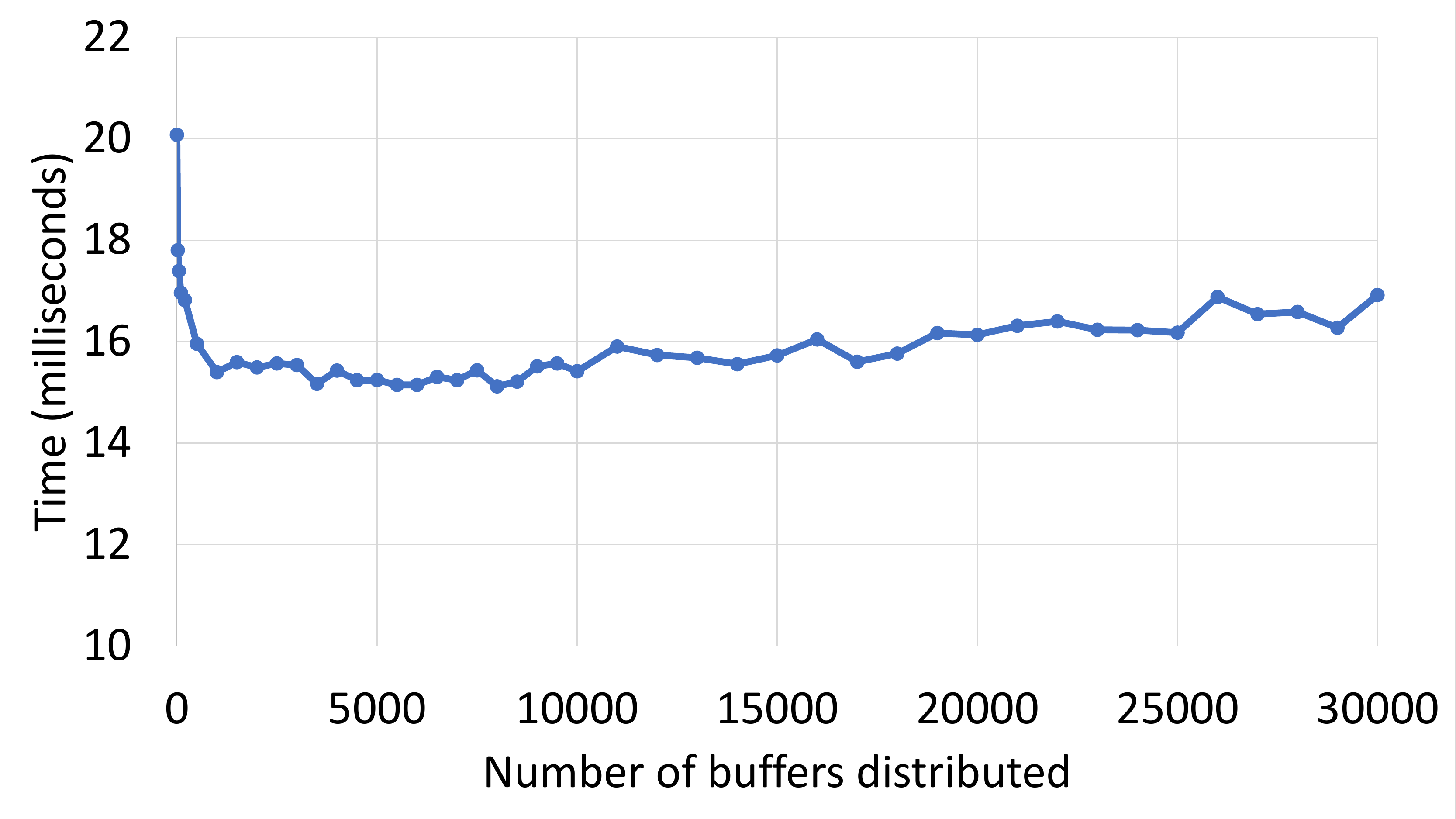}
	\caption{Query answering with 100GB data, vs. number of node distributed}
	\label{fig:disnode100}
\end{figure}

\begin{figure}[tb]
	\centering
	\includegraphics[page=2,width=0.9\columnwidth]{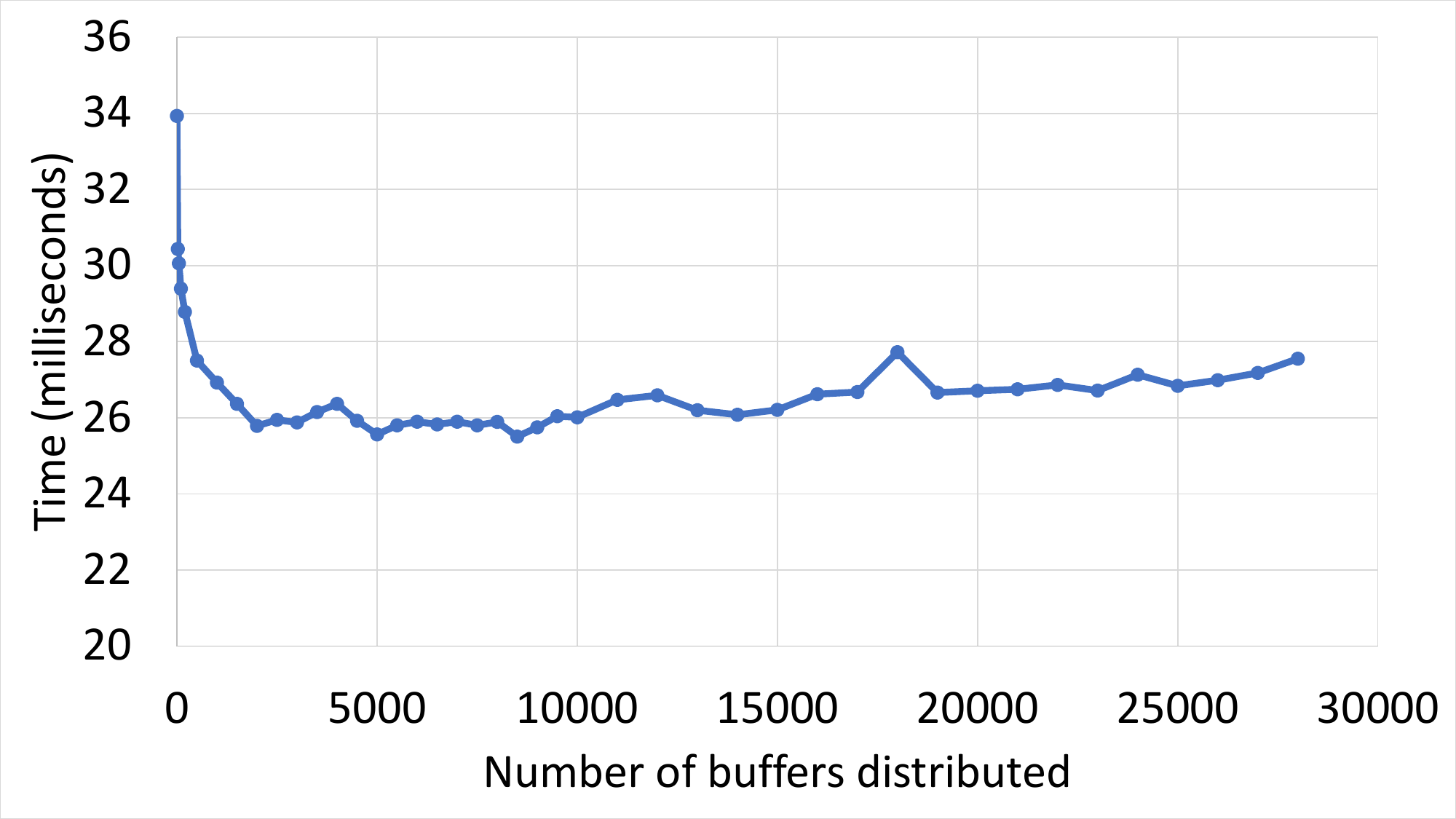}
	\caption{Query answering with 200GB data, vs. number of node distributed}
	\label{fig:disnode200}
\end{figure}

\textcolor{blue}{Figure~\ref{fig:nogray} illustrate the influence of different data attribute strategy. 
Because of the imbalance between nodes, the cost of DPiSAX and DRESS strategy are higher than DMESSI strategy.}

\begin{figure}[tb]
	\centering
	\includegraphics[page=1,width=0.9\columnwidth]{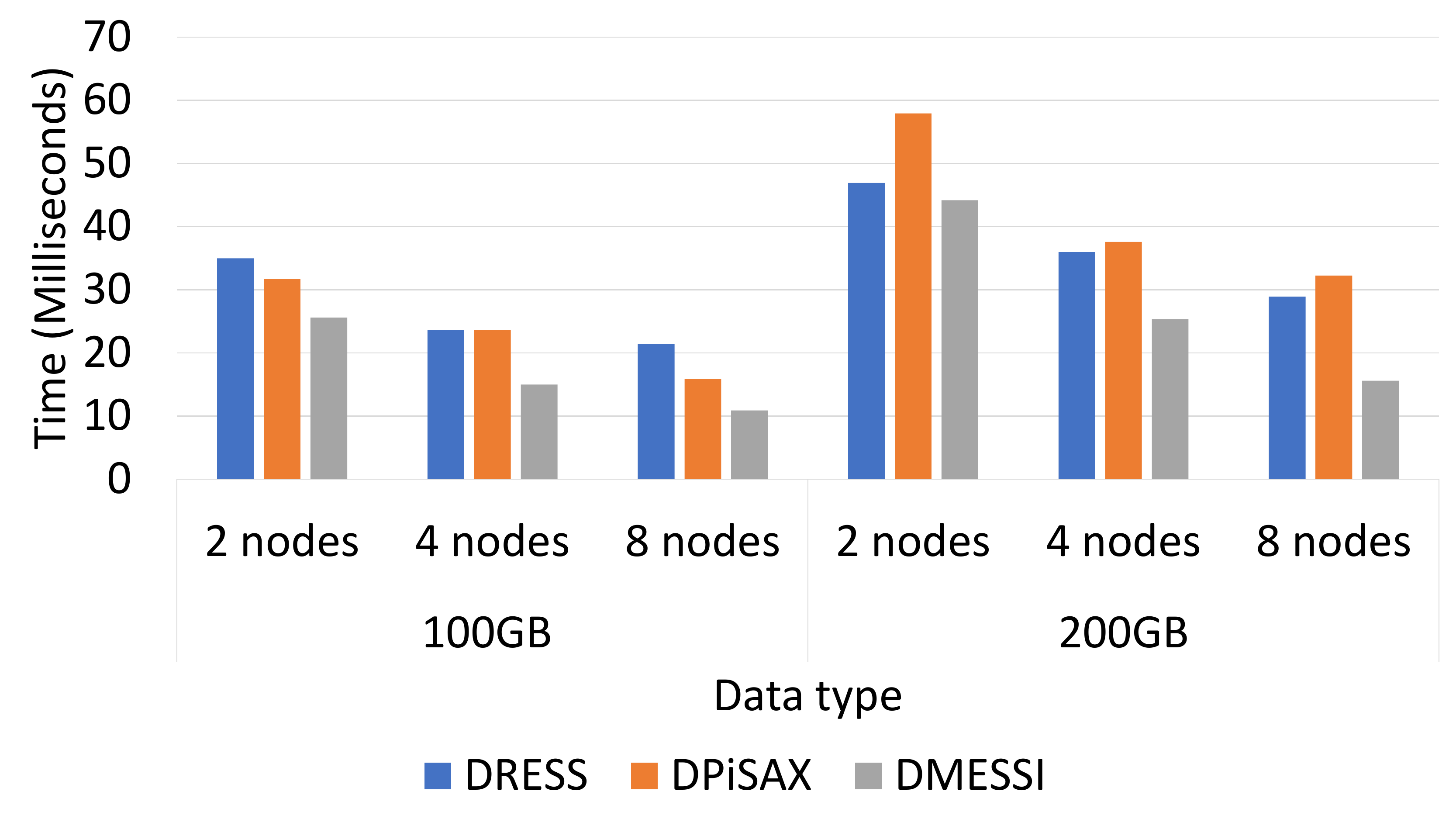}
	\caption{Query answering, vs. different attribute strategy}
	\label{fig:nogray}
\end{figure}

\textcolor{blue}{Figure~\ref{fig:std} shows the standard deviation of query answering time on different nodes with a single query. 
	We can see that the query answering time between different nodes is more balance for DMESSI.}

\begin{figure}[tb]
	\centering
	\includegraphics[page=1,width=0.9\columnwidth]{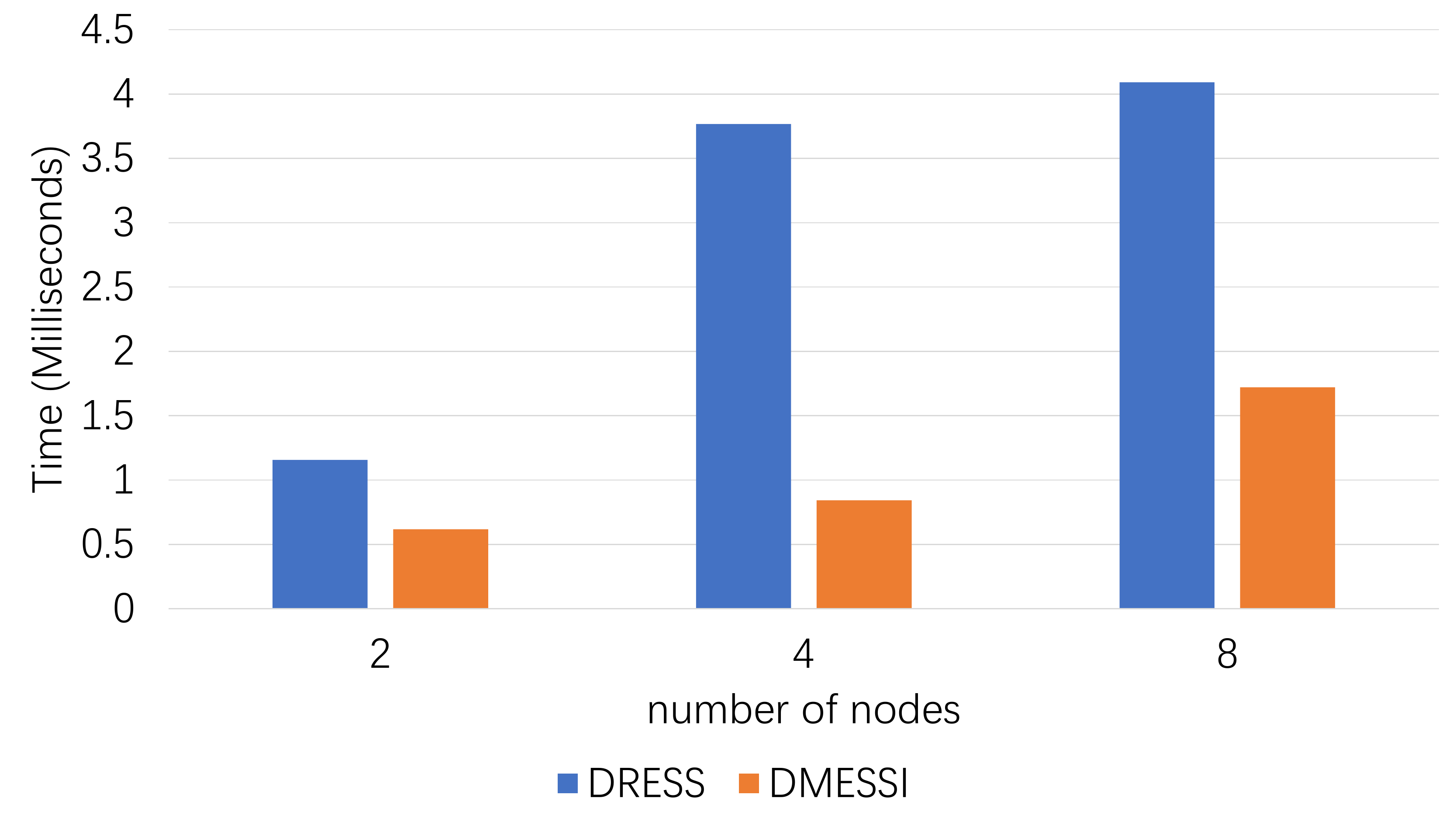}
	\caption{Time standard diviation for individual query.}
	\label{fig:std}
\end{figure}

} 

\section{Experimental Evaluation}
\label{section:evaluation}

\noindent{\bf Setup.} 
Experiments conducted on a cluster of 16 SR645 nodes, connected through an HDR 100 Infiniband network. 
Each node has 128 cores (with no hyper-threading), {200GB RAM (available to users out of the 256GB physical memory)}, and runs Red Hat Enterprise Linux 
r8.2. 
All evaluated algorithms written in C and compiled using MPICC, Intel(R) MPI Library for Linux 
OS, v2021.2.

\noindent{\bf Algorithms.} 
{Our experimental analysis includes the entire range of Odyssey's data distribution strategies with $k$ replication groups, 
\Partial-$k$, $k \in \{1, 2, 4, \ldots, N_{sn}\}$, as well as the density-aware data partitioning algorithm (\DA). 
Recall that \Partial-$N_{sn}$, or \ES\, corresponds to no replication, 
and \Partial-$1$, or \Full\, corresponds to full replication. 
}
Additionally, our analysis evaluates Odyssey's queries scheduling algorithms:
(i) static scheduling assigning equally sized query sets to nodes (\St); 
(ii) dynamic scheduling using a coordinator (\Dn); and 
(iii) predictions-based scheduling, including: 
static without ordering (\PSU), static with ordering (\PS), and dynamic (\PD).
Moreover, we evaluate Odyssey's work-stealing mechanism using both
\Dn\ and \PD, 
resulting in algorithms  \WS\ and \WSP, respectively.
The latter is our best 
scheduling algorithm 
(cf. paragraph ``Queries scheduling'').
We note that that Odyssey's query scheduling and work-stealing mechanisms can 
be used together only with the \Full\ or \Partial\ data distribution strategies that
provide some replication. 


We compare Odyssey to: 
(i) MESSI~\cite{messijournal}, where we run the MESSI index independently in each system node;
(ii) MESSI SW BSF, where we extend the previous solution by enabling system-wide sharing of the BSF values; and
(iii) DPiSAX~\cite{dpisaxjournal}, where we implement (in C) the DPiSAX data partitioning strategy, 
and (for fair comparison) implement query answering in each node using MESSI.

\noindent{\bf Datasets.} 
We evaluated Odyssey's strategies and algorithms using real and synthetic datasets, of varying sizes (refer to Table~\ref{table:datasets}). 
The synthetic data series, called \emph{Random}, were generated as random-walks (i.e., cumulative sums) of 
steps that follow a Gaussian distribution (0,1).
This type of data has been extensively used in the past~\cite{conf/sigmod/Faloutsos1994,
journal/kais/Camerra2014,conf/kdd/Zoumpatianos2015,DBLP:journals/vldb/ZoumpatianosLIP18,lernaeanhydra,lernaeanhydra2}, 
and models the distribution of stock market prices~\cite{conf/sigmod/Faloutsos1994}.
Our five real datasets come from the domains of seismology (\emph{Seismic}), astronomy (\emph{Astro}), deep learning (\emph{Deep}), 
image processing (\emph{Sift}), {and information retrieval (\emph{Yan-TtI})}. 
\emph{Seismic} 
contains seismic instrument recordings 
and consists of 100M data series of size 256~\cite{url/data/seismic}. 
\emph{Astro} represents celestial objects 
and consists of 100M data series of size 256~\cite{journal/aa/soldi2014}. 
\emph{Deep}
~\cite{url/data/deep1b} contains 1B Deep vectors of size 96 
extracted from the last layers of a convolutional neural network.
\emph{Sift}~\cite{SIFT} is comprised of image descriptors and  { Yandex Text-to-Image (\emph{Yan-TtI})~\cite{pmlr-v176-simhadri22a} contains 1B vectors that  include image- and textual-embeddings in the same space; it represents typical cross-modal information retrieval tasks.} 

\begin{table}[tb]
		\caption{{Details of datasets used in experiments.}}
	\centering
	{\small
	\begin{adjustbox}{center}
  		\begin{tabular}{||c | c | c | c | c||} 
  			\hline
  			Dataset & \# of series & Length (floats) & Size (GB) & Description  \\ [0.5ex]
  			\hline\hline
  			Seismic & 100M & 256 & 100 & seismic records \\ 
  			\hline
  			Astro & 270M & 256 & 265 & astronomical data \\ 
  			\hline
  			Deep & 1B & 96 & 358 & deep embeddings \\ 
  			\hline
  			Sift & 1B & 128 & 477 & image descriptors \\ 
  			\hline
  			{Yan-TtI} & 	{1B} & {200} & {800} & {image and text} \\ 
  			\hline
  			Random & 	{100M-1600M} & 256 & 	{100-1600} & random walks \\ 
  			\hline
  		\end{tabular}
  	\end{adjustbox}
} 

	\vspace*{-0.4cm}	
	\label{table:datasets}
\end{table}

\noindent{\bf Evaluation Measures.} 
During each experiment, $E$, and for each {\em node}, $\mathit{sn}$, we measure (i) the {\em buffer time} required to calculate
the iSAX summaries and fill-in the receive buffers, 
(ii) the {\em tree time} required to insert the items of the receive buffers in the index tree,
and (iii) the {\em query answering time} required to answer the queries assigned to $\mathit{sn}$.
The sum of these times 
constitute the {\em total time} that  $\mathit{sn}$ works during $E$;
also, buffer and tree times constitute the time required to create the index, called {\em index time}.
To compute 
all the above times during $E$, we take the maximum among the corresponding times of each node participating in $E$. 
%
We report the average times of $10$ experiments.
%

\begin{figure}[tb]
	\centering
	\begin{subfigure} [b]{0.45\textwidth}
		\centering
		\includegraphics[width=\textwidth]{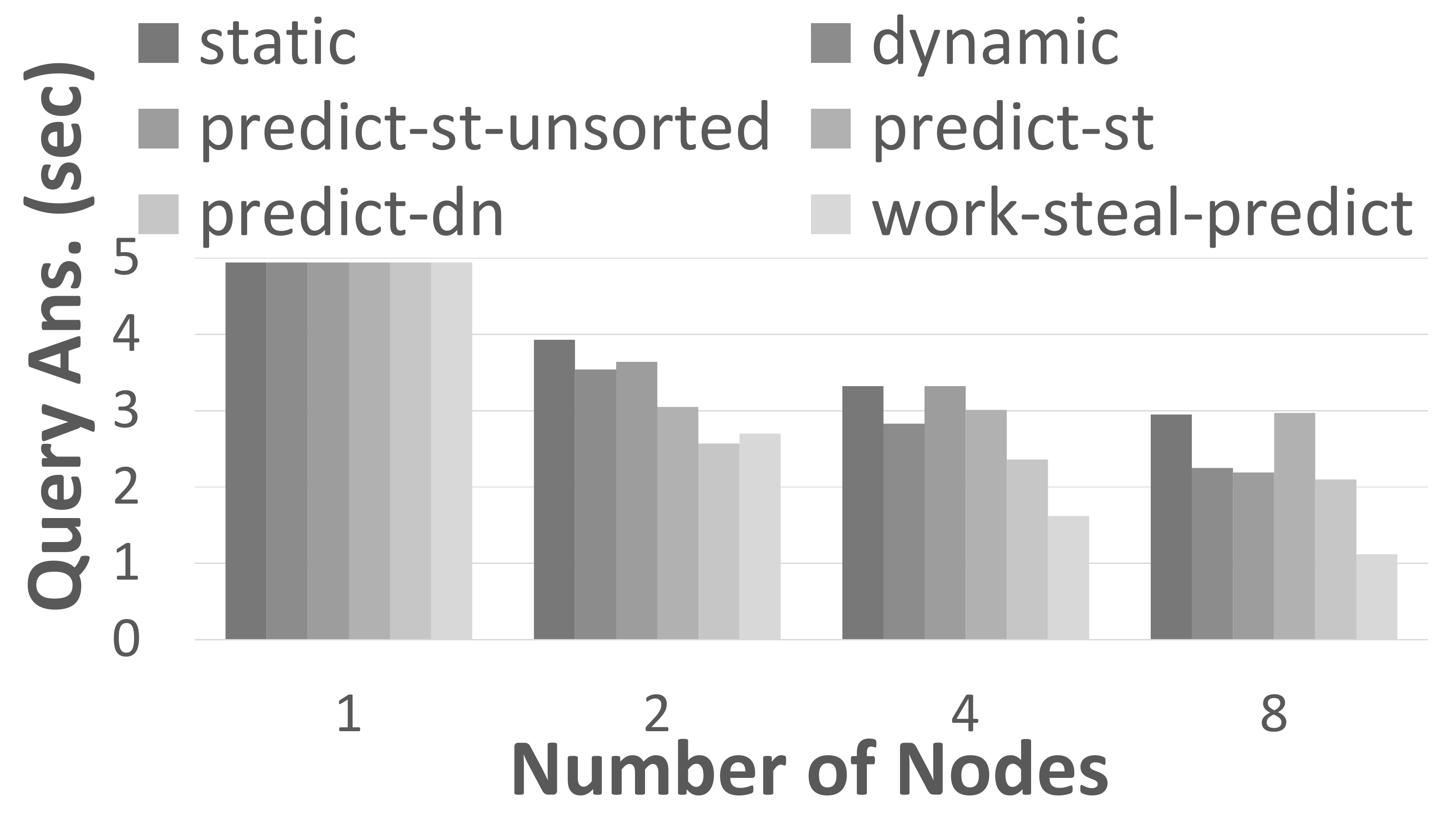}
		\caption{\Full\ replication}
		\label{fig:seismic_fdr_100:a}
	\end{subfigure}
	\begin{subfigure} [b]{0.45\textwidth}
		\centering
		\includegraphics[width=\textwidth]{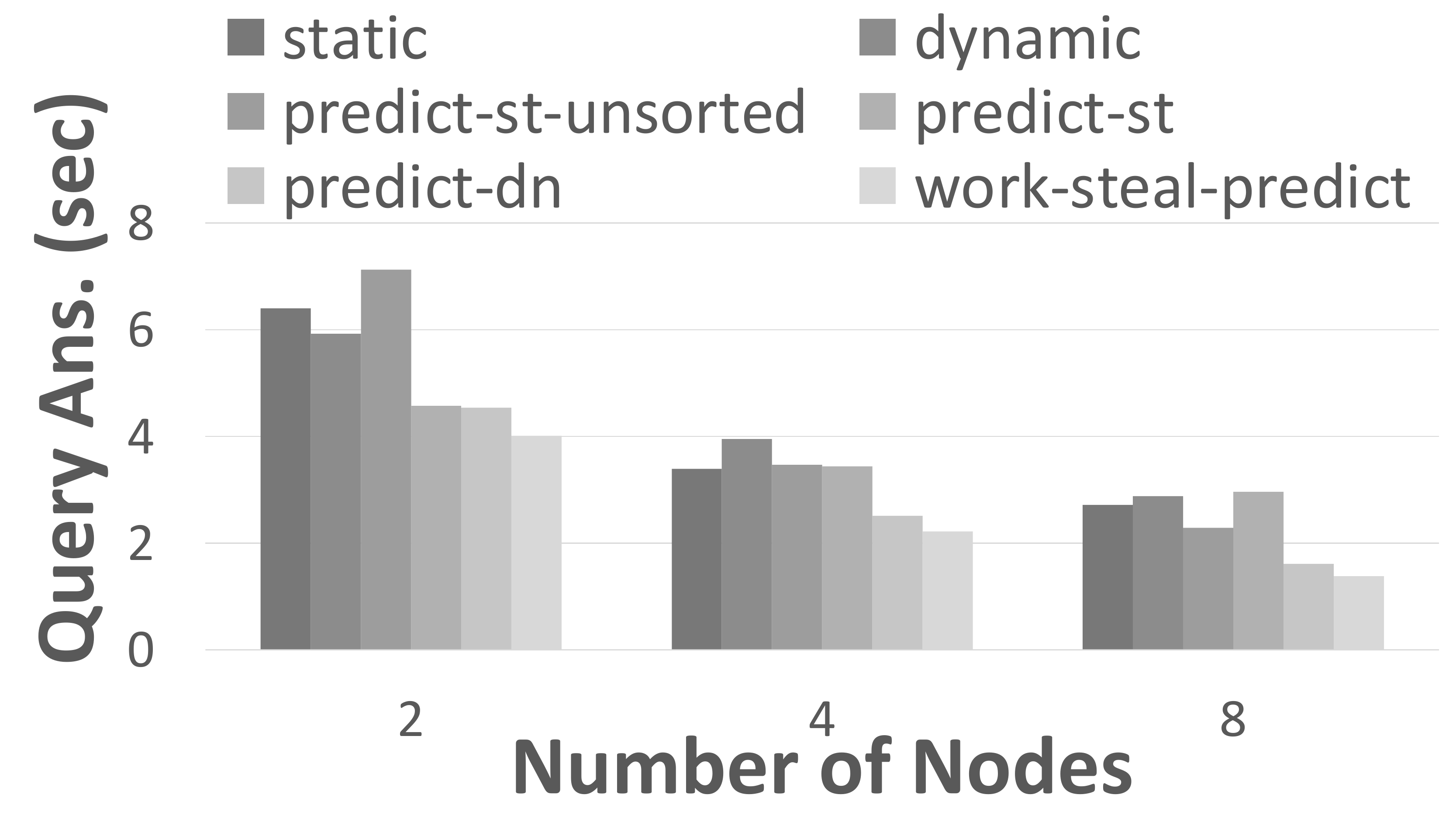}
		\caption{{\Partial-$2$ replication}}
		\label{fig:seismic_p2_100:a}
	\end{subfigure}
	\caption{{Odyssey's scheduling algorithms (Seismic).}}
	\label{fig:seismic_100:a}
\end{figure}

\noindent
{\bf Query scheduling.}
To compare Odyssey's queries scheduling algorithms, 
the full replication strategy
is selected, to avoid measuring any overheads 
resulting from the partial replicated strategies. Recall that scheduling algorithms 
can't be used together with the no replication strategies.
We experimented with both Random (synthetic dataset) and Seismic (real dataset), 
and all of our algorithms positively affected performance in comparison with \St.
Moreover, for the synthetic dataset, we have seen no remarkable differences between 
all our scheduling algorithms, since the randomness when producing the data series of both 
the dataset and the queries set, results in queries with almost the same effort to be answered.
We present the results for Seismic, where the effort for answering 
queries varies. 
Specifically, Figure~\ref{fig:seismic_100:a} shows that as the number of nodes increases,
\PD\ 
is the best scheduling policy in all cases and it is up to $150\%$
better than \St.

\noindent
{\bf Work-stealing.}
Figure \ref{fig:seismic_fdr_100:a} shows that \WSP\ greatly outperforms 
(up to almost 2x) \PD\ for large number of nodes when using \Full\ replication, i.e. our work-stealing technique 
positively affects performance on these cases.
{The same is true for \Partial-$2$ replication, but to a lesser extent.} 
Recall (from Section~\ref{sec:wsalg}) that this happens since all the algorithms that do not 
use the work-stealing technique suffer from load-imbalance issues. 
Specifically, when a query set contains a few (significantly less than the 
number of nodes) queries that require significantly more effort to get answered (than
the majority of queries), then as the number of nodes increases more nodes remain idle
at the end of the corresponding query answering phase, since no such difficult query
is assigned to them. 

\begin{figure}[tb]
	\centering
	\begin{subfigure} [b]{0.45\textwidth}
		\centering
		\includegraphics[width=\textwidth]{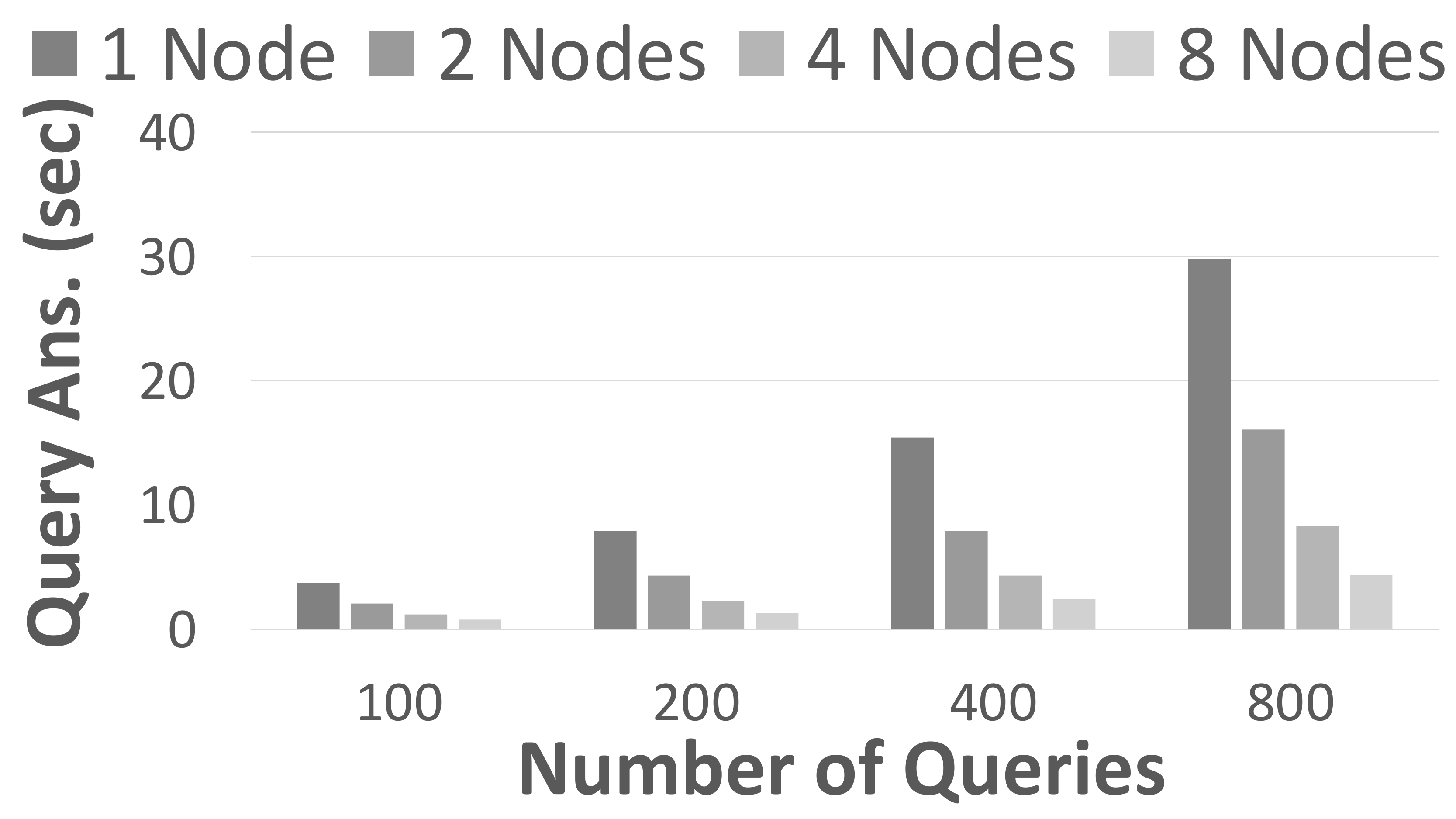} 
		\caption{\Full\ replication}
		\label{fig:queries-scalability-full}
	\end{subfigure}
	\begin{subfigure} [b]{0.45\textwidth}
		\centering
		\includegraphics[width=\textwidth]{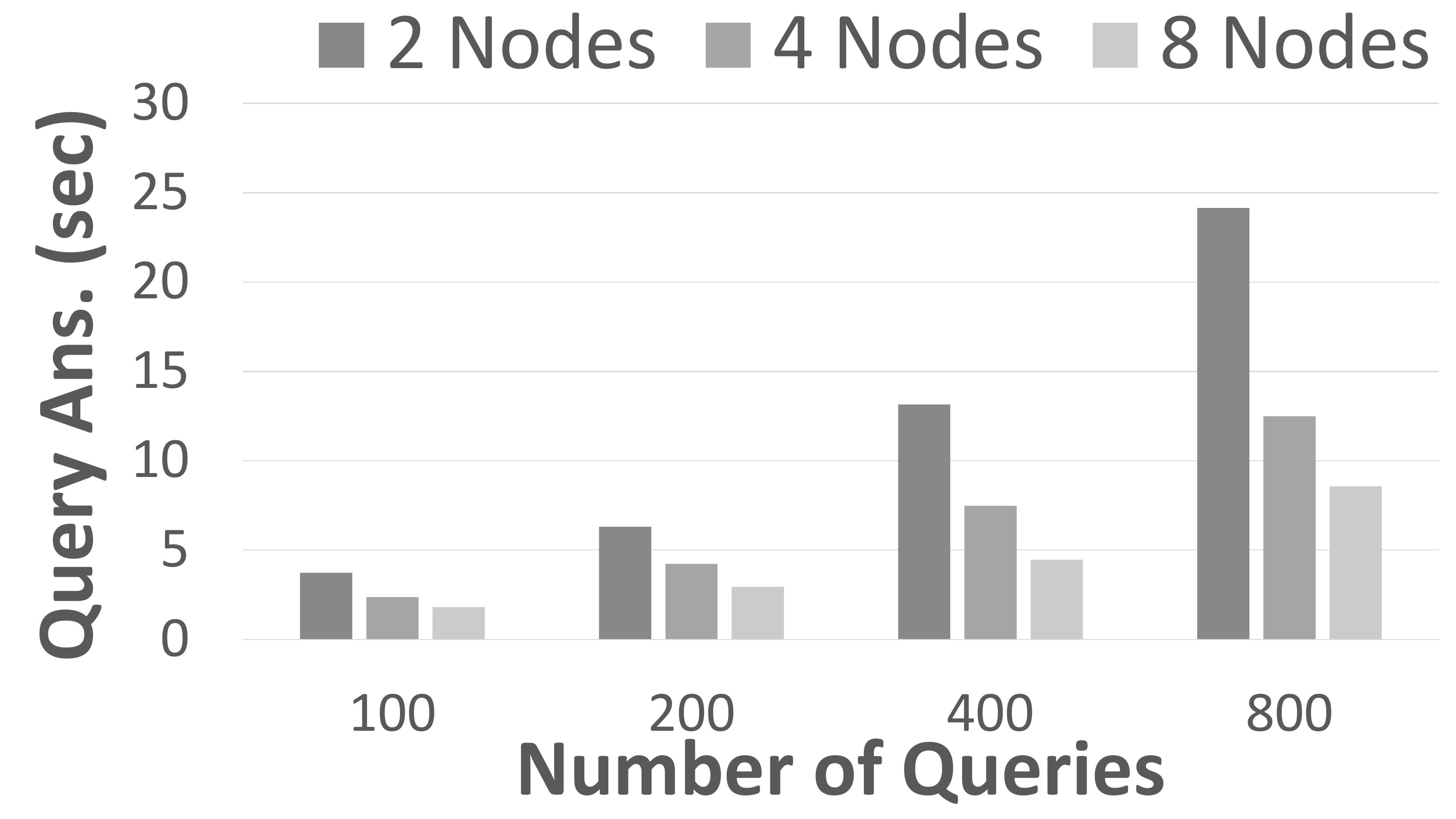}
		\caption{\Partial-$2$ replication}
		\label{fig:queries-scalability-partial}
	\end{subfigure}
	\caption{Query answering scalability as the number of queries increase (Random).}
\end{figure}

\begin{figure}[tb]
	\centering
	\begin{subfigure}[b]{0.45\textwidth}
		\includegraphics[width=\textwidth]{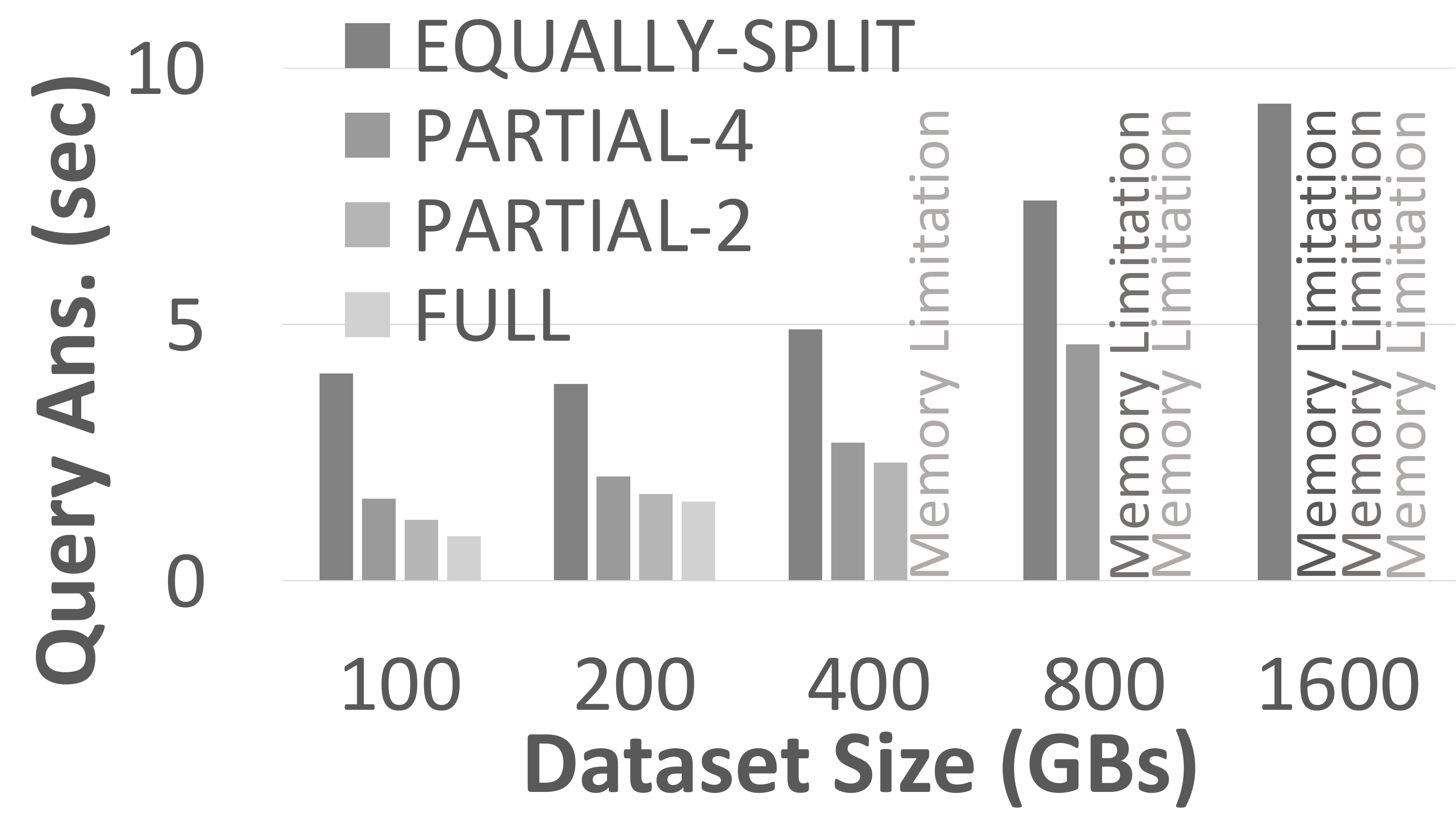}
		\caption{{Random}}
		\label{fig:random_scalability}
	\end{subfigure}
	\begin{subfigure}[b]{0.45\textwidth}
		\includegraphics[width=\textwidth]{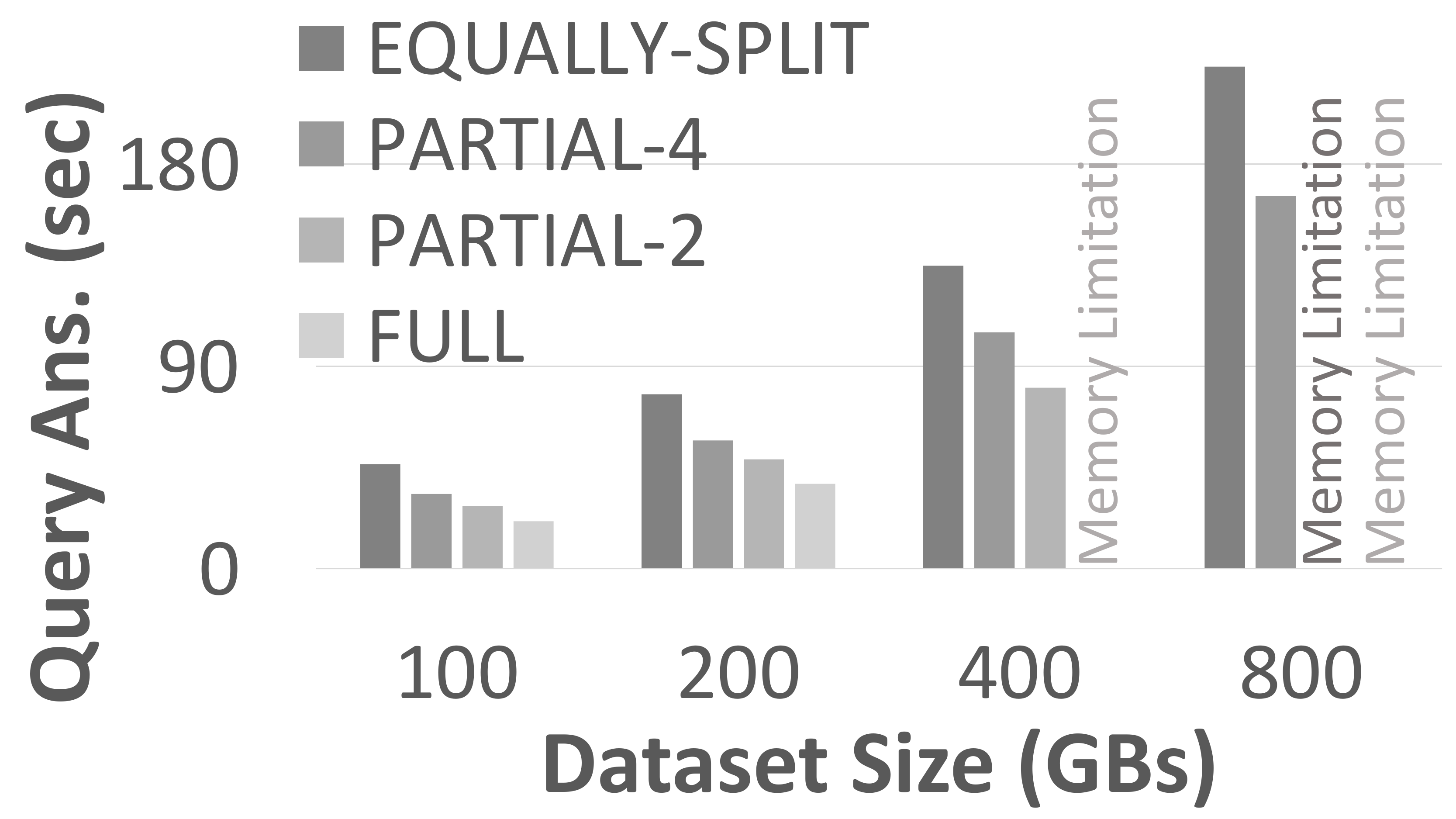}
		\caption{{Yan-TtI}}
		\label{fig:yandex_scalability}
	\end{subfigure}
	\caption{{Query time for 100 queries vs data size (8 nodes).}}
	\label{fig:scalability}
\end{figure}

\begin{figure}[tb]
	\centering
	\begin{minipage}[b]{0.45\textwidth}
	\centering
	\includegraphics[width=\textwidth]{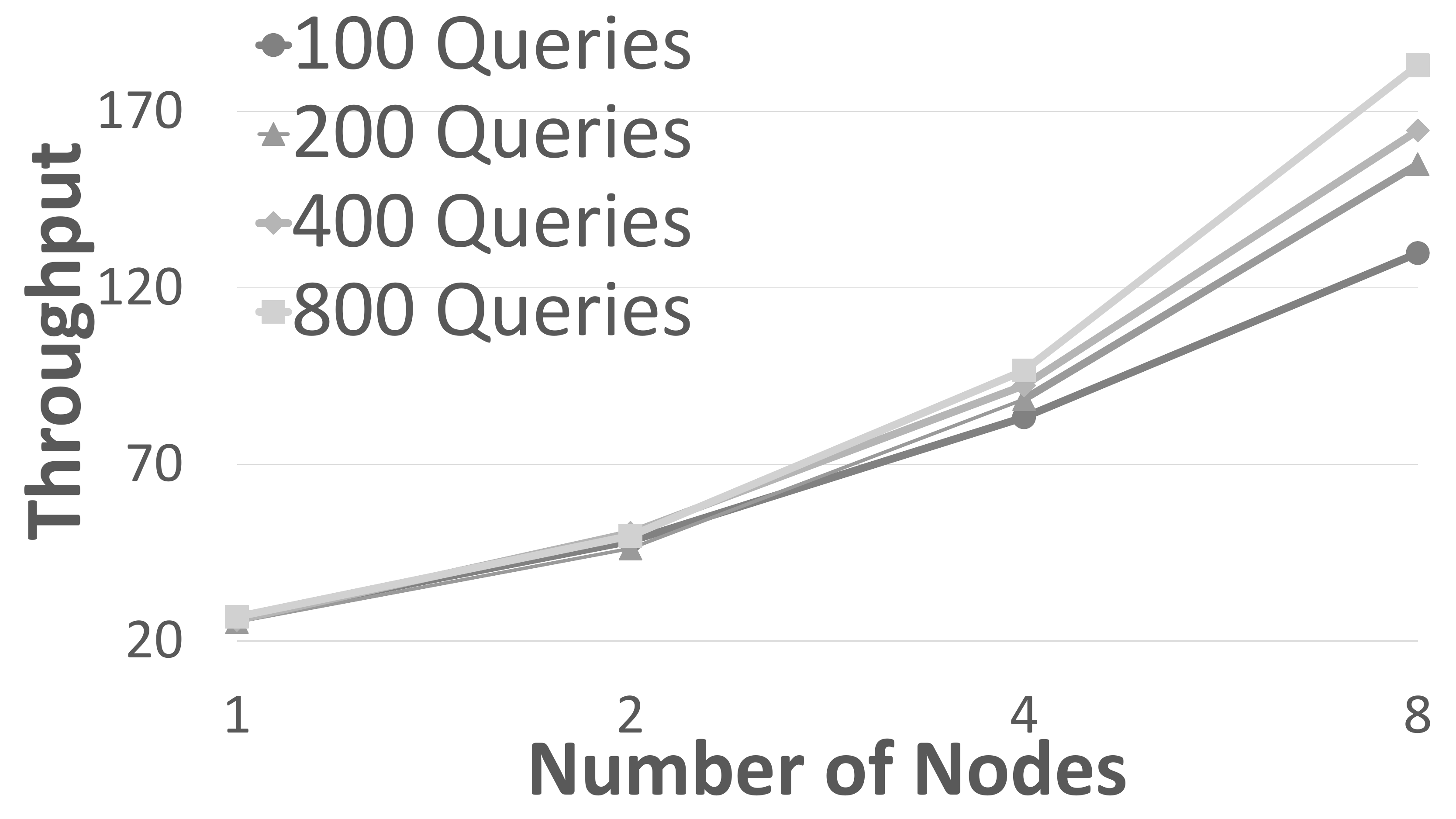}
	\caption{Query throughput (Random, \Full\ replication).}
	\label{fig:queries-throughput}
	\end{minipage}
	\begin{minipage}[b]{0.45\textwidth}
	\centering
	\includegraphics[width=\textwidth]{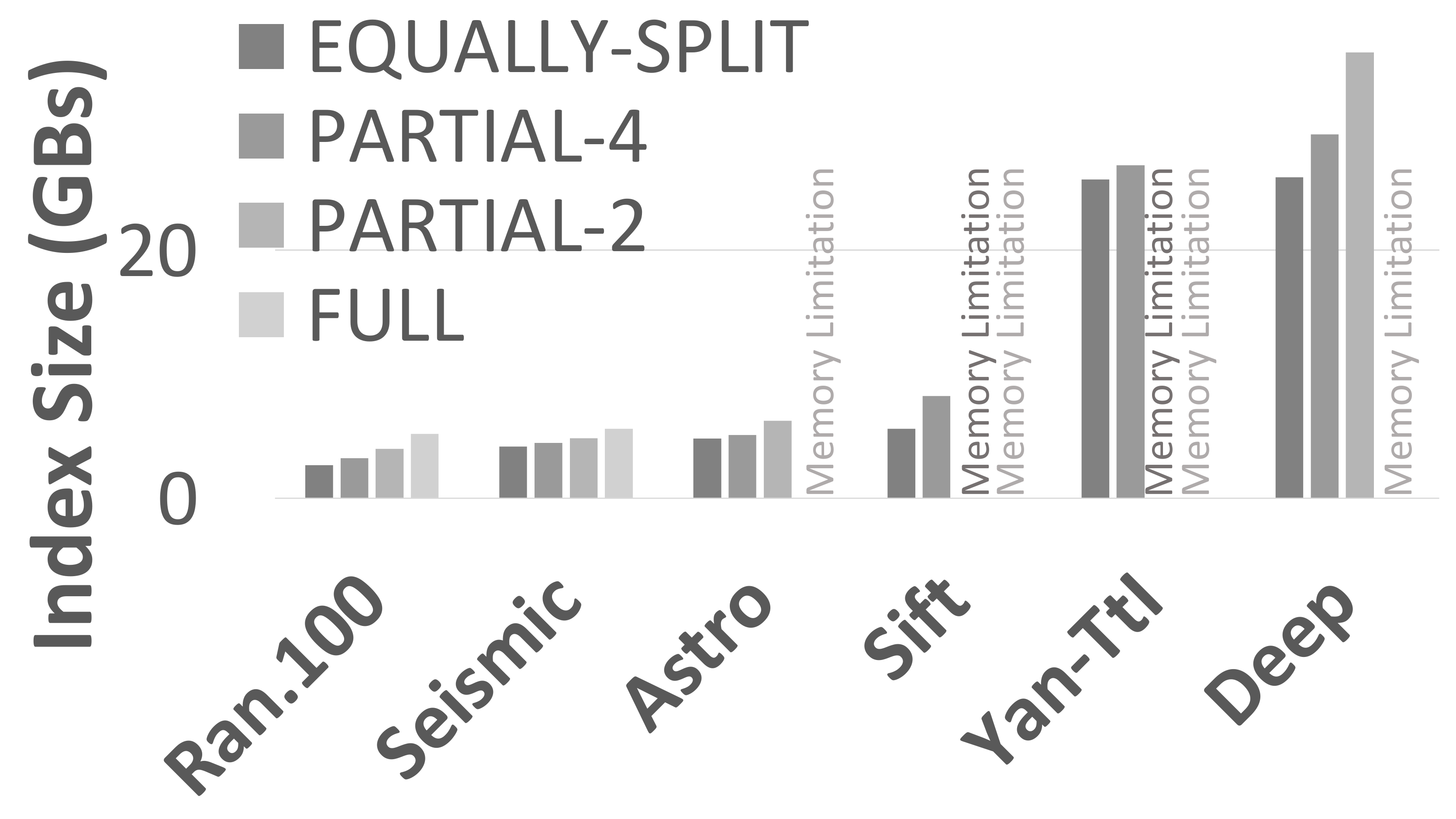}
	\caption{{Index size.}}
	\label{fig:index_size}
	\end{minipage}
\end{figure}

\noindent{\bf Query Scalability.}
To evaluate the scalability of Odyssey's algorithms with increasing number of 
queries, we conducted experiments with \WS\ using synthetic 
and real 
datasets.
In Figure~\ref{fig:queries-scalability-full}, we present the results for the Random dataset (results with the other tested datasets are similar)
with \Full\ replication, for a total of $100$, $200$, $400$, and $800$ queries. 
As we can see, \WS\ scales almost perfectly with the
increasing number of queries, since the time to execute $100$ queries in $1$ node
is the same with the time to execute $j * 100$ queries in $j$ nodes, $j \in \{ 2, 4, 8\}$.
We have observed the same trend for the \Partial\ scheduling algorithms
(Figure~\ref{fig:queries-scalability-partial}). 
Note that \Partial\ replication
can be applied only with two or more nodes.
{Additionally, we present in Figure~\ref{fig:scalability} scalability experiments, by increasing the dataset size, for Random (between 100-1600GB) and Yan-TtI (between 100-800GB). 
We measure the total query answering time for 100 queries, when using 8 nodes. 
Note that we could not execute all replication strategies for all dataset sizes, due to the memory capacity of our nodes. 
The results show that query answering time scales gracefully as we increase the dataset size, while increasing the replication degree leads to better performance.}
Moreover, we observe that Odyssey's query answering algorithm achieves good scalability 
as the number of nodes increases. 
This is better illustrated in Figure~\ref{fig:queries-throughput},
which presents the \WS\ throughput on the Random dataset. 

\begin{figure}[tb]
	\centering
	\begin{subfigure} [b]{0.24\textwidth}
		\centering
		\includegraphics[width=\textwidth]{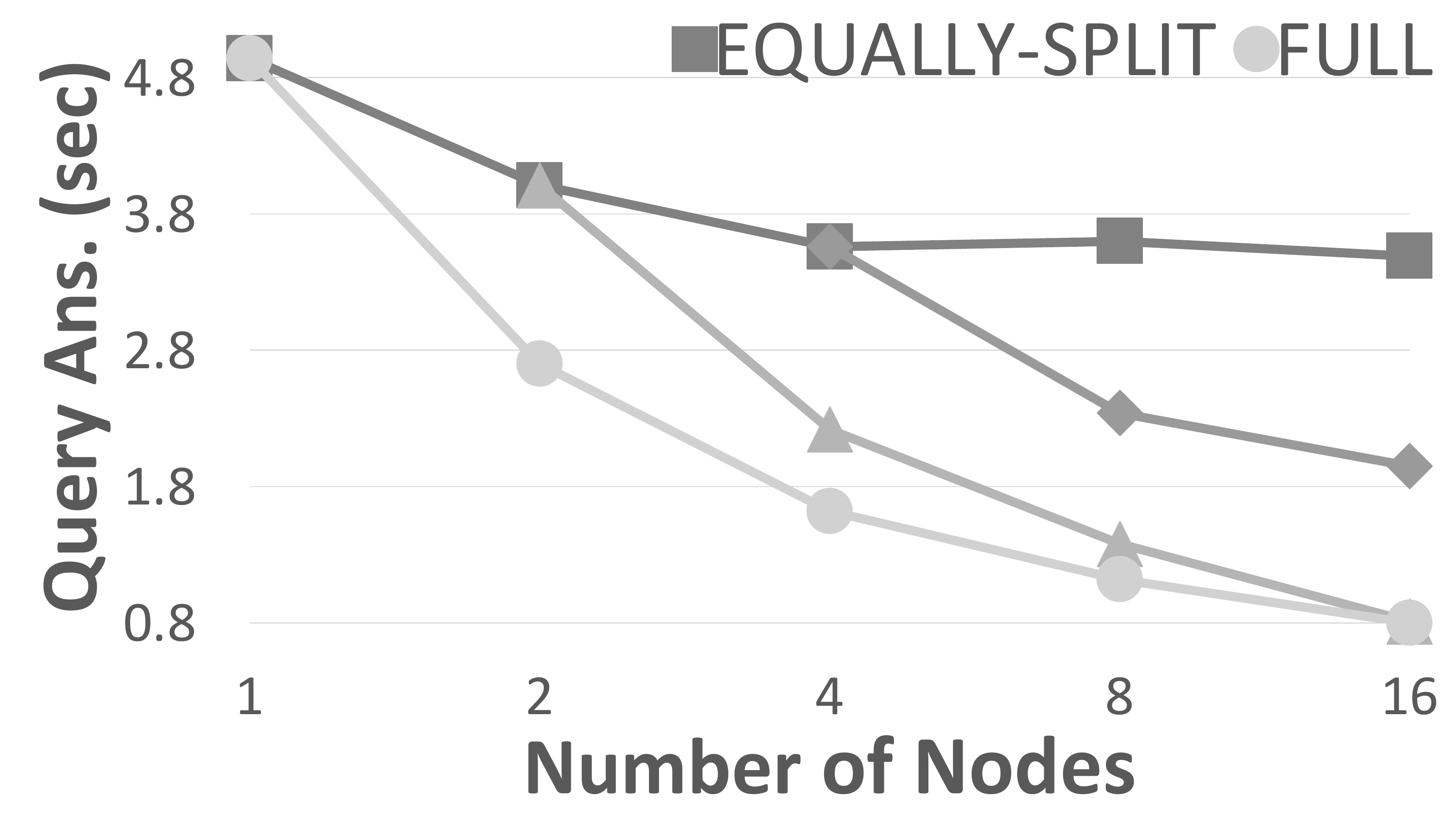}
		\caption{100 Queries}
		\label{fig:seismic_pdr_qa:100}
	\end{subfigure}
	\begin{subfigure} [b]{0.24\textwidth}
		\centering
		\includegraphics[width=\textwidth]{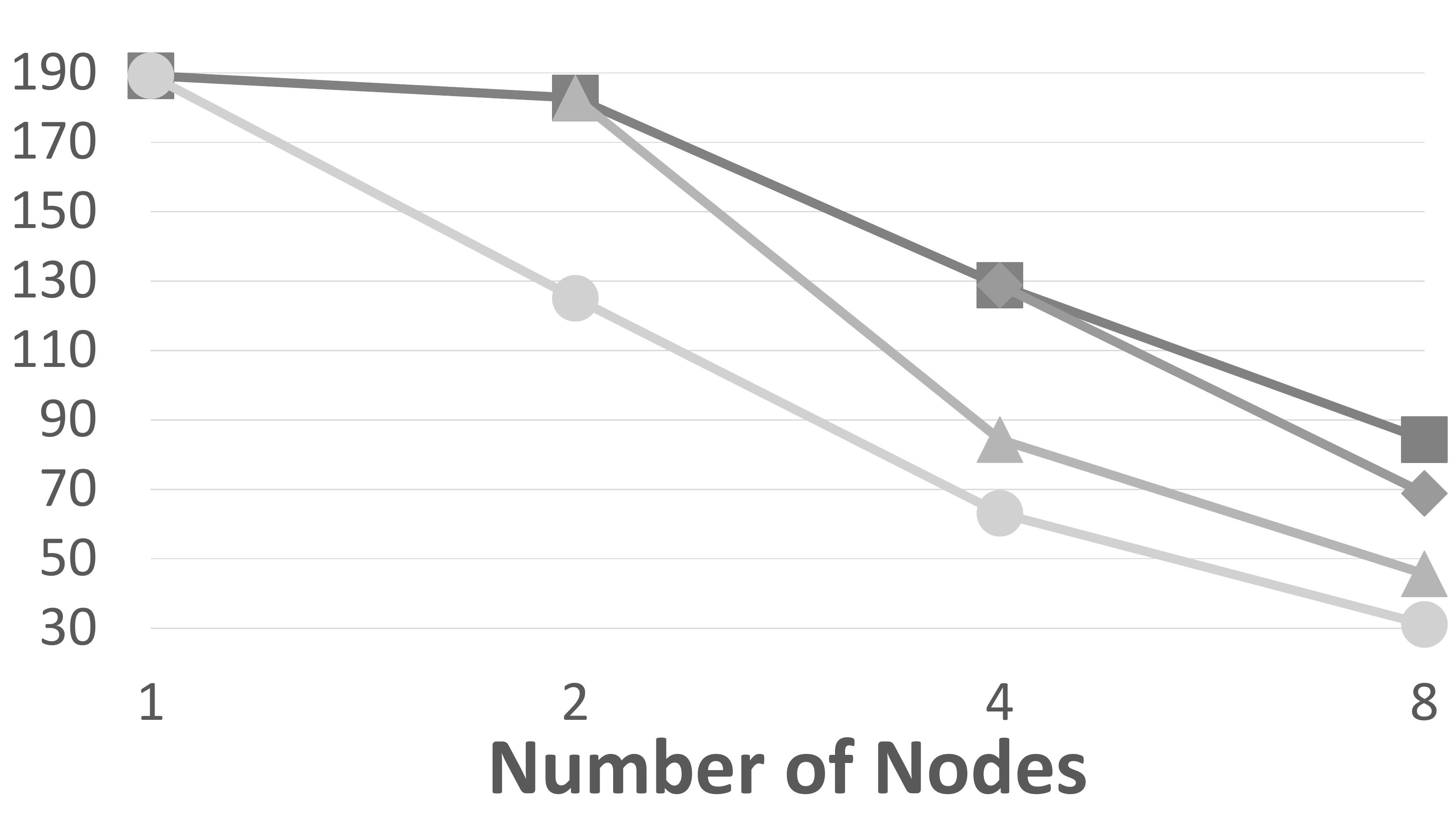}
		\caption{800 Queries}
		\label{fig:seismic_pdr_qa:800}
	\end{subfigure}
	\begin{subfigure} [b]{0.24\textwidth}
		\centering
		\includegraphics[width=\textwidth]{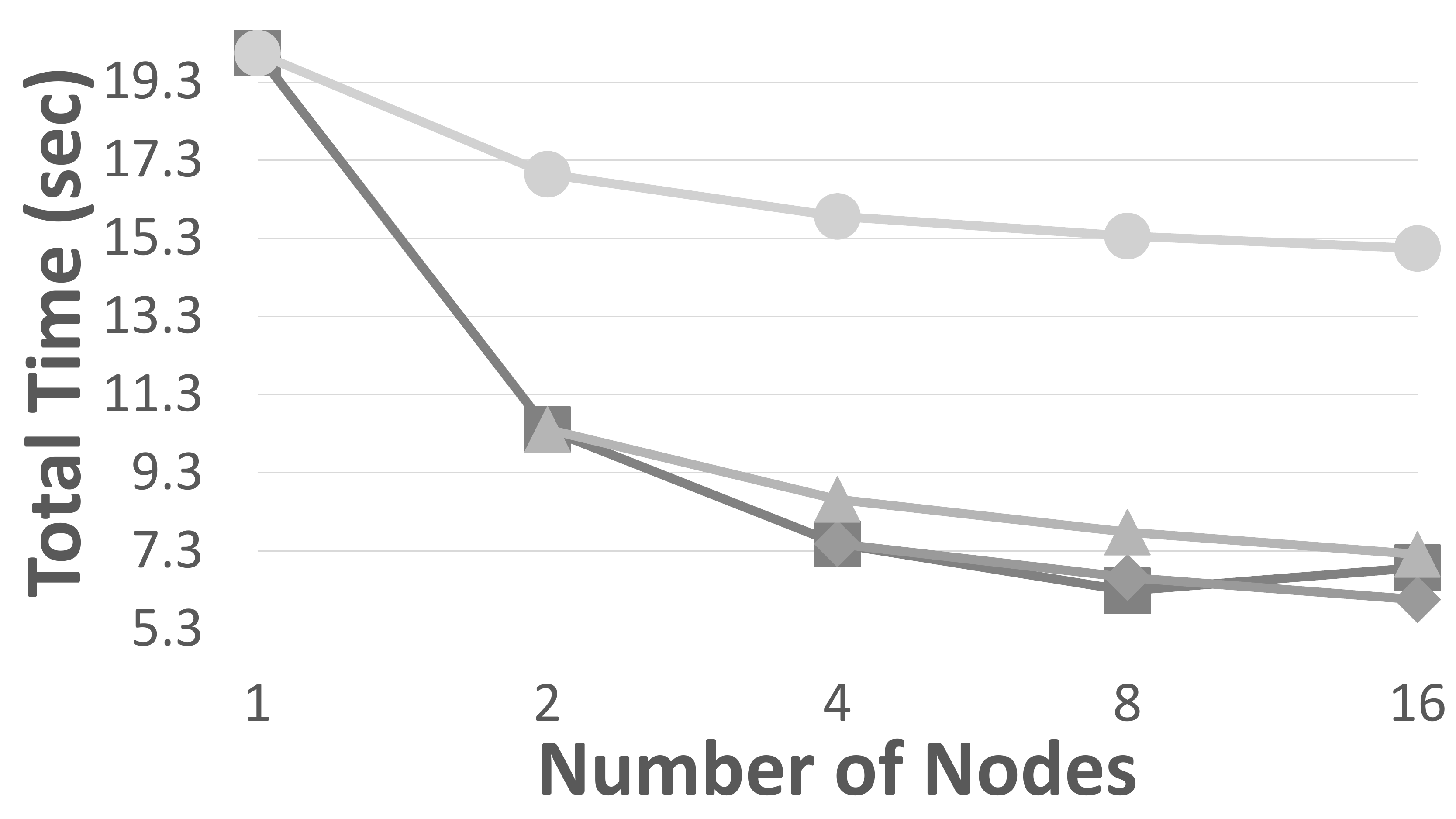}
		\caption{100 Queries}
		\label{fig:seismic_pdr_total:100}
	\end{subfigure}
	\begin{subfigure} [b]{0.24\textwidth}
		\centering
		\includegraphics[width=\textwidth]{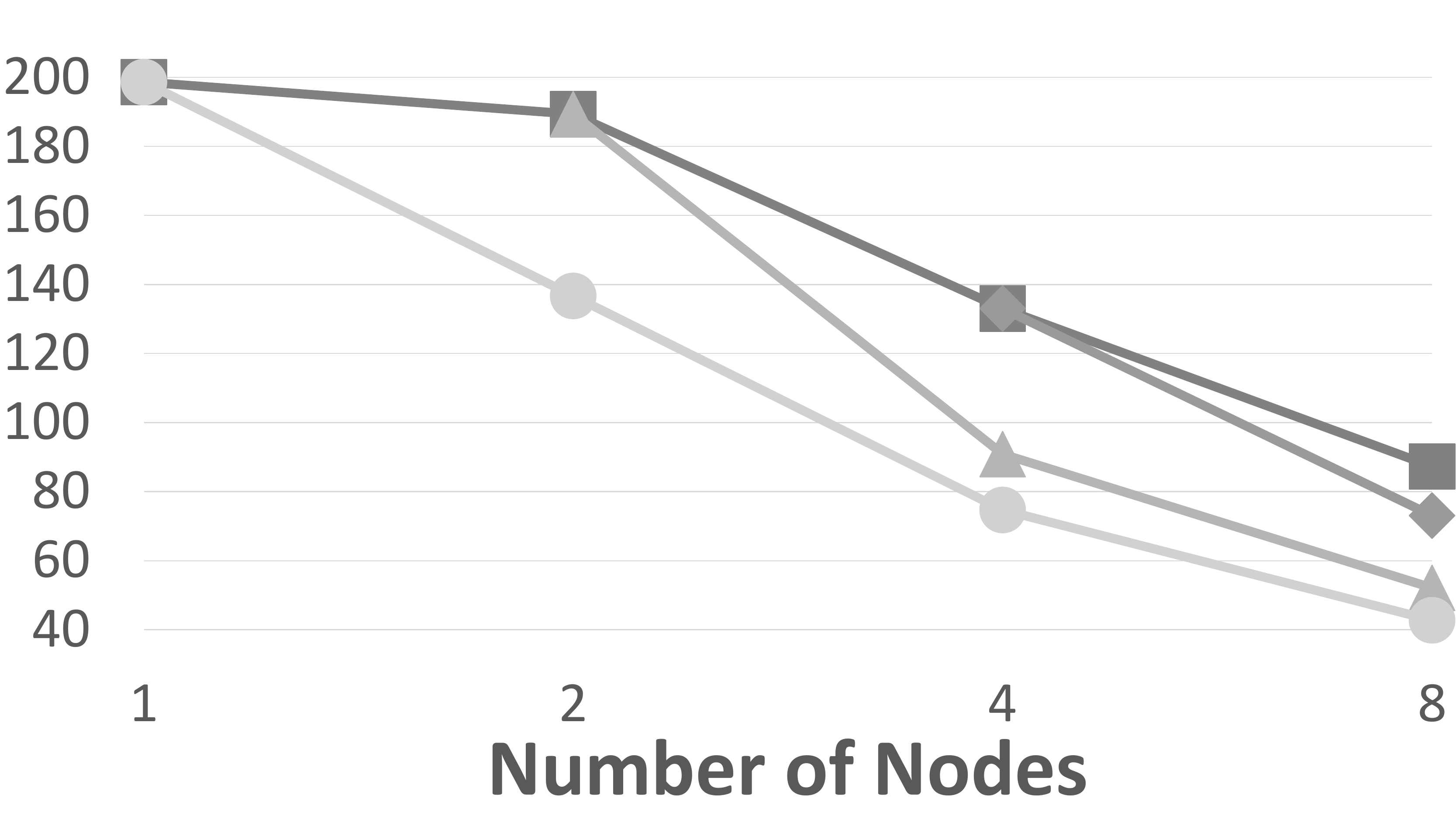}
		\caption{800 Queries}
		\label{fig:seismic_pdr_total:800}
	\end{subfigure}
	\caption{Comparison of Odyssey's replication strategies, using \WSP\ with Seismic.}
	\label{fig:seismic_pdr_total}
\end{figure}

\begin{figure*}[tb]
	\centering
	\begin{subfigure} [b]{0.24\textwidth}
		\centering
		\includegraphics[width=\textwidth]{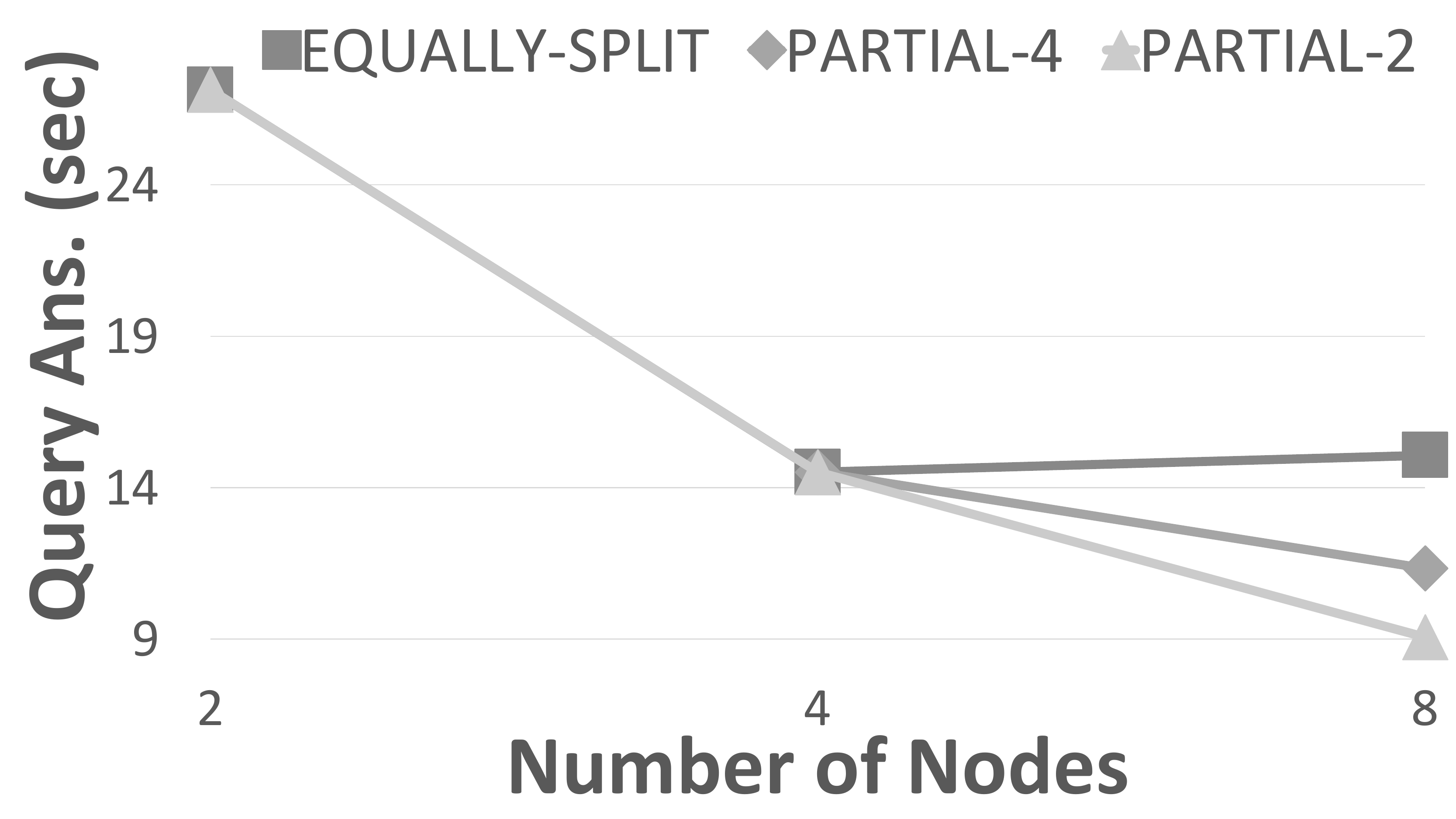}
		\caption{Astro}
		\label{fig:astro_qa_100}
	\end{subfigure}
	\begin{subfigure} [b]{0.24\textwidth}
		\centering
		\includegraphics[width=\textwidth]{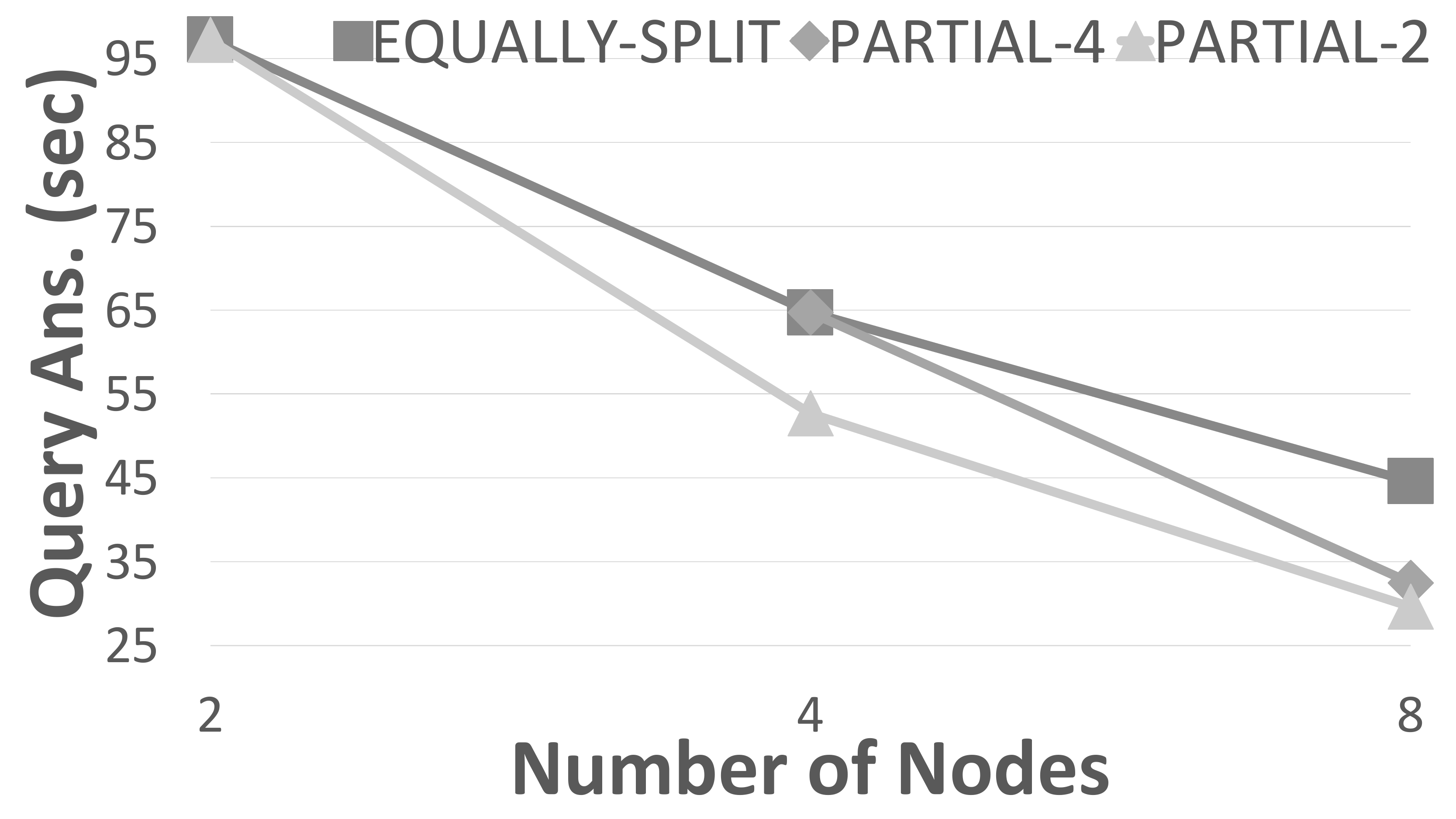}
		\caption{Deep}
		\label{fig:deep_qa_100}
	\end{subfigure}
	\begin{subfigure} [b]{0.24\textwidth}
		\centering
		\includegraphics[width=\textwidth]{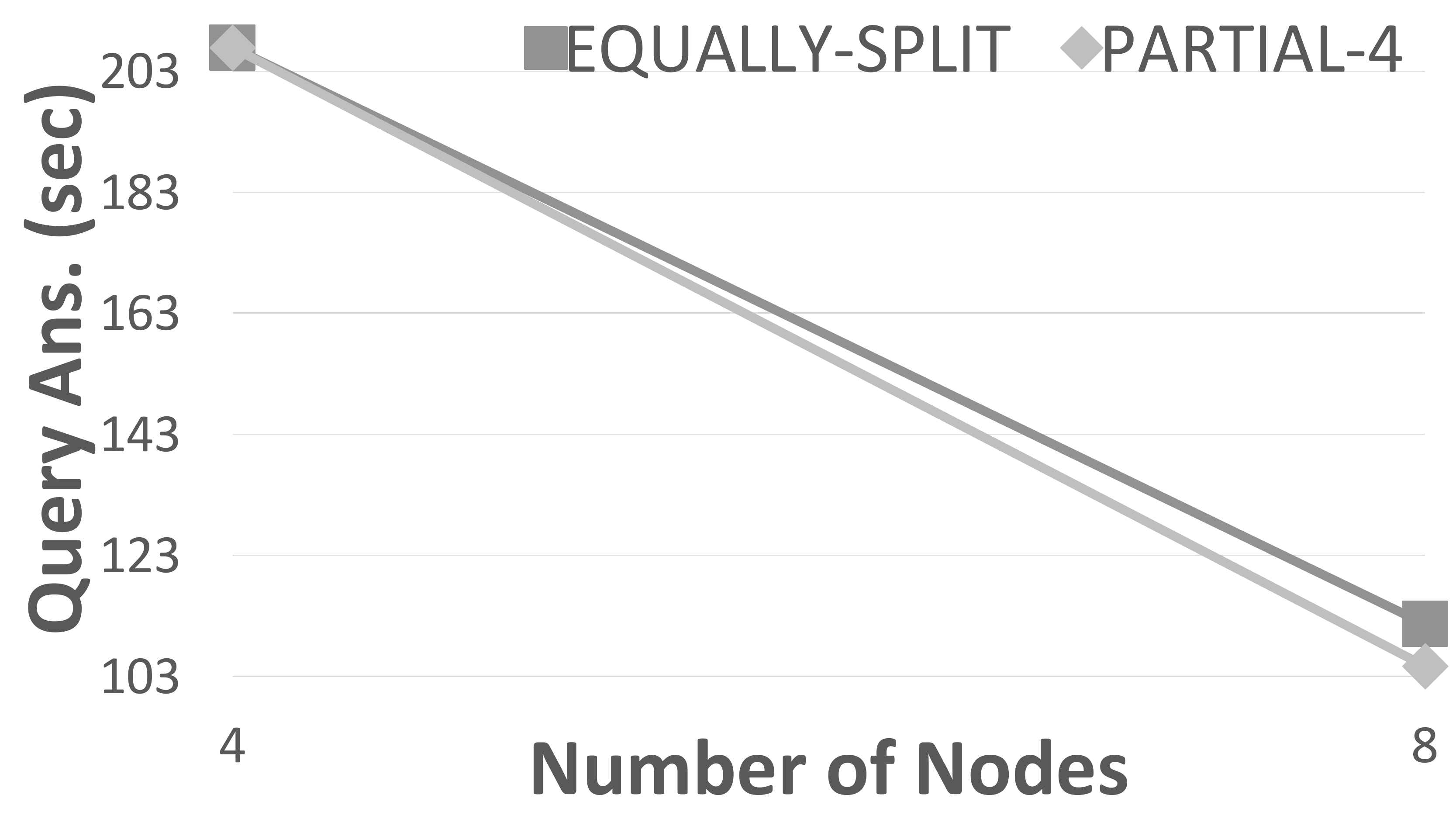}
		\caption{Sift}
		\label{fig:sift_qa_100}
	\end{subfigure}
	\begin{subfigure} [b]{0.24\textwidth}
		\centering
		\includegraphics[width=\textwidth]{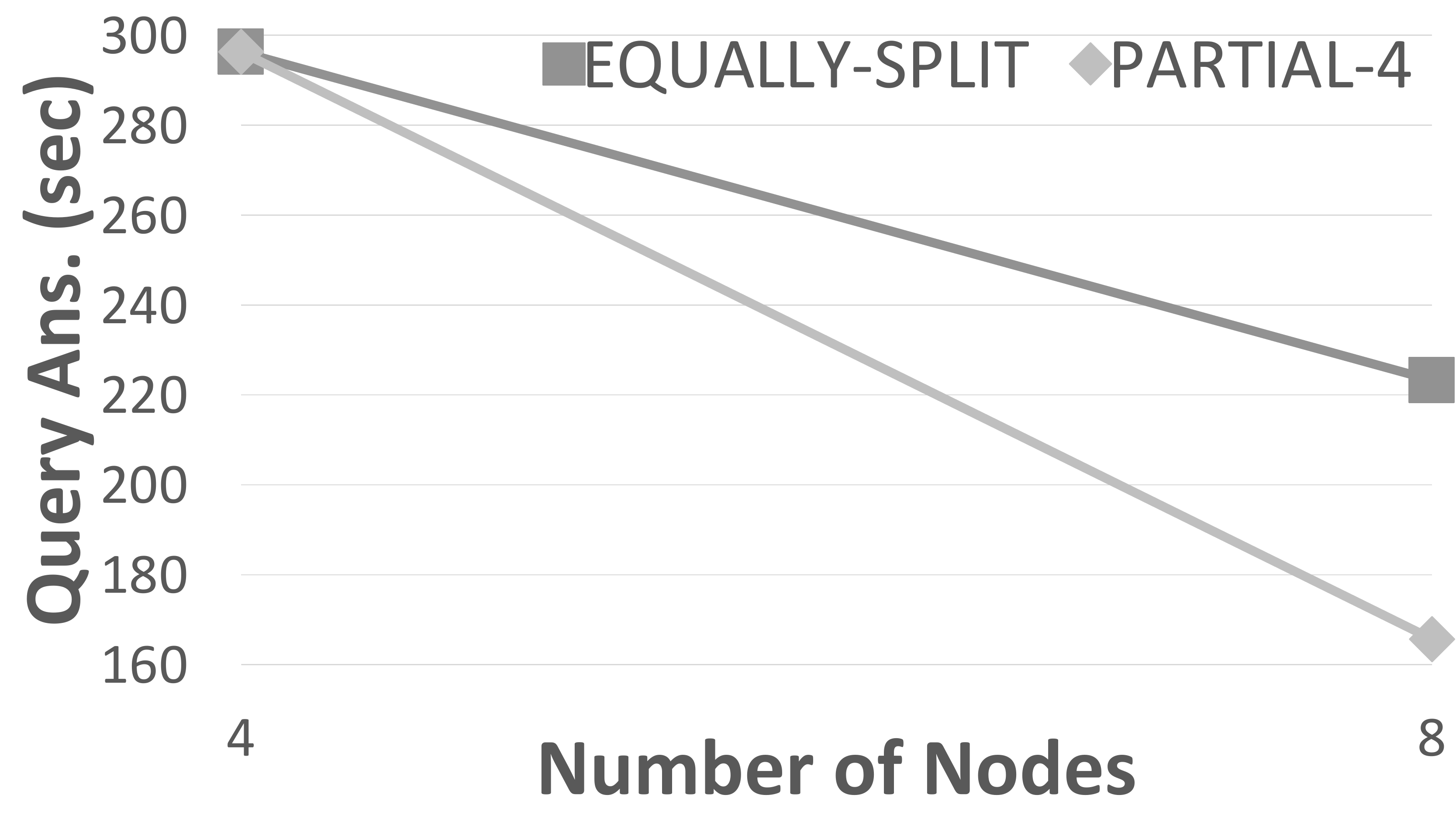}
		\caption{Yan-TtI}
		\label{fig:yandex_qa_100}
	\end{subfigure}
	\caption{{Comparison of Odyssey's replication strategies, using \WSP\ with real datasets, using 100 queries.}}
	\label{fig:real_partial_data_replication}
\end{figure*}

\noindent{\bf Replication.} 
We study now Odyssey's different replication strategies using 
the Seismic dataset and \WSP\ that is our best scheduling algorithm, to avoid any 
overhead incurred by load-imbalances between nodes.
Specifically, we test \ES, 
{\Partial-$4$}, {\Partial-$2$} and \Full, 
for varying number of queries.
Figures~\ref{fig:seismic_pdr_qa:100}-\ref{fig:seismic_pdr_qa:800} present the query answering time\footnote{We report results with 16 nodes only for the small workload, 
because the scheduler of our cluster does not allow long-running jobs on more than 8 nodes.},
where we observe that the more a dataset is replicated, the less time is required to answer queries, and
this is consistent for all number of queries. So, the \Full\ replication strategy has the smaller queries 
answering time. 
On the other hand, Figures~\ref{fig:seismic_pdr_total:100}-\ref{fig:seismic_pdr_total:800} present the 
total execution time, which 
includes also the time for index tree construction. 
Interestingly, for small query numbers ($100$), 
we observe exactly the opposite:
a larger amount of data replication, results in bigger total 
time, with \Full\
having now the bigger index tree construction time.
This happens because the increased index tree construction time dominates in the total time.
However, as the number of queries increases, the differences between the total execution time
of algorithms become smaller. Remarkably, for large enough number of queries (e.g., $800$),
the increased index tree construction cost is amortized by the smaller query answering time,
having \Full\ replication strategy performing better than \ES.
This analysis reveals an interesting trade-off (regarding the level of replication) 
between the query answering cost and the index tree construction cost, while the latter 
can be amortized using a large enough set of queries.
{Figure~\ref{fig:real_partial_data_replication} shows the results of the query answering experiment with 100 queries for the rest of the real datasets. 
We observe similar trends to those of Seismic (Figure~\ref{fig:seismic_pdr_qa:100}).}
{Overall, when query answering needs to be optimized, we recommend that Odyssey is used with the highest possible replication degree (given the dataset size and compute-cluster characteristics). 
}

\begin{figure}[tb]
	\centering
	\begin{subfigure} [b]{0.224\textwidth}
		\centering
		\includegraphics[width=\textwidth]{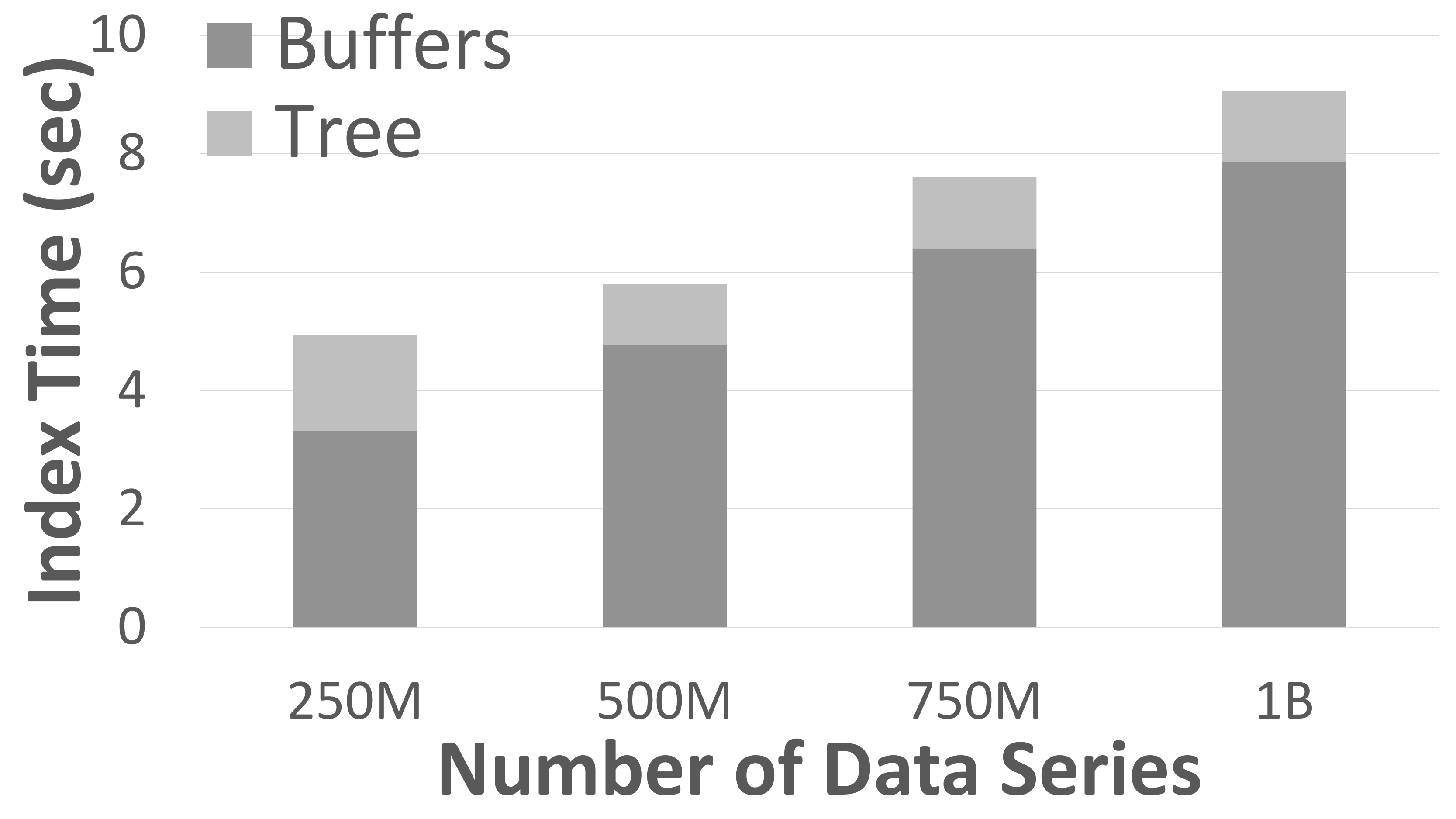}
		\caption{}
		\label{fig:index_scalability}
	\end{subfigure}
	\begin{subfigure} [b]{0.24\textwidth}
		\centering
		\includegraphics[width=\textwidth]{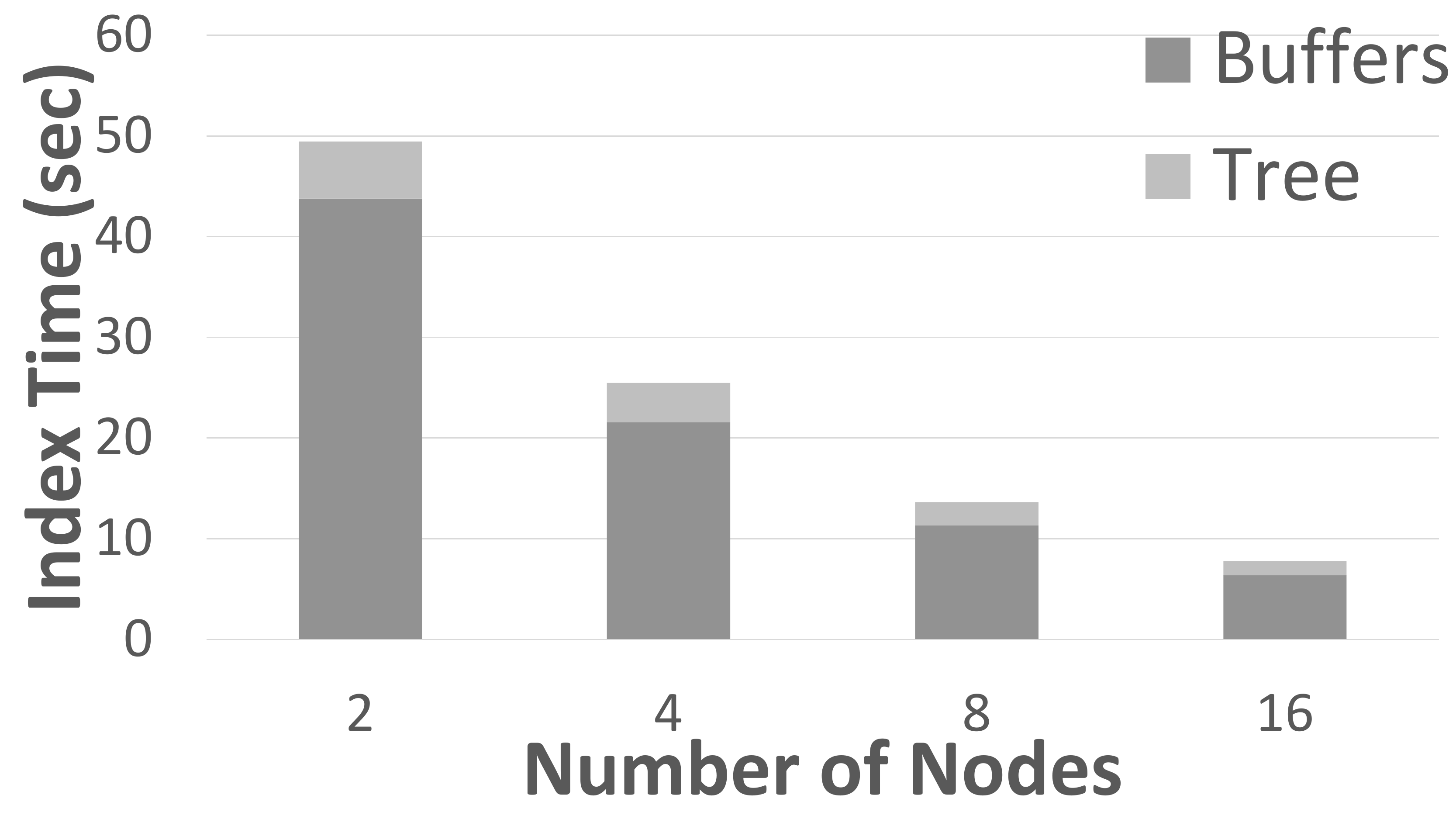}
		\caption{}
		\label{fig:index_creation}
	\end{subfigure}
	\begin{subfigure} [b]{0.24\textwidth}
		\centering
		\includegraphics[width=\textwidth]{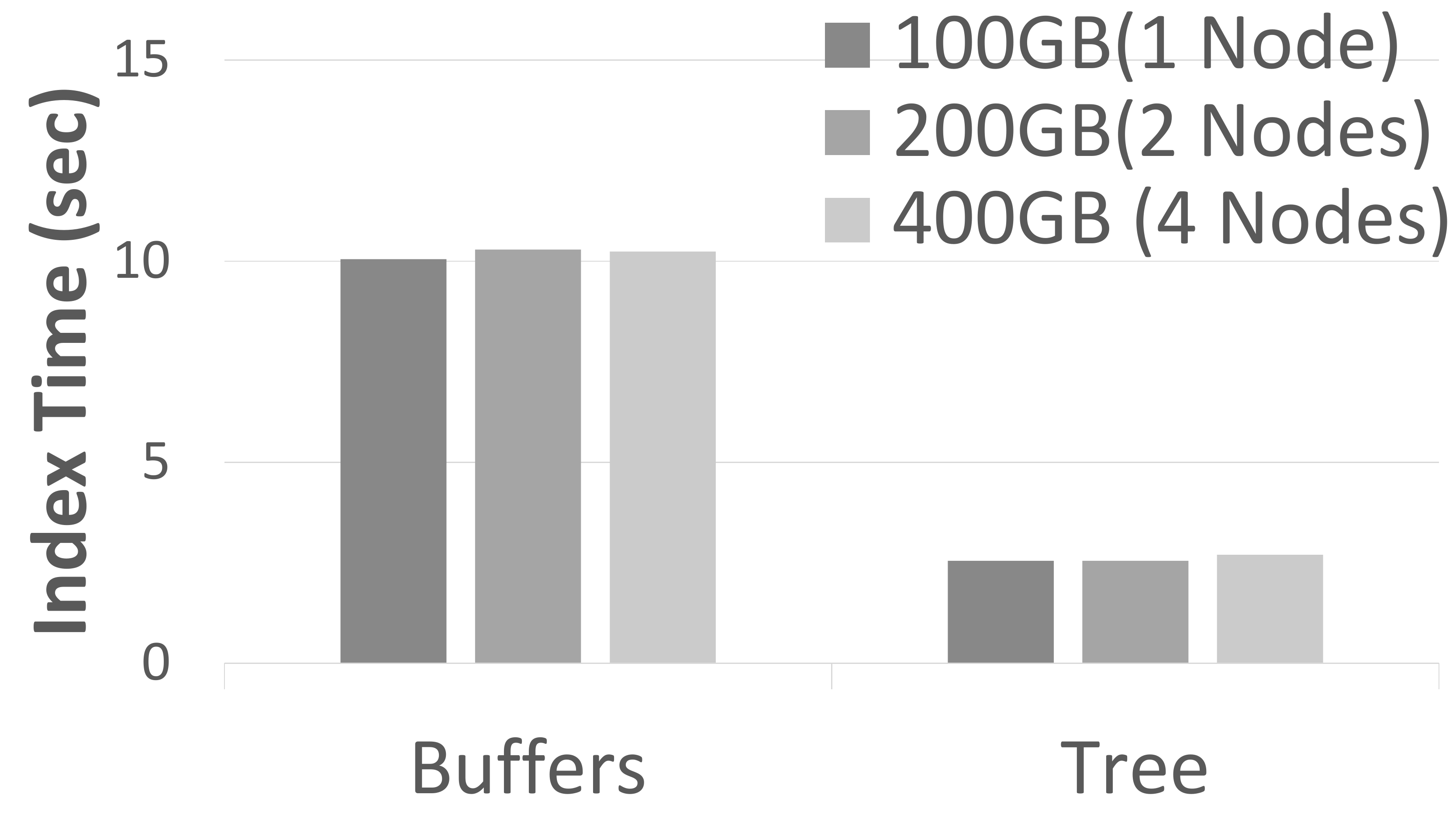}
		\caption{}
		\label{fig:random_index_scal}
	\end{subfigure}
\begin{subfigure} [b]{0.24\textwidth}
	\centering
	\includegraphics[width=\textwidth]{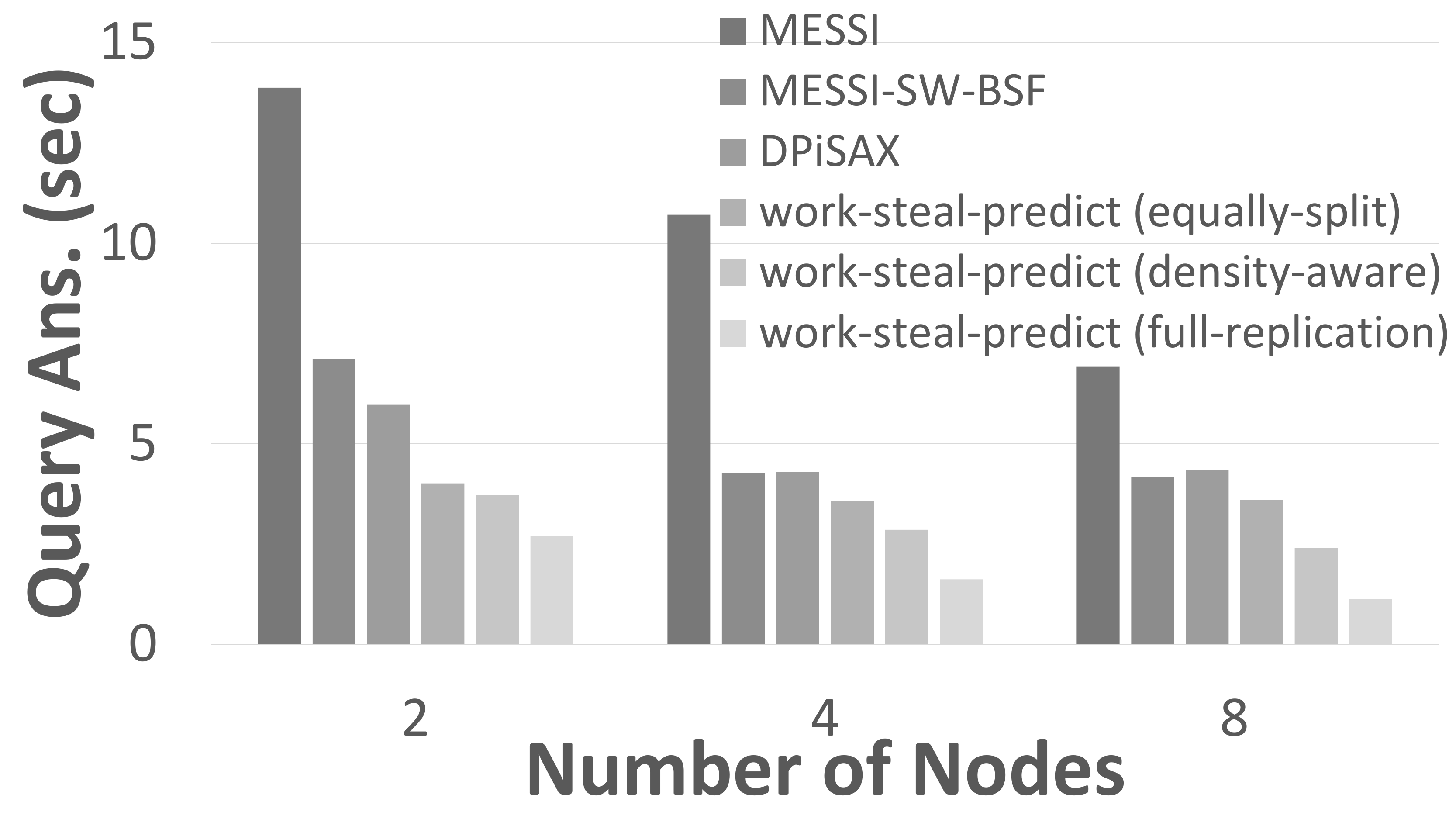}
	\caption{}
	\label{fig:random_others}
\end{subfigure}
	\caption{
	(a) Index scalability on Deep using \ES, as the dataset size increases, with $16$ nodes.
	(b) Index scalability on Deep using \ES, as the number of nodes increases. 
	(c) Index scalability on the Random dataset as both the dataset size and the number of 
		nodes increase linearly, using \ES. 
	(d) Comparison of \WSP\ with Odyssey's different data partitioning schemes, against 
		other implementations, using Seismic.}
	\label{fig:todo}
\end{figure}

\remove{
In the previous set of experiments, we showed that using these work-stealing methods could 
offer a load-balanced query answering scheme. But both the scheduling algorithms and the 
work-stealing algorithm are based on the fact that all the nodes have build the same iSAX index, 
thus having the same data. This assumption could be feasible for some systems that the memory 
capacity is high enough, but we cannot oversee the drawbacks of this scheme. 
We extended our algorithms to be able to operate with less memory requirements, 
by creating sets of nodes that have the same data, and applying the ideas we presented 
earlier to each node set (or node group) separately. Although this idea has trade-offs 
regarding the Query Answering time (because now every node group has to communicate 
with other to find the best answer for a query), it manages to keep the nice performance 
of the algorithm while also ensuring scalability for big datasets. Also, as a side-effect 
our scheme, which is a distributed partially replicated index, achieves a degree of speedup 
for index creation. In Figure~\ref{fig:todo} we present this trade-off, as we show both index 
creation and total query answering time for different replication rates, starting from no 
replication at all(meaning that each node gets a discrete part of the dataset), leading to 
full replication, as we presented above.
}

\noindent{\bf Index Scalability.}
{We present in Figure~\ref{fig:index_size} the total index size in GBs, for every replication strategy when using 8 nodes, for all real datasets we used and for Random 100GB (Ran.100). 
In all cases, the index size is very small compared to the size of the dataset.} 
Figures~\ref{fig:index_scalability} and~\ref{fig:index_creation} illustrate the index creation time
of Odyssey for our 1B series Deep dataset using \ES, as the dataset size increases on a system with $16$ nodes 
and as the number of nodes increases (while using the full size datasets), respectively.
In both cases, 
we observe optimal speedup regarding index creation. 
%
Additionally, Figure~\ref{fig:random_index_scal} presents the scalability of Odyssey on the Random dataset
as both the dataset size and the number of nodes increase linearly, again using \ES. 
As shown, Odyssey achieves
perfect scalability since the corresponding buffer times and index times remain almost constant. 

\noindent{\bf Data partitioning and comparison to competitors.}
Figure~\ref{fig:random_others} presents (i) a comparison of \WSP\, Odyssey's best performing algorithm, against 
\DMESSI, \DMESSIBSF, and \DPISAX; and (ii) the performance of Odyssey's different data partitioning 
schemes, i.e., \ES\ and \DA, as well as the \Full\ replication strategy, using Seismic.
Interestingly, \DMESSI\ performs significantly worse that all the other implementations, showing that 
by simply executing multiple instances of a SotA single-node algorithm like MESSI on a
multi-node system (in order to scale its applicability on larger dataset sizes) does not perform
well on real datasets; thus, more sophisticated approaches are required. 
On the other hand, Odyssey's \WSP\ with \Full\ replication strategy is significantly better than all 
its competitors. Specifically, it is up to $6.6$x, $3.7$x and $3.8$x faster than \DMESSI, \DMESSIBSF, and \DPISAX,
respectively.
Moreover, regarding Odyssey's data partitioning techniques, Figure~\ref{fig:random_others} shows that
\WSP\ with the \DA\ partitioning performs better than \ES.

\begin{figure}[tb]
	\centering
	\begin{minipage}[b]{0.45\textwidth}
	\raggedleft
	\includegraphics[width=\textwidth]{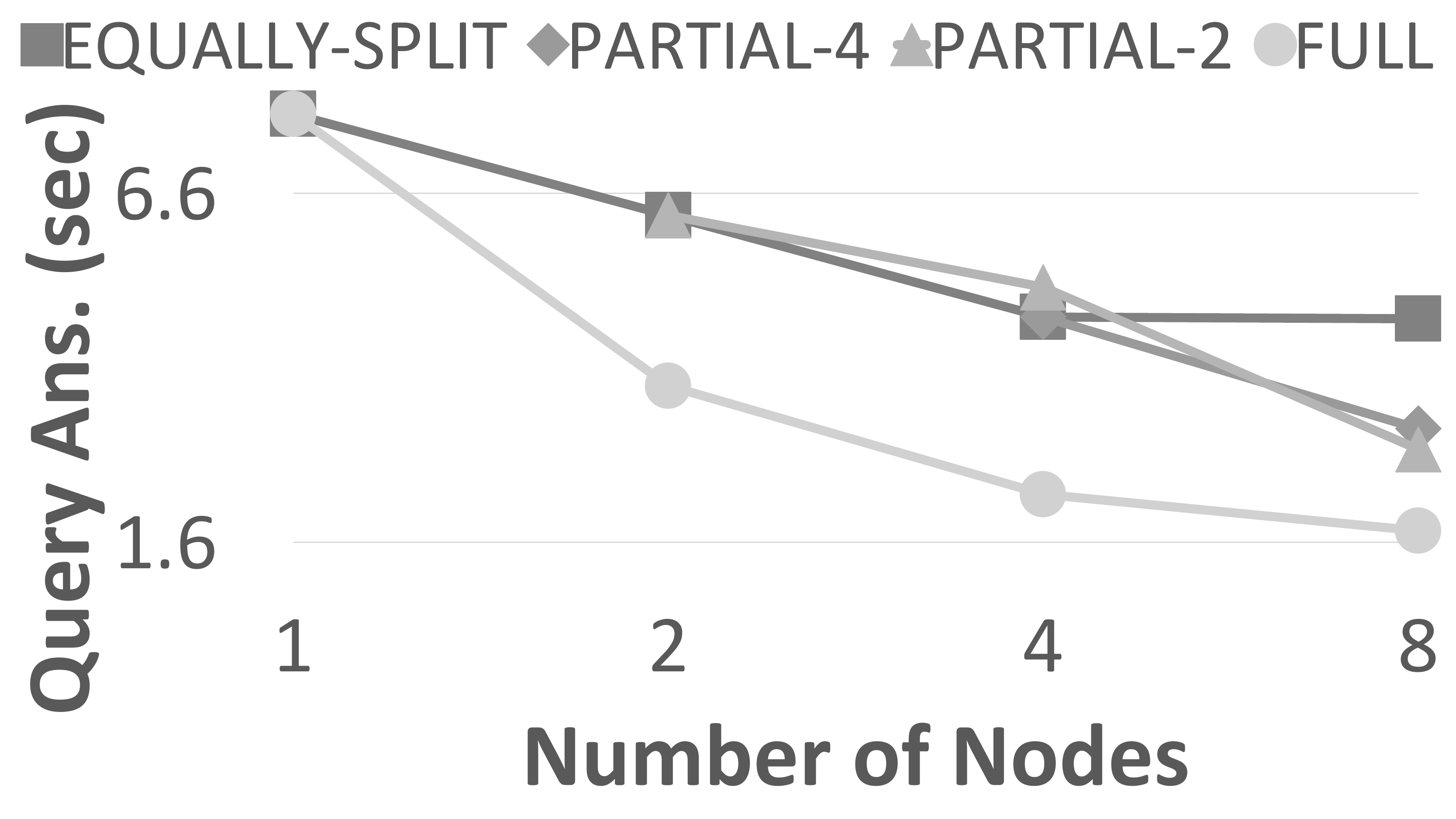}
	\caption{{\emph{10-NN} query \\answering (Random 100GB)}}
	\label{fig:knn}
	\end{minipage}
	\begin{minipage}[b]{0.45\textwidth}
	\raggedright
	\includegraphics[width=\textwidth]{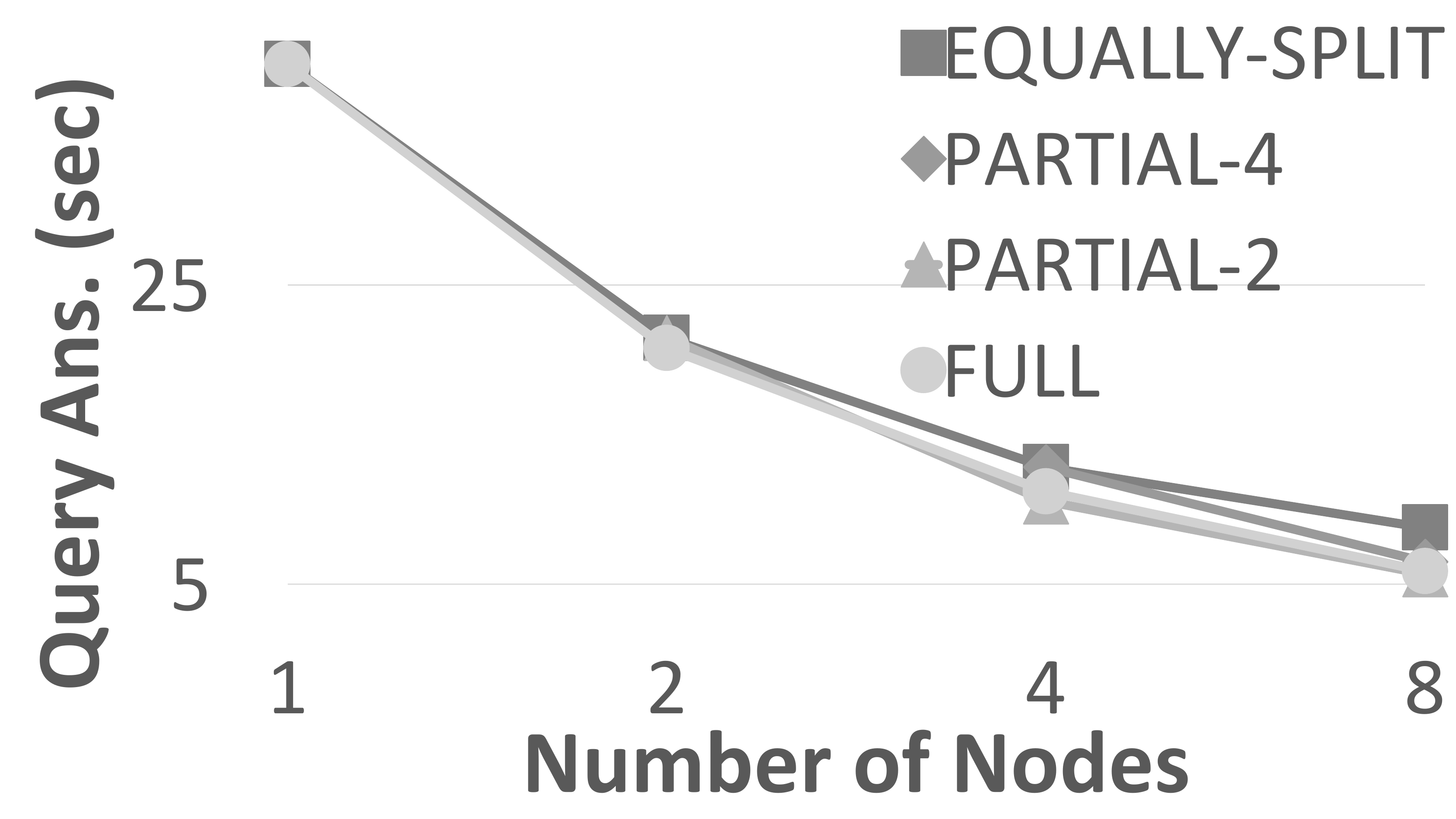}
	\caption{{DTW with 5\% warping (Random 100GB).}}
	\label{fig:dtw10_random}
	\end{minipage}
\end{figure}

{
\noindent{\bf Extensions to \emph{k-NN} and DTW.}
Finally, we present experiments with \emph{k-NN} queries, and the DTW distance, where we measure the query answering time for 100 queries as we increase the number of nodes, when using different replication strategies. 
We evaluated all replication strategies when varying $k$ between 1 and 20 for \emph{k-NN}, and when varying the warping window size between 1\%-15\% of the series length for DTW.
Figure~\ref{fig:knn} shows the \emph{k-NN} results for $k=10$, and Figure~\ref{fig:dtw10_random} shows the DTW results for 5\% warping (results with the rest of parameter values are similar). 
As expected, query answering times are in both cases higher than before, while using more nodes and higher replication degrees improves performance in the same way we have observed in previous experiments.
Results with Seismic exhibit similar trends and are omitted for brevity.
}

\section{Conclusions}
In this work, we presented Odyssey, a novel {\it distributed} data-series processing 
framework that takes advantage of the full computational capacity of modern clusters 
comprised of multi-core servers. 
%
Odyssey addresses a number of challenges in designing an efficient and 
highly-scalable {\it distributed} data series index, including efficient scheduling, load-balancing, and flexible
partial replication, and successfully  navigates the fundamental trade-off between data scalability 
and good performance during query answering. 
In future work, we plan to extend Odyssey 
to support subsequence similarity search~\cite{ulissejournal}, 
as well as approximate similarity search. 

\section*{Acknowledgments}
Work supported by NSFC Grant No. 62202450, EU Horizon 2020 Marie Sklodowska-Curie project No 101031688, and Hellenic Foundation for Research and Innovation (HFRI) under the ``Second Call for HFRI Research 
Projects to support Faculty Members and Researchers'' No 3684.
Numerical computations performed on the S-CAPAD/DANTE platform, IPGP, France.

\bibliographystyle{unsrtnat}
\bibliography{references,hydraref,edbttutorialref}

\end{document}